\documentclass[paper]{JHEP3} 

\usepackage{amsmath,amssymb}
\usepackage{aas_macros}

\newcommand{\SCX}{\ensuremath {\mathrm{SX}}}
\newcommand{\DCX}{\ensuremath {\mathrm{DX}}}
\newcommand{\NCX}{\ensuremath {\mathrm{NX}}}

\newcommand{\Tpi}{\ensuremath {E_{\pi}}}

\newcommand{\VEV}[1]{\ensuremath{\left\langle #1 \right\rangle}}
\newcommand{\Orderof}[1]{{\ensuremath{\mathcal{O}\!\left(#1\right)}}}
\newcommand{\sigmavof}[1]{\ensuremath{\langle \sigma_{{#1}}v \rangle}}
\newcommand{\me}{\ensuremath{m_{e}}}

\newcommand{\geff}{\ensuremath {g_{\mathrm{eff}}}}

\newcommand{\trit}{\ensuremath{\mathrm{T}}}
\newcommand{\Deut}{\ensuremath{\mathrm{D}}}
\newcommand{\deut}{\ensuremath{\mathrm{D}}}

\newcommand{\Hyd}{\ensuremath{\mathrm{H}}}

\newcommand{\he}[1]{\ensuremath{{}^{#1}\mathrm{He}}}

\newcommand{\be}[1]{\ensuremath{{}^{#1}\mathrm{Be}}}

\newcommand{\hef}{\ensuremath{{}^4\mathrm{He}}}
\newcommand{\het}{\ensuremath{{}^3\mathrm{He}}}
\newcommand{\lisx}{\ensuremath{{}^6\mathrm{Li}}}
\newcommand{\lisv}{\ensuremath{{}^7\mathrm{Li}}}
\newcommand{\bes}{\ensuremath{{}^7\mathrm{Be}}}

\newcommand{\fr}[2]{\ensuremath{\frac{#1}{#2}}}
\newcommand{\ie}{\textit{i.e.}}
\newcommand{\Kp}{\ensuremath{K^{+}}}
\newcommand{\Km}{\ensuremath{K^{-}}}
\newcommand{\Kpm}{\ensuremath{K^{\pm}}}
\newcommand{\Klong}{\ensuremath{K_{L}}}
\newcommand{\pip}{\ensuremath{\pi^{+}}}
\newcommand{\pim}{\ensuremath{\pi^{-}}}
\newcommand{\pipm}{\ensuremath{\pi^{\pm}}}
\newcommand{\pizero}{\ensuremath{\pi^{0}}}

\newcommand{\MP}{\ensuremath{M_{\mathrm{P}}}}

\newcommand{\eV}{\ensuremath{\mathrm{eV}}}

\newcommand{\MeV}{\ensuremath{\mathrm{MeV}}}
\newcommand{\GeV}{\ensuremath{\mathrm{GeV}}}

\newcommand{\seconds}{\ensuremath{\mathrm{sec}}}

\newcommand{\mbarn}{\ensuremath{\mathrm{mb}}}
\newcommand{\barn}{\ensuremath{\mathrm{b}}}
\newcommand{\fm}{\ensuremath{\mathrm{fm}}}
\newcommand{\cm}{\ensuremath{\mathrm{cm}}}

\usepackage{graphicx}
\usepackage{booktabs}
\usepackage{amsmath,amssymb}

\def\be{\begin{equation}}
\def\ee{\end{equation}}
\def\ba{\begin{eqnarray}}
\def\ea{\end{eqnarray}}
\def\ge{\mathrel{\raise.3ex\hbox{$>$\kern-.75em\lower1ex\hbox{$\sim$}}}}
\def\la{\mathrel{\raise.3ex\hbox{$<$\kern-.75em\lower1ex\hbox{$\sim$}}}}

\def\thesection{\arabic{section}}

\def\simgt{\mathrel{\raise.3ex\hbox{$>$\kern-.75em\lower1ex\hbox{$\sim$}}}}
\def\simlt{\mathrel{\raise.3ex\hbox{$<$\kern-.75em\lower1ex\hbox{$\sim$}}}}

\newcommand{\rad}{\mathrm{rad}}

\newcommand{\us}{U(1)$_S$}

\title{Metastable GeV-scale particles as a solution to the
  cosmological lithium problem} 

\author{Maxim Pospelov\\Department of Physics and Astronomy,  University of Victoria,
 Victoria, BC, V8P~1A1, Canada\\Perimeter Institute for Theoretical Physics, Waterloo,
 Ontario N2L~2Y5, Canada\\E-mail: \email{pospelov@uvic.ca}}
\author{Josef Pradler\\Perimeter Institute for Theoretical Physics, Waterloo,
 Ontario N2L~2Y5, Canada\\E-mail: \email{jpradler@perimeterinstitute.ca}}

\abstract{
  The persistent discrepancy between observations of \lisv\ with
  putative primordial origin and its abundance prediction in Big Bang
  Nucleosynthesis (BBN) has become a challenge for the standard
  cosmological and astrophysical picture.
  We point out that the decay of GeV-scale metastable particles $X$
  may significantly reduce the BBN value down to a level at which it
  is reconciled with observations.
  The most efficient reduction occurs when the decay happens to
  charged pions and kaons, followed by their charge exchange reactions
  with protons. Similarly, if $X$ decays to muons, secondary electron
  antineutrinos produce a similar effect.
  We consider the viability of these mechanisms in different classes
  of new GeV-scale sectors, and find that several minimal extensions
  of the Standard Model with metastable vector and/or scalar particles
  are capable of solving the cosmological lithium problem.
  Such light states can be a key to the explanation of recent cosmic
  ray anomalies and can be searched for in a variety of high-intensity
  medium-energy experiments.}

\begin{document} 

\section{Introduction}

Rapid progress in observational cosmology during the last decade
brought about the measurements of many cosmological parameters,
including a precise determination of the baryon-to-photon
ratio~$\eta_b$~\cite{Dunkley:2008ie} by the WMAP satellite
experiment. This puts the predictions of the Big Bang Nucleosynthesis
(BBN) theory on firm footing and allows for less ambiguous comparison
with observations. The current status of standard BBN (SBBN) with the
input from WMAP can be summarized as follows: there is no dramatic
(order of magnitudes) disagreement between predictions and
observations, but there is no ideal concordance either. For a series
of recent reviews see {\em e.g.}
\cite{Iocco:2008va,Cyburt:2008kw,Steigman:2007xt,Jedamzik:2009uy}.

There are at least two quantitative problems that look worrisome:
different measurements of the deuterium abundance, although on average
consistent with the SBBN prediction, exhibit a significant scatter.
This scatter may be the sign of underestimated systematic errors or
the manifestation of significant astration, thus hinting on a higher
primordial value for the deuterium abundance. In contrast to
deuterium, the scatter of \lisv/H data points along the so-called
Spite plateau \cite{Spite:1982dd} is rather small, which for a long
time thought to be a compelling argument for the primordial origin of
\lisv\ in these observations. As is well-known, this value is a factor
of 3-5 lower than the SBBN prediction,
$\lisv/\Hyd=(5.24^{+0.71}_{-0.67})\times 10^{-10}$~\cite{Cyburt:2008kw},
which is the essence of the lithium problem.

How serious are these problems of SBBN? It is entirely possible at
this point that future highest quality observations of D/H in quasar
absorption clouds would render D/H in accordance with SBBN in
combination with less scatter. Moreover, more elaborate stellar
evolution models with ab-initio calculations of lithium diffusion may
point to a systematic and uniform reduction of the SBBN value by a
factor of 3 or so.  At this point, it is too early to declare SBBN
being in serious trouble.  However, it is also tempting to speculate
that some subtle particle physics interference in the early Universe
may have resulted in distorted abundances of the primordial elements,
and possibly led to the reduction of lithium abundance.

The primordial value of the lithium abundance is given by the
freeze-out BBN value of \bes+\lisv, with atomic \bes\ decaying to
lithium at the later stages of cosmological evolution. The current
lithium problem stems from the overproduction of \bes\ at $T\sim
40$~keV.  \bes\ cannot be destroyed by protons directly, but instead
is depleted via the following chain of exoergic reactions
\begin{align}
\label{eq:li-depletion-chain}
  \bes + n \to \lisv \to \hef + \hef.
\end{align}
At the second step \lisv\ is destroyed by proton reactions, which
remain faster than the Hubble rate until $T\sim 10$ keV.

Different classes of modified BBN models where an additional reduction of
lithium can happen were analyzed in the literature over the years. As
was first pointed out by Reno and Seckel \cite{Reno:1987qw}, a
(moderate) injection of "extra neutrons" around the time of formation
of \lisv\ and \bes\ leads to an overall depletion of \lisv+\bes\ by
intensifying the destructive chain (\ref{eq:li-depletion-chain}).
This was emphasized again after the CMB determination of $\eta_b$ in
Ref. \cite{Jedamzik:2004er}, where it was demonstrated that any
particle physics source is capable of reducing \lisv+\bes\ as long it
leads to the injection of $O(10^{-5})$ neutrons per nucleon at
relevant temperatures. Perhaps the most natural source for a neutron
excess is the decay of massive particle species $X$.  Independently
from the motivation of reducing the \lisv\ abundance, a lot of work
has been invested into BBN models with decaying or annihilating
particles releasing a significant amount of energy into the primordial
plasma~\cite{Kawasaki:2004yh,Kawasaki:2004qu,Jedamzik:2006xz,Steffen:2006hw,Cyburt:2006uv,Cyburt:2009pg,Freitas:2009jb}.

So far, most of the analyses have concentrated on the injection of
energy by relics with masses comparable to the electroweak scale. This
is largely motivated by theoretical arguments in favor of new physics
at and below the TeV-scale, and by the possibility of having dark
matter in the form of weakly interacting massive particles (WIMPs). It
can be easily shown that it is unlikely that the residual WIMP
annihilation is responsible for the reduction of \bes+\lisv, simply
because the total energy injected via such mechanism is way below the
required levels \cite{Reno:1987qw,Jedamzik:2004ip}.  Known examples
that "work", \textit{i.e.} scenarios in which the \lisv\ abundance is
reduced while other elements are still agreeing with observations,
typically deal with unstable weak-scale particles.  These include some
supersymmetric scenarios with the delayed decays of charged sleptons
to gravitinos \cite{Jedamzik:2005dh,Cyburt:2006uv,Cumberbatch:2007me}.
The source of extra neutrons in these models is linked to the presence
of nucleons among the decay products.  An alternative plausible
mechanism for reducing \lisv\ is the catalytic suppression of the
\bes\ abundance by the capture of massive negatively charged particles
\cite{Pospelov:2006sc,Bird:2007ge,Jittoh:2007fr,Kusakabe:2007fv}, that
are again linked to the weak scale.  In this paper we investigate
whether the suppression of the lithium abundance can be triggered by
the decays of metastable GeV and sub-GeV scale electrically-neutral
particles. We address two types of models: the WIMP-type where
particle decays were preceded by the depletion through the
annihilation, and the super-WIMP type, where the abundance of decaying
particles is set by the thermal leakage of Standard Model (SM) states
into an initially vacuous super-WIMP sector.

During the last two years, the GeV-sector phenomenology experienced
some degree of revival due to its possible connection to the
enhancement of the leptonic fraction of cosmic rays in the pair
annihilation of dark matter
WIMPs~\cite{ArkaniHamed:2008qn,Pospelov:2008jd}. Particularly
noteworthy are the enhancement of the positron fraction seen by the
PAMELA satellite experiment \cite{Adriani:2008zr} and the
harder-than-expected spectrum for electrons and positrons detected by
the FGST instrument \cite{Abdo:2009zk}.  In such scenarios the
GeV-scale particles are designated as "mediators" connecting the dark
matter and Standard Model sectors \cite{Pospelov:2007mp}, allowing to
seclude the dark matter by choosing the the SM-mediator coupling to be
very small, but at the same time keeping the galactic annihilation
rate enhanced over the "standard" WIMP scenario.

The motivation for GeV and sub-GeV scale mediators comes from the
following consideration: the lightness of a gauge boson mediating an
attractive force in the dark sector, $V(r) = - \alpha_D/r\times
\exp(-m_Vr)$, compared to the characteristic WIMP momentum inside the
galaxy, $m_D v_{\rm gal}$, ensures a Coulomb/Sommerfeld enhancement of
the annihilation cross section relative to its freeze-out
value. Choosing $m_V \la m_D v \sim 100~{\rm MeV}\div 1~{\rm GeV}$,
results in a $\pi \alpha_D/ v$ enhancement of the cross sections at
the relevant velocities. Moreover, once recombination to WIMP-bound
states becomes kinematically possible, $m_V < m_D \alpha_D^2/4$, this
process dominates numerically over the direct two-body
annihilation. The resulting annihilation cross section for fermionic
WIMPs can be enhanced over the freeze-out value by two-to-three orders
of magnitude \cite{Pospelov:2008jd}:
\be \fr{\langle \sigma v \rangle_{\rm
    gal}}{\langle \sigma v \rangle_{\rm f.o.}}  \sim \fr{7
  \pi\alpha_D}{v_{\rm gal}} \sim O(10^2-10^3).
\label{recombination}
\ee
The numerical enhancement due to the bound-state effect over the naive
Sommerfeld value is about a factor of 7, accompanied by a possible
additional enhancement due to an increased lepton multiplicity in the
final state with angular momentum $J=1$. We note in passing that the
important effect of WIMP-onium formation was missed in a recent
re-analysis of Ref. \cite{Feng:2010zp}, that led to the erroneous
conclusion that the enhancement factor remains smaller than 100 over
the whole parameter space; larger factors were found
in~\cite{Cirelli:2010nh}.

Is it reasonable to expect that the same models that fit
the PAMELA and FGST signals \cite{ArkaniHamed:2008qn,Pospelov:2008jd}
are also responsible for the suppression
of \lisv\ abundance?  Even with the inclusion of some very generous
enhancement factors, the energy injection during BBN triggered by the
annihilation of electroweak-scale WIMPs remains rather small.
Moreover, models designed to explain the PAMELA signal tend to
minimize the fraction of baryons/anti-baryons in the final state
\cite{ArkaniHamed:2008qn,Pospelov:2008jd,Fox:2008kb,Nomura:2008ru}.
All that, together with previous investigations of BBN with
annihilation-induced energy injection, tend to indicate that WIMP
annihilation itself cannot be used as a solution to the lithium
problem. Therefore, the only chance of altering the BBN predictions
for \bes+\lisv\ within this class of models is if the GeV-scale
particles from the mediator sector are themselves relatively
long-lived, and their decays happen during or after BBN.
Interestingly enough, it turns out that the minimal ways of coupling
the Standard Model to light mediators often implies the longevity of
particles in the GeV sector \cite{Batell:2009di}.

The main mechanism by which the decays of the GeV-scale relics in the
early Universe can influence the outcome of the BBN is the injection
of light mesons such as pions and kaons as well as muons that all lead
to the extra source of $p\to n$ conversion.  In this work we explore
such scenarios, finding the "required" number of injected $\pi^-$,
$K^-$, and $\mu^{-}$ triggering $p\to n$ conversion in the right
amount, as well as the "optimal" lifetime--abundance window for such
injection. In order to have a consistent cosmological picture, the
abundance of parent GeV-scale relics should be small enough so that
they carry only a small fraction of the energy density of Universe
during BBN, but still provide a noticeable number of mesons and muons
per nucleon. We show that, in fact, many models with GeV-scale relics
fulfill this requirement, including some variants of the models
designed to fit PAMELA and FGST signals.  We also find that both the
WIMP and super-WIMP modifications of a secluded \us-model is capable
of reducing the lithium abundance to the observable level.
  
The structure of this paper is as follows. The next section contains
the analysis of the injection of $\pi$, $K$, and $\mu$ particles
vs.~the timing of injection that is needed for reducing
\lisv. Section~\ref{sec:metastable-gev-scale} investigates a variety
of different models for decaying GeV and sub-GeV scale relics, and
finds the regions of parameter space that lead to the depletion of the
lithium abundance. Our conclusions are summarized in
Sec.~\ref{Sec:Conclusions}. Appendix~A contains relevant details with
physics input that went into our BBN code.

\section{Meson and neutrino injection during BBN}

We begin this section by presenting an overview of the physics
processes triggered by the decays of meta-stable GeV-scale relics $X$
during BBN. Thereby we shall estimate timescales and interaction
efficiencies of the crucial reactions which eventually lead to the
reduction of the overall $\lisv$ BBN prediction.  Subsequently, the
various final states in the decay of $X$ are considered in detail.

The central parameter entering the discussion is the Hubble expansion
rate $H$ as it normalizes any interaction rate during nucleosynthesis.
The most relevant epoch for BBN corresponds to the time bracket of
$100\ \seconds \lesssim t \lesssim 1000\ \seconds$ or, equivalently,
to a temperature range of
\begin{align}
\label{eq:BBN-temp-window}
  1.2 \lesssim T_9\lesssim 0.4  ,
\end{align}
where $T_9$ is the photon temperature in units of $10^9$~K.
Thus, in the interesting regime after the annihilation of the
electron-positron pairs, the Hubble rate is given by
\begin{align}
\label{eq:hubble-rate}
  H & = \sqrt{\frac{\pi^2 \geff}{90}} \frac{T^2}{\MP} 
  \simeq (2.8\times 10^{-3}\, \seconds^{-1})\, T_9^2 \ \qquad (T\lesssim \me),
\end{align}
where $g_{\mathrm{eff}}\simeq 3.36$ counts the radiation degrees of
freedom; $\MP \simeq 2.43\times 10^{18}\, \GeV$ is the reduced Planck
scale.

The synthesis of \bes, proceeding via $\hef + \het \to \bes\ +\gamma$,
occurs in a more narrow temperature interval, $T_9\simeq 0.8\div 0.4$,
in which the Hubble rate is $H\sim 10^{-3}$ sec$^{-1}$. This occurs
immediately after the opening of the deuterium ``bottleneck'' at
$T_9\sim 0.8$, when significant quantities of \hef\ and \het\ are
formed, with a helium mass fraction of $Y_{p}\simeq 0.25$ and a number
density of \het\ per proton of $Y_{\het}\simeq 10^{-5}$.  The rate for
\bes\ formation per alpha particle,
\begin{align}
  \Gamma_{\bes} 
  \simeq (10^{-3}\, \seconds^{-1})\ T_9^{7/3} e^{-12.8/T_9^{1/3}}
  \left( \frac{Y_{\het}}{10^{-5}} \right)\quad (T_9\lesssim 0.8)
\end{align}
always remains slower than the Hubble rate and quickly becomes
completely inefficient due to a strong exponential Coulomb
suppression.
On the other hand, a non-standard ``thermal'' neutron abundance, at
the level comparable to $10^{-5}$ has the opportunity to reprocess some
fraction of \bes\ via $\bes + n \to \lisv + p $ since the rate for a
neutron capture per \bes\ nucleus is
\begin{align}
\label{eq:bes-n-capture}
  \Gamma_{\bes\to\lisv} \simeq (0.4\ \seconds^{-1})\,T_9^3 \left(
    \frac{Y_n}{10^{-5}} \right).
\end{align}
Indeed, looking at the fraction~$f_{\bes}$ of $\bes$ which can, in
principle, be converted within a Hubble time $\Delta t_H$,
\begin{align}
  f_{\bes}\simeq - \frac{1}{Y_{\bes}} \frac{dY_{\bes}}{dt}\Delta t_H \simeq -H^{-1}
  \Gamma_{\bes\to\lisv} \sim -10^{7}T_9 Y_n ,
\end{align}
shows that already $Y_n \simeq 5\times 10^{-6}$ at $T_9\simeq 0.5$ can
induce an $\Orderof{1}$-conversion of \bes.
$\lisv$ in the final state of this reaction is then quickly burned by
protons.

 The neutron participation in the lithium depleting chain does not
 influence the abundance of neutrons themselves, because of the small
 abundance of \bes. Instead, ``extra neutrons'' are depleted by
 protons via the $p + n\to \deut +\gamma$ reaction, which, at the
 relevant temperatures, remains faster than the rate for neutron
 decay. Comparing the relative changes of \bes\ and deuterium,
\begin{align}
  \frac{f_{\bes}}{f_{\deut}} = -\frac{Y_\deut}{Y_p}
  \frac{\Gamma_{\bes\to\lisv}}{\Gamma_{p\to\deut}} \sim -10^{5}\,
  ({\deut}/{\Hyd})
\end{align}
one can infer that $\Orderof{1}$-depletion of \bes\ due to an excess
of neutrons will be accompanied by an $\Orderof{1}$-rise in the
deuterium abundance since $\Deut/\Hyd \sim 10^{-5}$. Therefore, the
viability of the "extra neutron" solution to the lithium problem
should be judged form more accurate quantitative investigations of
such scenarios.

The decay of GeV and sub-GeV relic particles with a rate comparable to
(\ref{eq:hubble-rate}) at $T_9\sim 0.5$ will not lead to a significant
population of energetic photons and electrons.  The reason for that is
well-known: very abundant and energetic photons and electrons degrade
the energy of the decay products well below the nuclear dissociation
thresholds (see {\em e.g.} \cite{Kawasaki:2004qu}).  Since we
explicitly assume the absence of nucleons and anti-nucleons in the
final state, the main effect on the freeze-out abundances of light
elements will come from injection of mesons and neutrinos, that are
capable of triggering the $p\to n$ conversion and reducing \lisv+\bes.
In the following subsections we consider these mechanisms in turn,
starting from the simple estimates of the required number of mesons
and neutrinos.

\subsection{Estimates on meson injection}
\label{sec:timescales} 

Among the the decay products of the GeV-scale metastable states only
the long-lived mesons and neutrinos will have a chance to interact
with the light elements. Among those, the most important are charged
pions, \pipm, as well as charged and long-lived kaons, \Kpm\ and
\Klong, respectively. Their masses and lifetimes are given
by~\cite{Amsler:2008zzb}
\begin{align}
 m_{\pipm} & = 139.6\ \MeV , & \tau_{\pipm} & = 2.60\times 10^{-8}\ \seconds , \\
 m_{\Kpm} & = 493.7\ \MeV , &  \tau_{\Kpm}  & = 1.24\times 10^{-8}\ \seconds ,\\
 m_{K^0,\bar{K}^{0}} & = 497.6\ \MeV , &  \tau_{\Klong} & = 5.12\times 10^{-8}\ \seconds .
\label{eq:KL-prop}
\end{align}
The decay rate of meson $i$ is related to its lifetime at rest by a familiar 
time-dilatation formula: 
\begin{align}
  \label{eq:meson-decay-rate}
  \Gamma^i_{\mathrm{dec}} = \left\langle \frac{m_i}{E_i}\right\rangle
  \frac{1}{\tau_i} \simeq \mathrm{few}\times 10^7\ \seconds^{-1}
\end{align}
where in the last equality we have taken $\gamma \simeq 1$.  More
accurately, one has to consider the average value $\left\langle
  {m_i}/{E_i}\right\rangle$ over the "lifetime" trajectory of injected
mesons.  The averaging procedure is significantly different from the
usual thermal average when particles obey a Maxwell-Boltzmann
distribution.  It is determined by the efficiency of various energy
degradation mechanisms, of which the most important are Coulomb
scattering on electron and positrons and inverse-Compton-type
scattering on background photons.  Together they determine the
thermalization (or stopping) rate $\Gamma^i_{\rm stop}$. The stopping
rate may have a direct impact on the the reaction rates with nuclei
$\Gamma_N^i$, which are of the primary interest in this paper.

If injected sufficiently early, the charged mesons are effectively
stopped within time intervals shorter than their lifetimes $\tau_i$,
and before having a chance of participating in a nuclear reaction.
The comparison of stopping and decay rate is given by the ratio
\begin{align}
\fr{\Gamma^i_{\mathrm{stop}}}{\Gamma^i_{\mathrm{dec}}} = \fr{\tau^i_{\mathrm{dec}}}{\tau^i_{\mathrm{stop}}}
\simeq
  \left(\Gamma^i_{\mathrm{dec}}\int^{E_0}_{E_{\mathrm{min}}} \frac{dE}{|dE_i/dt|} \right)^{-1},
\end{align}
where $\tau_{\mathrm{stop}}$ measures the energy degradation time from
injection energy $E_0$ to some energy $E_{\mathrm{min}}$, below which
the kinetic energy of mesons is irrelevant. $dE_i/dt$ represents the
energy degradation rate of the particle $i$ traversing the BBN
plasma. Requiring complete stopping would correspond to the choice of
$E_{\mathrm{min}} \sim 3T/2$, but a more relevant parameter is some
characteristic nuclear energy scale. For reactions leading to
dissociation of \hef, $E_{\mathrm{min}}$ would correspond to
$E_{\mathrm{min}} \sim 20$ MeV.
In the temperature window (\ref{eq:BBN-temp-window}) and for injection
energies of $E_0 \lesssim 1\ \GeV$, Coulomb interactions on $e^{\pm}$
give the dominant contribution to $dE_i/dt$ (for details see {\em
  e.g.}  \cite{Kawasaki:2004qu}). Since the number densities of
electrons and positrons drop exponentially, the inverse Compton
scattering on background photons becomes the dominant energy loss
mechanism for $T<25$ keV.  Taking into account both channels for
energy loss, one typically finds that
\begin{eqnarray}
  \fr{\tau^i_{\mathrm{dec}}}{\tau^i_{\mathrm{stop}}} < 1\quad \mathrm{for}\quad T_9>(0.3\div 0.4).
\end{eqnarray}
This point is illustrated in Fig.~\ref{stopping}, where the dividing
line of ``1'' separates the two regimes of complete and incomplete
stopping. The exponential sensitivity to temperature is reflected in
almost vertical contours.  Figure \ref{stopping} tells us that for $t
\lesssim 10^3\ \seconds$ the charged mesons are thermalized before
they decay.  Thus, for $i= \pipm$ (and analogously for $\Kpm$)
injected around $ T_9 \sim 0.5$ the following hierarchy of scales is
applicable:
\begin{align}
  H \ll \Gamma_p^\pi \ll \Gamma_{\mathrm{dec}}^\pi  \lesssim \Gamma_{\rm stop}^\pi.
  \label{hierarchy}
\end{align}

The rate for charge exchange reactions with the most abundant nuclear species,
\textit{i.e.} the rate of proton-to-neutron conversion, is given by
\begin{align}
  \label{eq:hadint-rate}
  \Gamma_p^\pi= n_p \sigmavof{}_{pn}^\pi \simeq ( 3\times 10^{2}\
  \seconds^{-1} )\ \,\frac{T_9^3\sigmavof{}_{pn}}{1\ \mbarn}.
\end{align}
The averaging procedure is again determined along the trajectory of
injected mesons and includes both the "in-flight" and "at-rest"
contributions.  However, for the purpose of a simple estimate,
Eq.~(\ref{eq:hadint-rate}) uses the normalization on a typical size of
a pion induced reaction at the threshold. In case of incomplete
stopping, this will underestimate the proton-neutron conversion
because of the delta-resonance enhancement of the charge-exchange
reaction. Equation~(\ref{eq:hadint-rate}) immediately leads us to the
probability of charge exchange reaction of a stopped pion (kaon) on
protons during the meson lifetime, per each injected $\pi^-(K^-)$ at
$T_9\sim 0.5$:
\begin{align}
  P_{p\to n}^\pi = \int_{t_{\rm inj}}^\infty
  \exp(-\Gamma_{\mathrm{dec}}(t-t_{\rm inj})) \Gamma_p dt \simeq
  \Gamma_p^\pi \tau_{\pipm} \sim O(10^{-6})
  \label{probability}
\end{align}

Since every charge exchange reaction leads to the creation of an extra
neutron, the estimate (\ref{probability}) represents the {\em
  efficiency} of producing neutrons from negatively charged kaons and
pions.  Notwithstanding a rather crude nature of this estimate, the
probability (\ref{probability}) tells us that an injection of
$\Orderof{10}$ negatively charged pions per proton should be
equivalent to the injection of $\Orderof{10^{-5}}$ extra neutrons, and
thus capable of reducing the overall \lisv+\bes\ abundance by an
$\Orderof{1}$ factor.

\FIGURE[t]{
\includegraphics[width=0.6\textwidth]{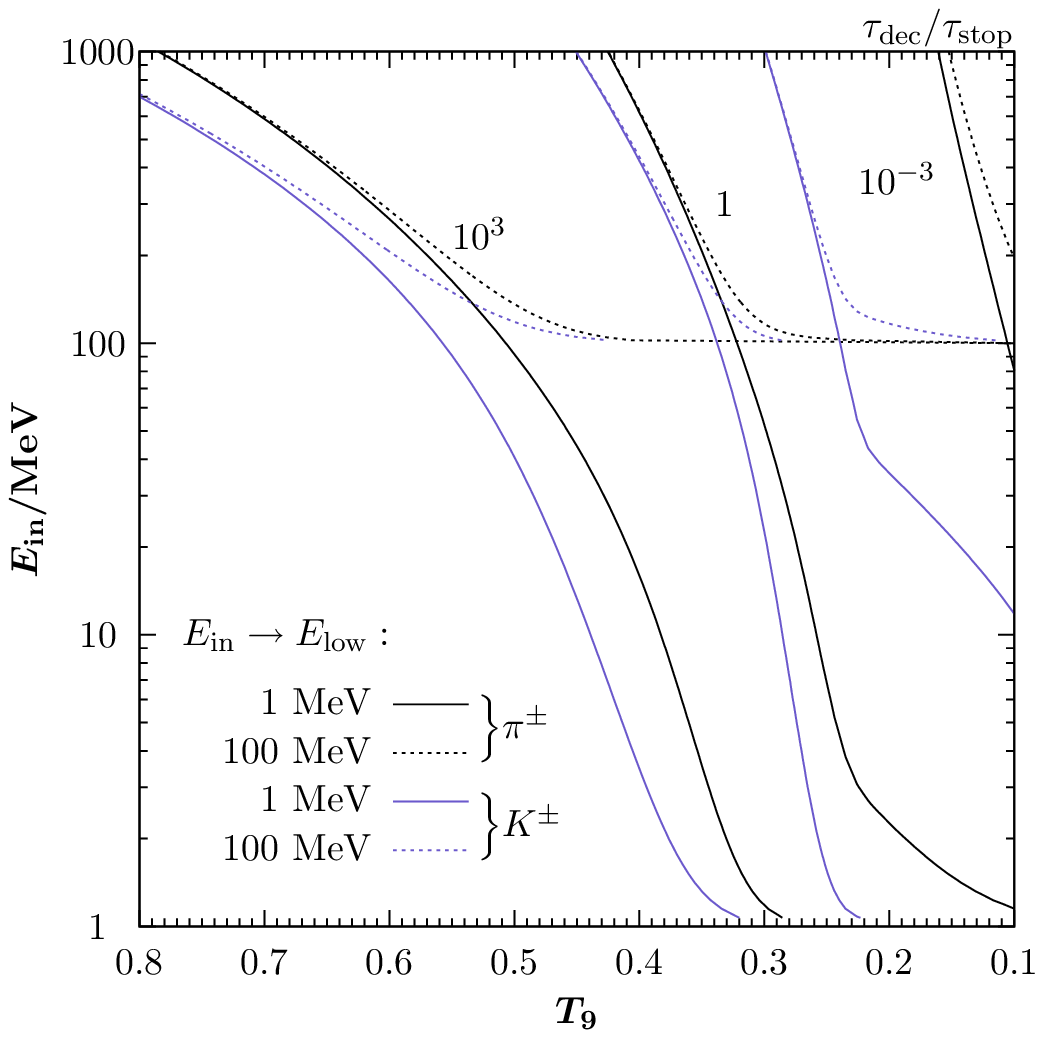}
\caption{\small Stopping of charged pions and kaons in the plasma as a
  function of temperature $T_9$, initial injection energy
  $E_{\mathrm{in}}$, and the minimal energy below which the kinetic
  energy of mesons can be neglected, $E_{\mathrm{low}}$. The contour
  $\tau_{\mathrm{dec}}/\tau_{\mathrm{stop}}=1$ is the dividing line
  between efficient (left) and incomplete (right) stopping.}
\label{stopping}
}

Since for kaons the charge exchange cross sections are even higher
\cite{Reno:1987qw}, one can achieve an adequate suppression of
\lisv+\bes\ even with the injection of one $K^-$ per baryon.
The case of $X$-decays into neutral kaons is special with respect to
other meson final states in the sense that $K_L$ has a relatively long
lifetime (\ref{eq:KL-prop}) but, unlike $\pipm$ or $\Kpm$, is not
stopped by electromagnetic interactions. From the conservation of
isospin and from charge independence of strong interactions, we
nevertheless expect the impact of $K_L$ on the BBN predictions to be
similar to that of $K^{-}$.

In our treatment, the extra meson species are included in the set of
Boltzmann equations, along with the population of parent particles $X$
that decay into mesons and have an abundance with the simple
exponential time-dependence, \be Y_X(t) = Y_X^0\exp(-t/\tau_X), \ee
with $\tau_X$ being the $X$-lifetime.  We choose to normalize the
$X$-abundance on the total number of baryons, $Y_X \equiv n_X/n_b$; in
the following we shall often drop the superscript on $Y_X^0$ for
simplicity.  When the fast stopping of charged particles is operative
at early times, the amount of the injected energy and therefore the
actual mass of the $X$-relic are not entering the problem (with $X\to
K^0\bar K^0$ being the exception.)  For later times, the incomplete
stopping brings the dependence on the injected energy. We write the
Boltzmann equation for meson species $i$ in the following form
\begin{equation}
  \fr{d{Y}_{i}}{dt} \simeq \sum_{j} \xi_i^{(j)}  Y_j \Gamma^{j}_{\mathrm{inj}} - Y_i
  \Gamma^i_{\mathrm{dec}} - Y_N \Gamma_N^i.
  \label{BE}
\end{equation}
where $\Gamma^{j}_{\mathrm{inj}}$ is the rate of $i$-injection from
source $j$ with multiplicity~$\xi_i^{(j)}$.  When considering only the
primary meson production from $X$-decays such as in the case of
$\Kpm$, $\sum_{j} \Gamma^{j}_{\mathrm{inj}} \simeq
\Gamma_{\mathrm{dec}}^X = 1/\tau_{X}$.  Given the hierarchy of
interactions (\ref{hierarchy}), the last term in (\ref{BE}) is small.
Both $\Gamma^i_{\mathrm{dec}}$ and $\Gamma_N^i$ include the effects of incomplete stopping 
that depends on the initial energy injection $E_0$ and background temperature at the time of injection,
\be
\Gamma^i_{\mathrm{dec}}=\Gamma^i_{\mathrm{dec}}(E_0,T);~~~ \Gamma_N^i = \Gamma_N^i(E_0,T).
\label{correction}
\ee
Neglecting subtleties of incomplete stopping at this point, 
one can find an approximate solution to the Boltzmann equation for mesons 
in the quasi-static approximation $dY^{\mathrm{qse}}_i/dt =
0$,
\begin{align}
  Y^{\mathrm{qse}}_i = \sum_{j} {\xi_i^{(j)} Y_j\Gamma^{j}_{\mathrm{inj}}}/{
    \Gamma^i_{\mathrm{dec}}} \sim \Orderof{10^{-10}} \times  Y_X, 
    \label{qse}
\end{align}
where in the second relation we assumed multiplicities to be order
one, and took $\Gamma_{\rm inj} \sim 10^{-3}\ \seconds$. Even though
the equilibrium values for the meson abundances remain low at all
times, the $p\to n$ conversion rate for $Y_X \sim \Orderof{10}$ is
only five orders of magnitude slower than the Hubble rate, thus
causing $\Orderof{10^{-5}}$ protons be converted to neutrons within
one Hubble time. This is consistent with our estimate of efficiency
(\ref{probability}).

Finally, the pion(kaon)-induced transitions $\bes\to\lisv$ and
$\het\to\trit$ also lead to the depletion of \bes. However, these
processes are far less important compared to $p\to n$
interconversion. If we assume a fractional $O(10^{-5})$-conversion of
$p\to n$ due to $\pi^-$ charge exchange, a similar figure would stand
for the $\bes\to\lisv$ and $\het\to\trit$ conversion probabilities,
which is completely negligible in the final lithium abundance.

\subsection{Estimates on muon and neutrino injection}

In order to account for the injection of muons and neutrinos one has
to deal with quite different physics. The charge exchange reactions of
muons on nucleons are mediated by weak interactions and can be
neglected during the muon lifetime of two microseconds.  However, the
decays of muons, $\mu^- \to \nu_\mu \bar \nu_ee^-$, source energetic
electron antineutrinos $\bar \nu_e$, that survive for a long time,
which increases their probability of charge-exchange interaction with
protons.

To make our discussion more concrete, we shall assume that all
neutrinos originate from muons decaying at rest so that $E_\nu <
m_\mu/2$. In this case, the relation between the relevant rates is
quite different from~(\ref{hierarchy}): \be \Gamma^\nu_p, ~
\Gamma^\nu_{\rm stop} \ll H, \ee where $\Gamma^\nu_{\rm stop} $ is the
rate of antineutrino energy degradation due to scattering on
background neutrinos and electron-positron pairs.  This rate scales
with temperature and the energy of non-thermal neutrinos $E_\nu(T) $
injected at $T_{\rm inj}$ as follows:
\begin{align}
  \Gamma^\nu_{\rm stop} \sim G_F^2 E_\nu(T) T^4 \sim G_F^2 T^5
  \fr{\VEV{E_\nu(T_{\rm inj})}}{T_{\rm inj}},
\end{align}
where we disregard the difference between neutrino and
photon temperatures.  It is indeed much smaller than the Hubble
expansion rate, 
\begin{align}
  \fr{\Gamma^\nu_{\rm stop}}{H} \sim \left( \fr{T}{3~{\rm MeV} }
  \right)^3 \fr{\VEV{E_\nu(T_{\rm inj})}}{T_{\rm inj}} \sim 10^{-3},\qquad
  (T,~T_{\rm inj}\sim 30~{\rm keV}),
\end{align}
where 3 MeV scale enters as the decoupling temperature of background
neutrinos, and the energy of non-thermal antineutrinos is taken to be
the average energy in the muon decay, $ \VEV{E_\nu(T_{\rm inj})} =
\fr{3}{10} m_\mu \simeq 32 $ MeV.

The charged current rate for antineutrino interactions with protons is
given by \be \Gamma^\nu_p = n_p \sigma_{pn }^{\bar\nu} \simeq
10^{-41}~{\rm cm}^2 \times \fr{n_p E_\nu^2}{(10~{\rm MeV})^2} \simeq
(3.6\times 10^{-12}~{\rm sec}^{-1} )\times \fr{T_9^3E_\nu^2}{(10 ~{\rm
    MeV})^2}, \ee where $E_\nu \gg m_e$ is assumed.
The ratio of $\Gamma^\nu_p $ to the Hubble rate gives the efficiency
of producing extra neutrons from each neutrino injected with
$\Orderof{30\ \MeV}$ energy:
\begin{align}
  P^\nu_{p\to n} = \int_{t_{\rm inj}}^{\infty} \Gamma_p^\nu dt =
  \fr{1}{3} \fr{\Gamma^\nu_p(T_{\rm inj})}{H(T_{\rm inj})} \sim
  2\times 10^{-9}
  \label{estimate2}
\end{align}
This efficiency is several orders of magnitude smaller than
Eq. (\ref{probability}), and we conclude that $\Orderof{10^4}$ muon
decays per proton are required around $T_9 \sim 0.5 $ in order to
produce $\Orderof{10^{-5}}$ extra neutrons. At the same time,
$P^\nu_{p\to n} \ll P^\pi_{p\to n}$ introduces a natural
simplification to the problem, as the effects of secondary neutrinos
originating from pion and kaon decays can be ignored relative to the
direct influence of $\pi^-$ and $K^-$ via the charge exchange
reactions.  Moreover, as the efficiency (\ref{estimate2}) also
suggests, a direct nuclear-chemical impact of electron antineutrinos
on the lithium abundance via $\bes +\bar\nu_e\to \lisv + e^{+}$
($Q=-0.16\ \MeV$) is negligible.

Having checked that the efficiency of neutron production from pions,
kaons and muons (neutrinos) can be sufficient for the resolution of
the lithium problem, we now turn to more detailed calculations. Some
further details on the Boltzmann code that we use can be found in the
Appendix.

\subsection{Decays to pions }
\label{sec:treatment-pions}

Charged pions are a likely final state for almost any hadronic final
decay mode. For example, even if $X$ would exclusively decay into
\Kpm, \ie\ $\xi^{(X)}_{\pipm} = 0$ in Eq.~(\ref{BE}), pions would
still be populated by subsequent \Kpm-decays, $Y_{\Kpm}\simeq
Y_{\pipm}$, since $\xi^{(\Kpm)}_{\pipm} = \Orderof{1}$.

Charged pions have a chance to interact with protons and \hef\ before
decaying. Fully thermalized pions have the following charge exchange
reactions on nucleons with positive energy release:
\begin{align}
\label{pi-on-p}
\pim + p & \to n + \gamma : & & \!\!\!\!\!\!\!\!\!\!(\sigma
v)^{\pim}_{pn(\gamma)} \simeq 0.57\ \mbarn ,
&  \!\!\!\!\!\!\!\!\!\!Q & = 138.3\ \MeV ,\\
\label{pi-on-p-chx}
\pim + p & \to n + \pizero :&& \!\!\!\!\!\!\!\!\!\! (\sigma
v)^{\pim}_{pn(\pi^0)} \simeq 0.88\ \mbarn ,
&  \!\!\!\!\!\!\!\!\!\!Q & = 3.3\ \MeV ,\\
\pip + n & \to p + \pizero : &&\!\!\!\!\!\!\!\!\!\! (\sigma
v)^{\pim}_{np} \simeq 1.7\ \mbarn , & \!\!\!\!\!\!\!\!\!\!Q & = 5.9\
\MeV .
\end{align}
These reactions interconvert neutrons and protons and thereby increase
the $n/p$ ratio, because protons are far more numerous once neutrons
have been incorporated into~\hef.  The threshold value for the $\pim
+p$ reaction cross section can be extracted from the lifetime of the
pionic hydrogen atom (see {\em e.g.} the review
\cite{Gasser:2007zt}). The strength of the channels $\pi^0$ and
$\gamma$ at threshold can be inferred from the Panofsky
ratio~\cite{Flugel:1999gr}.  Our value at the (thermal) threshold \be
\label{ourvalue}
(\sigma v)^{\pim}_{{\rm th}}\equiv F_{p\pim} \left[ (\sigma
  v)^{\pim}_{pn(\gamma)}+(\sigma v)^{\pim}_{pn(\pi^0)} \right] \simeq
F_{p\pim}\times 1.45\ \mbarn, \ee is in good agreement with the one
used in the BBN literature \cite{Reno:1987qw} from where we also took
the cross section for $\pip + n$. This threshold value is subject to
the usual Coulomb enhancement that is accounted for with the
multiplicative factor $F_{p\pim}$ in (\ref{ourvalue}). For the
negatively charged pions and at the temperature range of interest
($T_9 \sim 0.5$) this translates into an enhancement of the reactions
rates by a factor of $F_{p\pim} \simeq 2$. Further details on Coulomb
enhancement are provided in the Appendix
\ref{sec:some-details-boltzm}.

The incomplete stopping of pions introduces an important modification
to the rates of these reactions, and to the efficiency of $p\to n$
conversion. To quantify this effect, we introduce the correction
factor $\kappa(E_0,T)$ as a function of primary kinetic injection
energy $E_0$ that includes both the effect of the time dilatation (and
effectively longer lifetimes of fast pions) and, more importantly, the
momentum dependence of the cross section that has a strong peak around
a pion  energy of $E_{\pi}\sim 180\ \MeV$:
\begin{align}
  \kappa(E_0,T) \equiv \fr{P_{p\to n}(E_0,T)}{P_{p\to n}(T)}
  =\fr{1}{\tau_\pi (\sigma v)_{{\rm th}}} \int_{0}^{\infty} dt (\sigma
  v)_{E_{\pi}(t)} \exp\left( -\int_0^{t} dt' \fr{1}{\tau_\pi \gamma(E_{\pi}(t'))}
  \right).
\label{kappaET}
\end{align}
The exponential factor in this expression is the survival probability,
that is the probability of finding a meson at time $t$ after its
injection at $t=0$.  Defined this way, the correction factor
$\kappa(E_0,T) =1$ in the limit of $\tau_{\rm stop}
\ll \tau_\pi$.  Notice also that due to the scale hierarchy, $H \ll
\tau_\pi^{-1}$, we can disregard the effects related to cosmological
expansion inside the integrals of Eq. (\ref{kappaET}), and set the
initial moment of injection to be $t=0$.

The kinetic energy $E_{\pi}$ of an injected pion, and the time $t$
along the ``lifetime'' trajectory are related via the rate of energy
loss: \be
\label{tE}
t(E_{\pi}) = \int^{E_0}_{E_{\pi}}\fr{dE'}{|dE'/dt|};\quad
\fr{dE}{dt} = \left( \fr{dE}{dt} \right)_{\rm Coul} + \left(
  \fr{dE}{dt}\right)_{\rm Comp}.  \ee As alluded before, $dE/dt$ is a
strong function of the background temperature.  Finding the explicit
dependence between time and kinetic energy of injected pions
numerically, and using the experimental data for the inelastic
$p+\pim$ reaction away from threshold, we find the correction factor
for the efficiency, $\kappa(E_0,T)$. Figures \ref{kappaT9} and
\ref{kappaE} plots this factor for the representative temperature and
initial energy slices.  One can see that due to incomplete stopping at
late times, the correction factor can be very large, possibly reaching
\be \kappa_{\rm max} \simeq \gamma_{\rm max} \fr{(\sigma v)_{\rm
    max}}{(\sigma v)_{\rm th}} \simeq 30, \ee where $\gamma_{\rm max}
\simeq 2.3$ corresponds to the pion momentum at delta-resonance, where
the ratio of cross sections reaches $(\sigma v)_{\rm max}/(\sigma
v)_{\rm th} \simeq 14$.

\FIGURE[t]{
\includegraphics[width=0.6\textwidth]{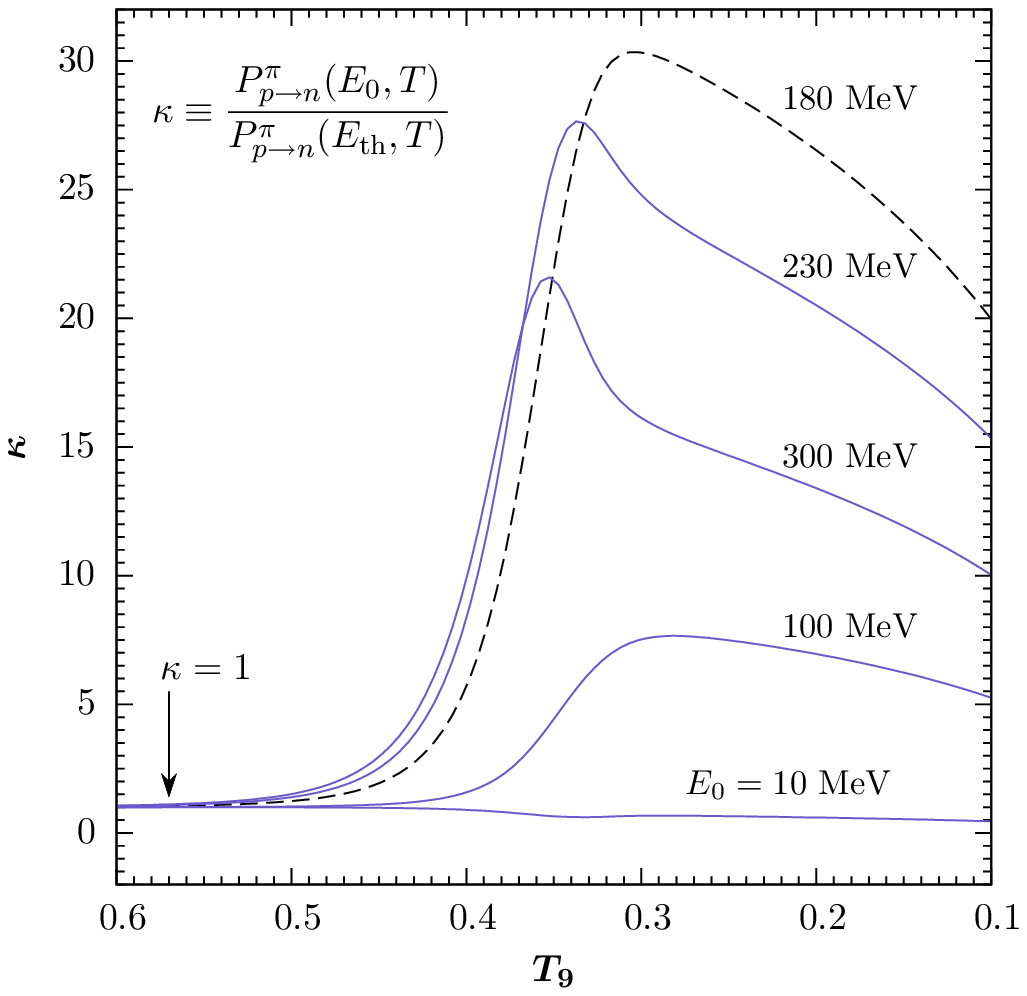}
\caption{\small The correction factor $\kappa(E_0,T)$, relating the
  efficiency of $p\to n$ conversion by pions in-flight to the thermal
  case, as a function of temperature for representative values of
  injection energy $E_0$; $E_{\mathrm{th}}=3T/2$. The dashed line
  shows the case of maximal efficiency due to pion injection at the
  delta-resonance energy. Formally, $\kappa <1$ is possible for low
  temperatures and small injection energies because the Coulomb
  corrections of a fully thermalized pion are substantial.}
\label{kappaT9}
}
\FIGURE[t]{\includegraphics[width=0.6\textwidth]{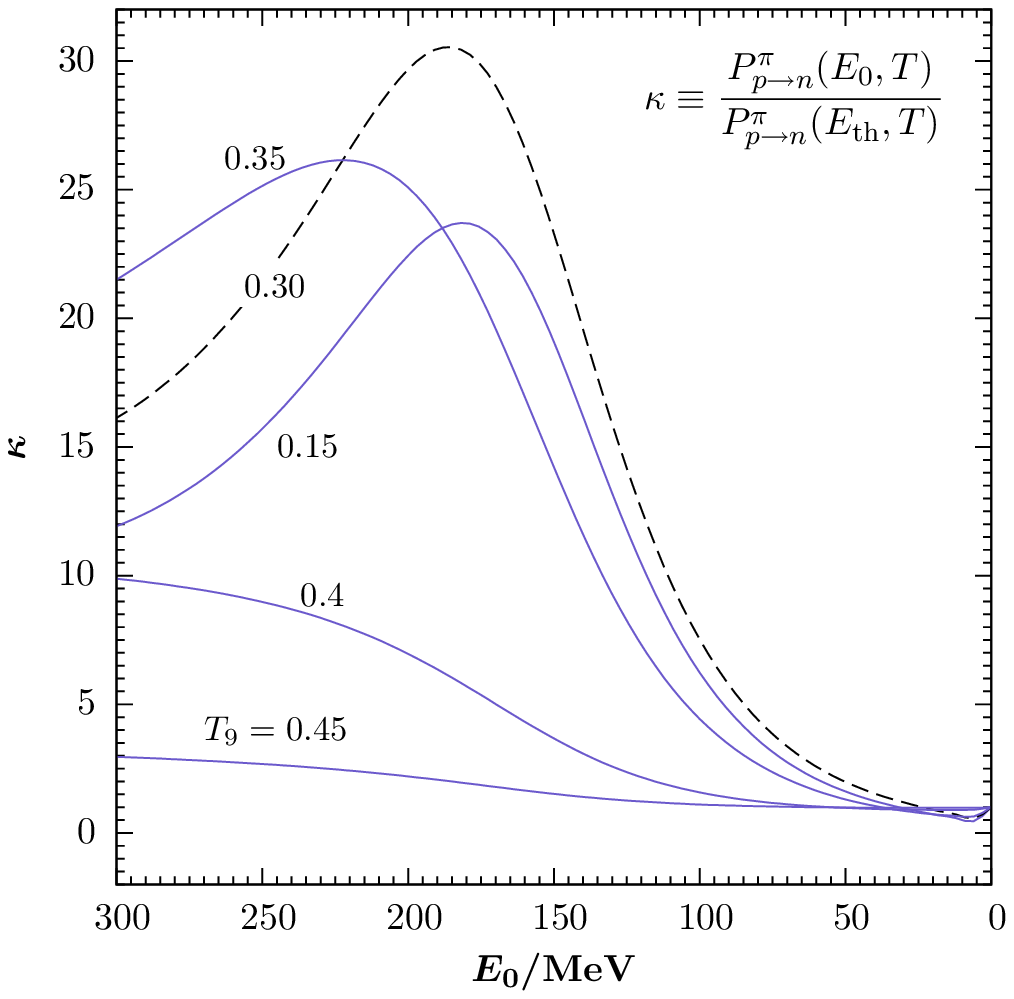}
  \caption{\small The correction factor $\kappa(E_0,T)$, relating the
    efficiency of $p\to n$ conversion by pions in-flight to the
    thermal case, as a function of injection energy $E_0$ for
    Representative values of temperature. The curves corresponding to
    late injection $T_9$ = 0.15 and 0.3 when stopping does not occur
    follow the familiar energy dependent profile of the pion-nucleon
    cross section with the broad delta resonance at 180~MeV. }
\label{kappaE}
}

At the next step we account for the possible reactions on \hef. 
Again, we separate our discussion into the reactions induced by 
stopped and in-flight pions. 
Charge exchange reactions of thermal $\pi^\pm$ on \hef\ are not possible
because of the deep binding of \hef\ compared to other mass-$4$
isomers. However, $\pi^\pm$ can be fully absorbed by \hef\ leading to
$\sim 100$ MeV energy release and a variety of nuclear final states.
Moreover, the reactions with thermal $\pi^+$ are very suppressed because of the 
Coulomb repulsion, and thus we concentrate on $\pi^-$ absorption. A
measurement of the ground state level width $\Gamma_{1S} = (45\pm 3)\
\eV $~\cite{Backenstoss1974519} of pionic helium then allows us to
obtain the total low-energy in-flight cross section $(\sigma v) \simeq
7.3\ \mbarn$; for further details see
Appendix~\ref{sec:pion-kaon-capture}. The branching ratios into the
different final states have been measured~\cite{Daum1995553} to be
$(\trit n):(\deut nn):(pnnn) = (17\pm 9)\%:(63\pm 26)\%:(21\pm
16)\%$. Adopting the central values (with $20\%\, pnnn$) we arrive at
\begin{align}
\label{eq:pi-he4-to-T}
  \pim + \hef & \to \trit + n\hphantom{2}: \quad (\sigma v)^{\pim}_{\trit n}
  \simeq 1.1\ \mbarn ,
  \quad  \,\,\,\,\, Q = 118.5\ \MeV ,\\
  \pim + \hef & \to \deut + 2n : \quad (\sigma v)^{\pim}_{\deut nn} \simeq
  4.1\ \mbarn ,
  \quad \,\, Q = 112.2\ \MeV , \\
  \pim + \hef & \to p + 3n \,\,: \quad (\sigma v)^{\pim}_{pnnn} \simeq 1.3\
  \mbarn , \quad Q = 110\ \MeV .
\end{align}
Before using (\ref{eq:pi-he4-to-T}) in our code,
we account for Coulomb corrections, which leads to the 
enhancement of $F_{\hef\pim} \simeq 3.5$.

The reactions of "in-flight" pions with \hef\ can also be
significantly enhanced by the presence of the delta-resonance.  Again,
one has to distinguish two types of reactions, inelastic scattering,
$\pi^\pm + \hef \to \pi + N$, and absorption, $\pi^\pm +\hef \to N $
where $N$ represents a variety of multi-nucleon/nuclear final
states. We use the results of experimental studies
\cite{Binon:1975zc,Steinacher1990413,Mateos:1998bc} to account for the
pion-\hef\ reactions across the delta-resonances. We also extrapolate
these results to the threshold regions, and calculate effective cross
sections to the various (exclusive) final states by averaging over the
pion ``lifetime'' trajectory while taking into account the respective
energy-dependent branchings. The details of this procedure are
summarized in Appendix~\ref{sec:inflight}.

Besides enhancing the number of free neutrons, the reactions on \hef\
have an additional important effect: they produce energetic $A=3$
elements that are able to participate in the non-thermal reactions
leading to \lisx. For example, reactions with thermalized $\pim$ in
Eq. (\ref{eq:pi-he4-to-T}) in $\sim 17\%$ of all cases contain \trit\
nuclei with an energy $E^{\mathrm{in}}_{\trit} \simeq 30\ \MeV$
injected into the plasma. This leads to a possible secondary source of
$\lisx$ via their fusion on ambient alpha particles
\cite{Reno:1987qw,Dimopoulos:1988ue},
\begin{align}
\label{eq:secondary-lisx}
  \trit+\hef|_{\mathrm{bg}} \to \lisx + n \quad Q = -4.8\ \MeV .
\end{align}
The number of produced \lisx\ per injected \trit\ (and likewise per
injected \het) can be found by tracking the \trit-degradation from
$E_{\trit,\mathrm{in}}$ until the threshold energy in the frame of the
thermal bath, $E^{\trit}_{\lisx,\mathrm{th}} \simeq 8.4\ \MeV$,
\begin{align}
\label{eq:Lisx-efficiency}
  N_{\lisx} \simeq
  \int_{E^{\trit}_{\lisx,\mathrm{th}}}^{E_{\trit,\mathrm{in}}} dE_{\trit}
  \frac{n_{\he4} \sigma_{\trit(\hef,n)\lisx} v_{\trit}}{|dE_{\trit}/dt|}
  ,
\end{align}
where $v_{T}$ is the velocity of \trit. 
As is well-known, the production of \lisx\ becomes more efficient at
late times, when its thermal destruction slows down. This occurs in
the regime of $\tau_\pi \ll \tau_{\rm stop}$, and most of the
energetic $A=3$ elements originate from \hef\ reacting with in-flight
$\pi^\pm$. This complicates the treatment of finding the final \lisx\
abundance as the branching to $A=3$ elements become energy dependent.
Moreover, non-absorptive inelastic pion-helium reactions result in
mass-3 injection spectra which are continuously distributed. We
account for all these effects with details presented in
Appendix~\ref{sec:inflight}.

In a final step we also account for potential effects on the light
element abundances coming from the ``visible'' energy injection in the
decays of the pions. Any primary electromagnetic energy deposition
$E_{\mathrm{inj}}$ is quickly dispersed in an electromagnetic cascade
with $E_{\mathrm{inj}}$ shared among a large number of photons.
Photons with energies less than $E_C \simeq
m_e^2/22T$~\cite{Kawasaki:1994sc} loose their ability to pair create
$e^{\pm}$ in scatterings on the background radiation. Those associated
``break-out'' photons can then destroy the light elements.  The
earliest time at which this happens can be determined by equating
$E_C$ to the nuclear binding energies $E_{b}$ against
photodissociation:
\begin{equation}
  t_{\rm ph} \simeq 
  \left\{ \begin{array}{lll}
      2\times 10^4\ \seconds &  \mathrm{for}\ \bes+\gamma\to\het+\hef  & (E_b = 1.59\,\MeV)\nonumber\\
      5\times 10^4\ \seconds &  \mathrm{for}\ \deut+\gamma\to n+ p &
      (E_b = 2.22\,\MeV)
\label{eq:t-dissoc}
\end{array}\right .
\end{equation}
We see that the electromagnetic energy injection plays no role in the
most interesting $X$-lifetime region $\tau_X\lesssim 10^{4}\ \seconds$
in which the lithium depleting chain~(\ref{eq:li-depletion-chain}) is
operative. Only for $\tau_X > 10^4\ \seconds$ do we expect an
influence on the element abundances, starting with the destruction of
\bes\ followed by \deut. The deepest bound element, \hef\ is only
destroyed for $t>10^6\ \seconds$.
Charged pions decay into $\mu^{\pm}$ which are, however, not
initiating an electromagnetic cascade since for $t\gtrsim 10^4\
\seconds$ their rate for Thomson scattering becomes smaller than the
muon decay rate. Thus, for simplicity, we assume that a total of
$E_{\mathrm{inj}} = \VEV{E_e} = (7/20)m_{\mu}$ per $\pipm$ is injected
in form of electromagnetic energy, where $\VEV{E_e}$ is the average
electron energy in the muon decay. Fortunately, though this neglects a
certain fraction of accessible kinetic energy of the muon, a more
accurate treatment is inconsequential for our further discussion. On
the same footing, we also neglect a direct energy production in
$X$-decay via {\em e.g.}  $X\to\pizero\pizero$. More details can be
found in Appendix~\ref{sec:electr-energy-inject}.

\FIGURE[t]{
\includegraphics[width=0.6\textwidth]{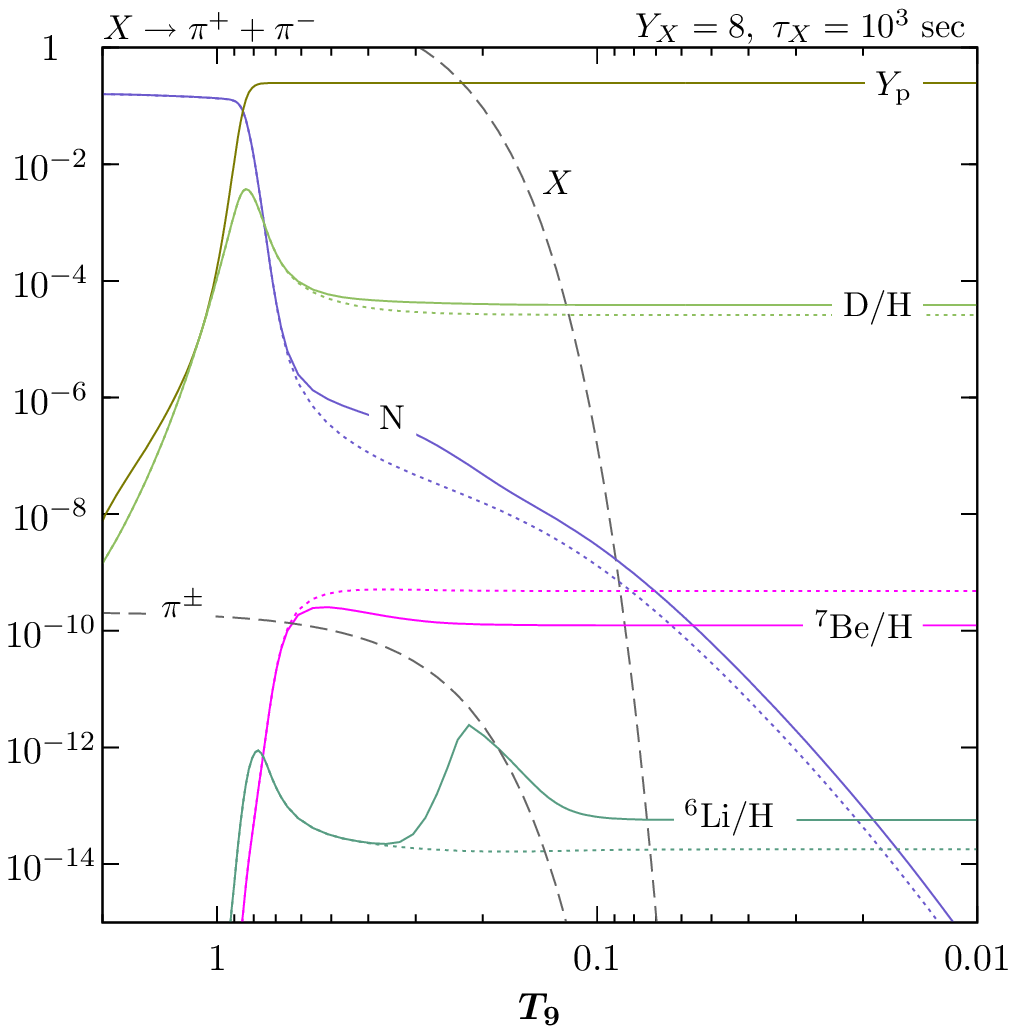}
\caption{\small Temperature evolution of light nuclei, meta-stable
  parent $X$ particles and daughter \pipm\ mesons for $Y_X = 8$ and
  $\tau_{X} = 10^3\ \seconds$ as input values and for injection not
  too far from threshold. The pion induced elevation of $n/p$ around
  $T_9\sim 0.5$ leads to the expected increase in the D/H and \lisx/H
  abundances, and to a decrease in \bes/H when compared with the
  respective SBBN predictions (dotted lines); $N$ is the neutron
  abundance normalized to baryons.}
\label{evolution}
}

The pion-enriched BBN was run for different input values of
$\xi^{(X)}_{\pipm} Y_X^0$, $\tau_X$ and the pion injection energy $E_0$.
In Figure~\ref{evolution} we plot an example of one possible choice
with $Y_X^0 = 8$, $\xi^{(X)}_{\pipm}=1$, $\tau_X = 10^3$~sec and negligible
kinetic energy of injected pions.  As expected, one can see a
noticeable increase of neutrons at lithium-relevant temperatures.
This leads to a decrease in the \bes\ abundance and also an increase
in D/H.  The secondary source of \lisx\, although leading to a modest
increase of \lisx\ over the SBBN prediction, turns out to be far below
the observable level of \lisx/H $\sim 10^{-11}$. This is because the
\lisx-burning reactions are very fast above $T_9>0.1$.

The exploration of the full parameter space yields the region in which
the primordial lithium over-production problem is solved. For this
solution, we require \lisv/H to stay in the interval
\begin{align}
\label{eq:LIobs}
  \lisv/\Hyd &= (1\div 2.5)\times 10^{-10} .
\end{align}
The ballpark of observations lie in the range $\sim 1.5\div 2.0$~(see
\textit{e.g.} \cite{Steigman:2007xt,Iocco:2008va} and references
therein), but also values on the upper end of the adopted range have
been reported \cite{2002A&A...390...91B,Melendez:2004ni}.  For
completeness, we also mention that the latest observations seem to
suggest what could be called a drop-off of stars from the most
metal-poor end of the Spite
plateau~\cite{Aoki:2009ce,2010arXiv1003.4510S} (towards lower values
of \lisv/H.) It is important to note that such a feature, as puzzling
as it may be, in itself does not alleviate the tension between the
SBBN prediction and the lithium observations, but rather enhances the
controversy.  The observational status on that issue is still very
much under dispute, as is best illustrated by recent
Ref.~\cite{Melendez:2010kw}. That work does not seem to confirm such a
``sagging tail'' in \lisv/H but rather finds two plateau values, $\sim
1.5$ and $1.9\times 10^{-10}$.  At this point, we choose the range
(\ref{eq:LIobs}) based on the overall range of lithium abundance in
the metal-poor stars along the Spite plateau. Should the mechanisms
for astrophysical depletion of the lithium abundance firm up, the
range (\ref{eq:LIobs}) must be shifted upward.

In addition to (\ref{eq:LIobs}), we shall also put the following
constraints on the remaining light elements which are produced in
observable quantities in BBN:
\begin{align}
\label{eq:Dobs}
  \deut/\Hyd & \leq 4\times 10^{-5} , \\
\label{eq:HEobs}
 0.24 & \leq Y_p \leq 0.26 , \\
\label{eq:LI6obs}
  \lisx/\Hyd & \leq 6\times 10^{-11}, \\
\label{eq:hetobs}
  \het/\deut &\leq 1.
\end{align}
The upper limit~(\ref{eq:Dobs}) corresponds to the highest reliable
determination~\cite{Burles:1997fa,Kirkman:2003uv} of D/H in a QSO
absorption system with a simple enough velocity structure. Given the
significant scatter in the various determinations of its primordial
value, the possibility of deuterium astration remains and values as
high as~(\ref{eq:Dobs}) cannot be convincingly excluded at the present
stage.  For \hef\ the inference of its primordial mass fraction is
potentially plagued by systematic uncertainties and values in the
range~(\ref{eq:HEobs}) have been witnessed over the years with most
recent determinations being on the higher
side~\cite{Izotov:2010ca,Aver:2010wq}. Observations of the isotopic
ratio $\lisx/\lisv$ in metal poor halo stars are extremely
difficult. Though a $5\%$ plateau value has been
claimed~\cite{Asplund:2005yt}, its inference has also been
challenged~\cite{Steffen:2010pe} so that we only impose
(\ref{eq:LI6obs}) as an upper limit. Finally, with \deut\ being more
fragile than \het, the ratio \het/\deut\ is a monotonically increasing
function of time so that the solar-system
value~(\ref{eq:hetobs})~\cite{1993oee..conf.....P} provides an upper
limit on the primordial value.

\FIGURE[t]{
\includegraphics[width=0.6\textwidth]{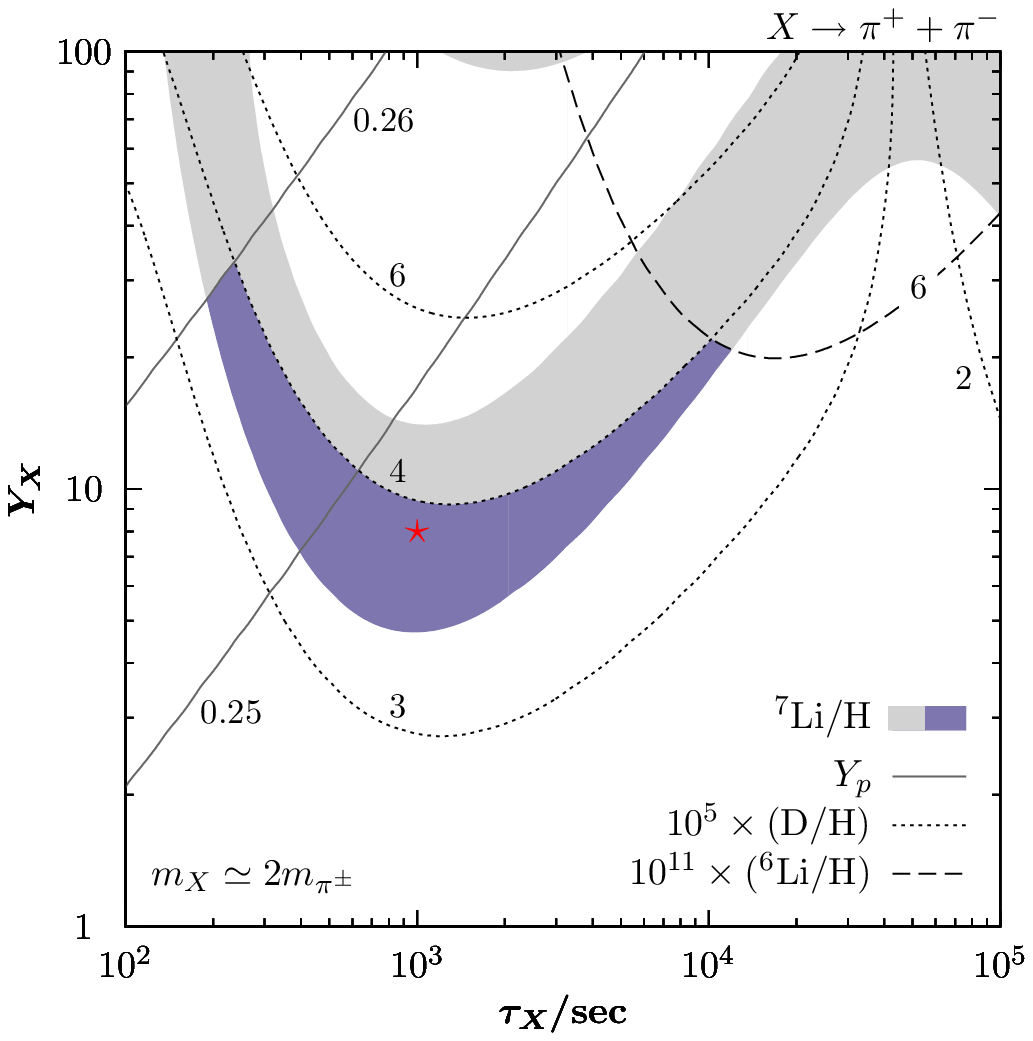}
\caption{\small Light element abundance yields in the plane of
  $X$-lifetime vs. $X$-abundance (prior to decay),
  $(\tau_X,\xi_{\pi}^{(X)}Y_X)$ with $\xi^{(X)}_{\pi}=1$ and
  negligible kinetic energy of injected pions.  The band shows the
  region $ 10^{-10}\leq \lisv/\Hyd \le 2.5\times 10^{-10} $ in which
  the BBN lithium prediction is reconciled with observationally
  inferred primordial values. In the dark (blue) shaded part all
  limits (\ref{eq:Dobs})-(\ref{eq:hetobs}) on the remaining light
  elements are respected. Contours of constant helium mass fraction
  $Y_p$ are shown by solid lines and of constant $\deut/\Hyd$
  abundance by dotted lines; the dashed line in the upper right corner
  corresponds to $\lisx/\Hyd = 6\times 10^{-11}$ with smaller values
  anywhere below. The star is the point in parameter space for which
  Fig.~\ref{evolution} was obtained.}
\label{figure1}
}

\FIGURE[t]{
\includegraphics[width=0.6\textwidth]{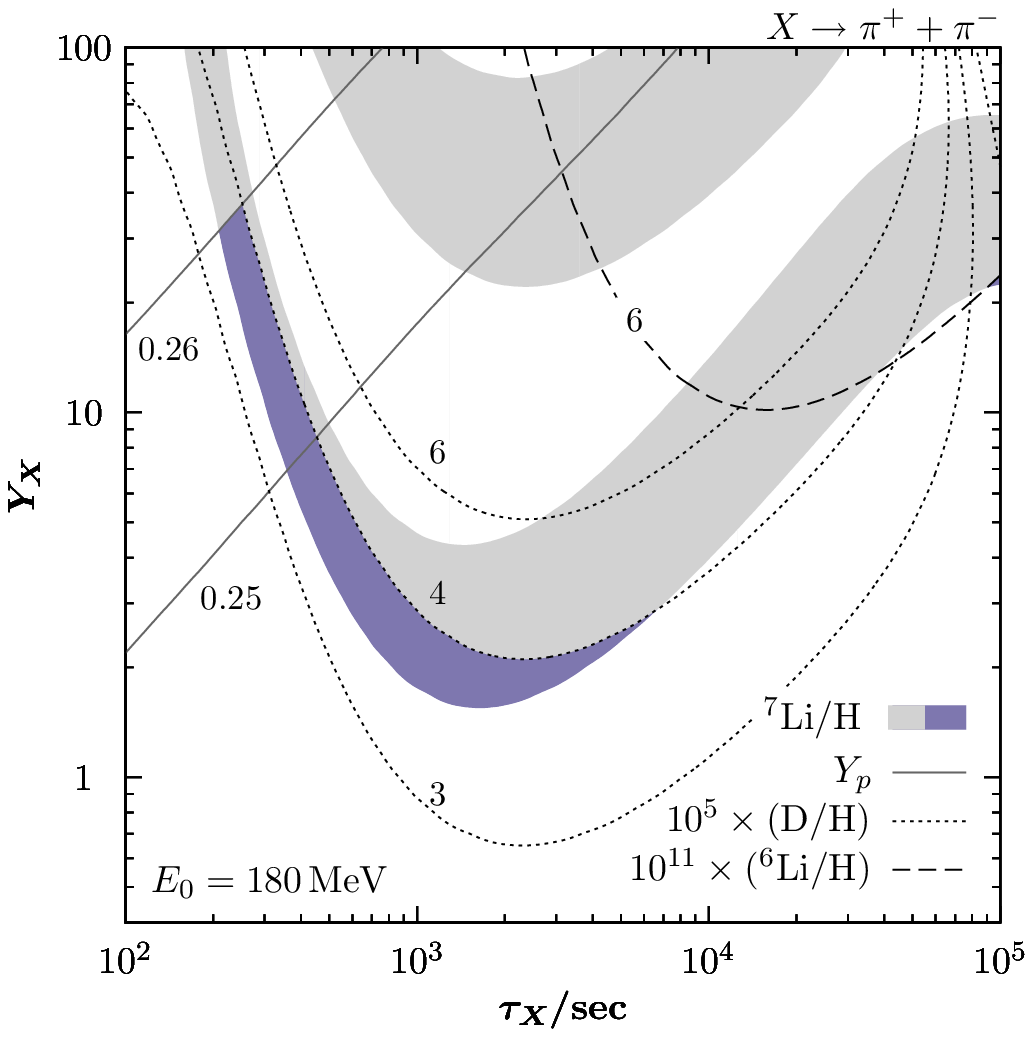}
\caption{\small Same as in Figure \ref{figure1}, but with a kinetic
  energy of injected pions $E_0 = 180$~MeV which approximately
  corresponds to the maximum of the pion-nucleon cross section.  }
\label{figure1b}
}

In Figure~\ref{figure1} we present a parameter scan in the
$(\tau_X,\xi^{(X)}_{\pi}Y_X)$-plane with $\xi^{(X)}_{\pi}=1$, in the
assumption of negligible kinetic energy or injected pions, so that all
reactions and decays occur at rest. The star is the point in parameter
space for which Fig.~\ref{evolution} was obtained. The shaded band
shows the region in which \lisv/H is within the observationally
favored range (\ref{eq:LIobs}).  More pions become available for
larger values of $Y_X$ so that (\lisv+\bes)/\Hyd\ decreases from its
SBBN value $\sim 5\times 10^{-10}$ at $Y_X= 1$ to $2.5\times 10^{-10}$
at the lower border of the shaded band. As expected, there is also an
associated increase of D/H from its SBBN value, $\sim 2.6\times
10^{-5}$, with contours of constant deuterium shown by the dotted
lines and running almost in parallel to constant lithium. The lowest
D/H value along the \lisv/H band is $ 3.25 \times 10^{-5}$ (for
$\tau_X\leq 10^4\ \seconds$).  The contours of constant \hef\ are
shown by the solid lines. The effect on \hef\ becomes stronger for
smaller $X$-lifetime values. As is well known, \hef\ exhibits the
strongest sensitivity on neutron/proton interconversions which take
place before the opening of the deuterium ``bottleneck''.  We further
observe that the secondary production of
$\lisx$~(\ref{eq:pi-he4-to-T}) becomes only important for large
lifetimes $\tau_X\gtrsim 10^4\ \seconds$ (dashed line). This, however,
is already the region in the parameter space in which the proposed
mechanism for lithium depletion by extra neutrons becomes
inefficient. By the same token, the photo-destruction of light
elements remains decoupled from the main effect as it only happens at
late times~(\ref{eq:t-dissoc}). The destruction of \bes\ and \deut\
can respectively be seen by the downward trend of the lithium band as
well as by the dotted curve labeled ``2''. Finally, we note that the
ratio~(\ref{eq:hetobs}) of $\het/\deut$ is never saturated throughout
the whole $(\tau_X,Y_X)$-plane. The dark (blue) shading of the band
corresponds to the region in which all constraints (\ref{eq:LIobs})
and (\ref{eq:Dobs})-(\ref{eq:hetobs}) are obeyed. It is this region,
for which the pion injection provides a satisfactory solution to the
lithium overabundance problem.

As was argued earlier, pions with kinetic energy close to the
delta-resonance may have a considerably larger efficiency of $p\to n $
conversion below $T_9 \simeq 0.4$.  To demonstrate this point, we take
the same assumptions as in Fig.~\ref{figure1} but assume a primary
pion injection energy of $E_0=180$~MeV. The results of the new scan
are presented in Fig.~\ref{figure1b}. The increased efficiency of
$p\to n$ conversion (along with stronger \pipm-\hef\ reactions) leads
to a lowering of the bands, \textit{i.e.}~less pions per baryon are
necessary for the required depletion of \lisv. Note, however, that the
overall effect is somewhat milder than initially expected. For a
lifetime $\tau_X = 10^3\,\seconds$ at which pions are still being
stopped, one finds a factor $\sim 2\div 3$ reduction in $Y_X$. For
larger lifetimes it only becomes about half an order of magnitude
(note the slightly different scales on the y-axes of
Figs.~\ref{figure1} and \ref{figure1b}.)  Though \bes\ is indeed
efficiently destroyed by strongly elevated neutron concentrations it
turns out that the limiting factor is the final step in the overall
\bes+\lisv\ depleting chain~(\ref{eq:li-depletion-chain}),
\textit{i.e.}  the subsequent \lisv-destruction via thermal proton
burning, $\lisv+p\to\hef+\hef$. This latter reaction effectively shuts
off for $T_9<0.3$ and the overall lithium abundance is produced in
form of \lisv.
 Residual
late $X$-decays at temperatures $T_9\lesssim 0.5$ also lead to the 
narrowing of the lithium band at small lifetimes
$\tau_X< 10^3\,\seconds$ and large $Y_X$. We
remark in passing that in that region, the accompanying increase in D
also leads to an enhancement in the $n$-sourcing thermonuclear DD and
DT fusion reactions.

\subsection{Decays to  kaons }
\label{sec:treatment-kaons}

Aiming at a reasonably accurate treatment of kaons is even more
difficult. Let us first focus on $\Kpm$, and assume for simplicity
that the kinetic energy is relatively small, so that one can neglect
the effects of the decay in flight.  In the case of kaon injection it
is important to include their hadronic decays into charged pions, as
well as the direct interaction of kaons with nucleons. The production
of charged pions in kaon decays is relatively easy to account for:
\begin{align}
  \Kpm & \to 
  \begin{cases}
     \pipm \pizero (\pizero) & 22.4\% \nonumber \\
     \pipm \pip \pim &5.6\%
  \end{cases}
\end{align}
which leads to a comparable population of $\Kpm$ and $\pipm$. As has
been already mentioned, there is no need to track muons (and the
associated neutrino yield) since $ P_{p\to n}^\nu \ll P_{p\to n}^\pi
$. For the average kinetic energy of produced charged pions we take
$E_0\sim 80$~MeV. Even in the two-body decay, \pipm\ have energies of
only $\sim110\ \MeV$ so that they fall somewhat lower than the
$\Delta$-resonance, and their incomplete stopping at late times will
not have a large effect on the efficiency of $p\to n$ conversion.

We now turn to the calculations of the direct capture of kaons on
nucleons.  Whereas the charge exchange reaction $\Km + p \to \bar{K}^0 +
n$ has $Q=-5.2\ \MeV$ and is thus not allowed kinematically on
threshold, \Km\ reactions on nucleons can proceed via the "$s$-quark
exchange reactions" with hyperons and pions in the final state:
\begin{align}
\label{eq:kprct}
  \Km + p &\to \Sigma^{\pm}\pi^{\mp},\, \Sigma^0 \pizero,\, \Lambda
  \pizero , \\
\label{eq:kprct2}
  \Km + n &\to \Sigma^{-} \pizero,\, \Sigma^0\pim,\, \Lambda\pim .
\end{align}
In order to obtain the inclusive $n\leftrightarrow p$ interconversion
cross sections, one then has to take into account the decays of the
strange baryons in the final states
\begin{align}
  \Sigma^{+} & \to 
  \begin{cases}
     p \pizero & 51.6\% \nonumber \\
 n \pip &48.3\%
  \end{cases} , &
  \Lambda & \to 
  \begin{cases}
     p \pim & 63.9\%\\
 n \pizero &35.8\%
  \end{cases} , \\
  \Sigma^{-} & \to \,\,\,\,\, n \pizero \quad \,\,99.8\% , &
  \Sigma^{0} & \to \,\,\,\,\, \Lambda \gamma \quad \,\,\,100\% .
\label{eq:Hyperon-branchings}
\end{align}

The cross section for each of the processes (\ref{eq:kprct}) and
(\ref{eq:kprct2}) can be obtained in terms of an isospin ($I$)
decomposition of the wave-functions of the reactants.  Together with
the knowledge of the relative phases $\phi$ of the individual isospin
amplitudes, the cross sections can then be inferred from the total
$\Km$-absorption cross sections of a given~$I$,
\begin{align}
  \label{eq:isospin-abs-cs}
 \sigma_I \simeq \frac{4\pi b_I}{k} \left| \frac{1}{1-i k A_I} \right|^{2} ,
\end{align}
where $A_I= a_I + i b_I$ are complex scattering lengths; $k$ is the
c.m.~momentum of the incoming state.
Taking into account the deviations from charge independence due to
$n$, $p$, and $\Kpm$ and $K^0$ mass differences
following~\cite{Dalitz:1960du,Martin:1970je} and using $A_{0} = (-1.74
+ i 0.70 )\ \fm$, $A_{1} = (-0.05 + i 0.63 )\ \fm$,
$\phi_{\mathrm{th}} = -52.9^\circ$ as well as a ratio $0.34$ of
$\Lambda\pi^{0}$ to total $I=1$ hyperon
production~\cite{Martin:1970je} one arrives at the values for
(\ref{eq:kprct}) and (\ref{eq:kprct2}).
In a final step, we use the branching
ratios~(\ref{eq:Hyperon-branchings}) to obtain the inclusive threshold
cross sections for proton/neutron interconversion,
\begin{align}
\label{KmN}
  \Km + p & \to n + X : \quad (\sigma v)^{\Km}_{pn} \simeq 32\ \mbarn ,\\
\label{KmN2}
  \Km + n & \to p + X : \quad (\sigma v)^{\Km}_{np} \simeq 13\ \mbarn ,
\end{align}
where in~(\ref{KmN}) the Coulomb correction has been factored
out. Whereas the first cross section is in good agreement with the
value previously used in the BBN context, the second one is a factor
of two smaller than in~\cite{Reno:1987qw}. In this regard, we remark
in passing that the latter reactions~(\ref{eq:kprct2}) are only due to
$I=1$ scatterings.

In principle, the nucleons for some of the final states of (\ref{KmN})
and (\ref{KmN2}) can have energies on the order of 30 MeV. While
protons will be stopped before inducing any nuclear changes, neutrons
with such energies are stopped primarily via interactions with $p$ and
\hef, and could split \hef\ nuclei in the collisions into some of its
constituents.  However, since \hef\ is one order of magnitude less
abundant than $p$, and since the maximum energy of $n$ is close to the
\hef\ desintegration threshold, we neglect such secondary effects.
In addition, we further note that due to isospin invariance we do not
need to consider hyperon production processes in $\Kp n$ scatterings.

As in the case of $\pim + \hef$, reactions of $\Km$ on helium have not
been considered in the BBN context.  Fortunately, the measurements of
Ref.~\cite{PhysRevD.1.1267} provide us with a detailed list of
branching ratios for $\Km$ absorption on \hef\ at rest. By accounting
for the decay modes~(\ref{eq:Hyperon-branchings}) we find the
following particle multiplicities per $\Km$ absorption
\begin{alignat}{2}
  \xi_{\het} &\simeq \xi_{\trit} \simeq 0.13 , & \xi_{n} & = 1.57 -
  0.34\lambda_{\deut} ,   \nonumber\\
  \xi_{\deut} & = 0.17 + 0.34\lambda_{\deut},\qquad & \xi_{p} &= 1.29 -
  0.34\lambda_{\deut} .
\label{eq:branchings-he4-Kminus}
\end{alignat}
When a final state is not resolved we have, for simplicity, assumed
that a fraction of $\lambda_{\deut}$ is released in form of \deut\ and
$(1-\lambda_{\deut})$ in form of nucleons, and we adopt
$\lambda_{\deut}=0.5$. Thus, for example, in the reaction $\Km+\hef\to
\Lambda (\Sigma^0) (pnn)$, which occurs with branching fraction
$22.5\%$, we assume $50\%$ $\Lambda (\Sigma^0) (Dn)$ and $50\%$
$\Lambda (\Sigma^0) (pnn)$. Thereby, we are neglecting a certain
fraction of mass-3 nuclei.  In this regard, note that---unlike in the
case of the threshold reaction $\hef+\pim$---an accurate computation of the secondary
\lisx-yield is more difficult as the \het/\trit\ injection energy
is now continuous.
However, as we have already seen in the previous section, \lisx\
production in excess of observationally constrained levels is an issue
only in the region where (\ref{eq:li-depletion-chain}) looses its
efficiency. Thereby, we shall make the simple assumption that on
average $\trit$ and $\het$ carry one third of the liberated energy,
$\VEV{E_{\trit}}\sim 30\ \MeV$ and $\VEV{E_{\het}}\sim 50\ \MeV$,
respectively.  (With our assumptions) the latter value is higher as
\het\ stems predominantly from processes with $\Lambda$ and not
$\Sigma$ production.
Finally, we note that the contribution of hyperfragments of
$_{\Lambda}^{4}\mathrm{He}$ is $\sim 2\%$~\cite{PhysRevD.1.1267} which
thus does not pose any further complication.
With the above multiplicities~(\ref{eq:branchings-he4-Kminus}) one
then obtains effective cross sections for each isotope in the final
state,
\begin{align}
  \Km +\hef \to N + \Pi: \quad (\sigma v)_N^{\Km} = \xi_N 
  (\sigma_{\hef}^{\Km} v) ,
\end{align}
with $\Pi$ symbolizing an arbitrary pionic final state and $N =
\het,\,\trit,\,\deut,\,n$ or $p$. The extraction of $(\sigma
v)_N^{\Km}$ is similar to the pion case and is given
by~(\ref{eq:sigma-Km-hef}) in Appendix~\ref{sec:pion-kaon-capture}.

In order to take into account the effects from electromagnetic energy
injection in kaon decays we recall from the previous section that
$\pipm$ as well as $\mu^{\pm}$ are unlikely to initiate an
electromagnetic cascade for $\tau_{X}\gtrsim 10^{4}\ \seconds$. The
multiplicities in the $\Kpm$-decay to $\mu^{\pm}$ (including muons
from $\pipm$ final states) and to $\pi^0$ are $\xi_{\mu}^{\Kpm}\simeq
1$ and $\xi_{\pi^0}^{\Kpm}\simeq 0.3$, respectively. For the
electromagnetic energy release in the muon decay we again take
$E_{\mathrm{inj}}^{\mu}\simeq m_{\mu}/3$ and for $\pi^0$ we assume
that, in addition to their rest mass, they carry on average
approximately one third of the energy released in the particular decay
channel, yielding $E_{\mathrm{inj}}^{\pi^0}\sim 250\ \MeV$. Taken
together, this amounts to a ``branching fraction''
$\mathrm{Br}_{K^{\pm}\to {\mathrm{vis}}}\sim 0.55$ and a total
electromagnetic energy injection of $E_{\mathrm{inj}} \sim 2\times
\mathrm{Br}_{K^{\pm}\to {\mathrm{vis}}} m_{\Kpm}$ per $X$-decay.  As
for simplicity we assume that the kaon kinetic energy is small, we do
not have to worry about the increase of the electromagentic deposition
by decaying in-flight kaons.
However, we emphasize that even in the case of slow kaons, and because
of the exponential sensitivity to $\tau_X$, $\Orderof{10}$ \Kpm\ per
baryon will have strong effect on nuclear abundances once a
photodissociation threshold is met for $\tau_X > 10^4\ \seconds$.

\FIGURE[t]{
  \includegraphics[width=0.6\textwidth]{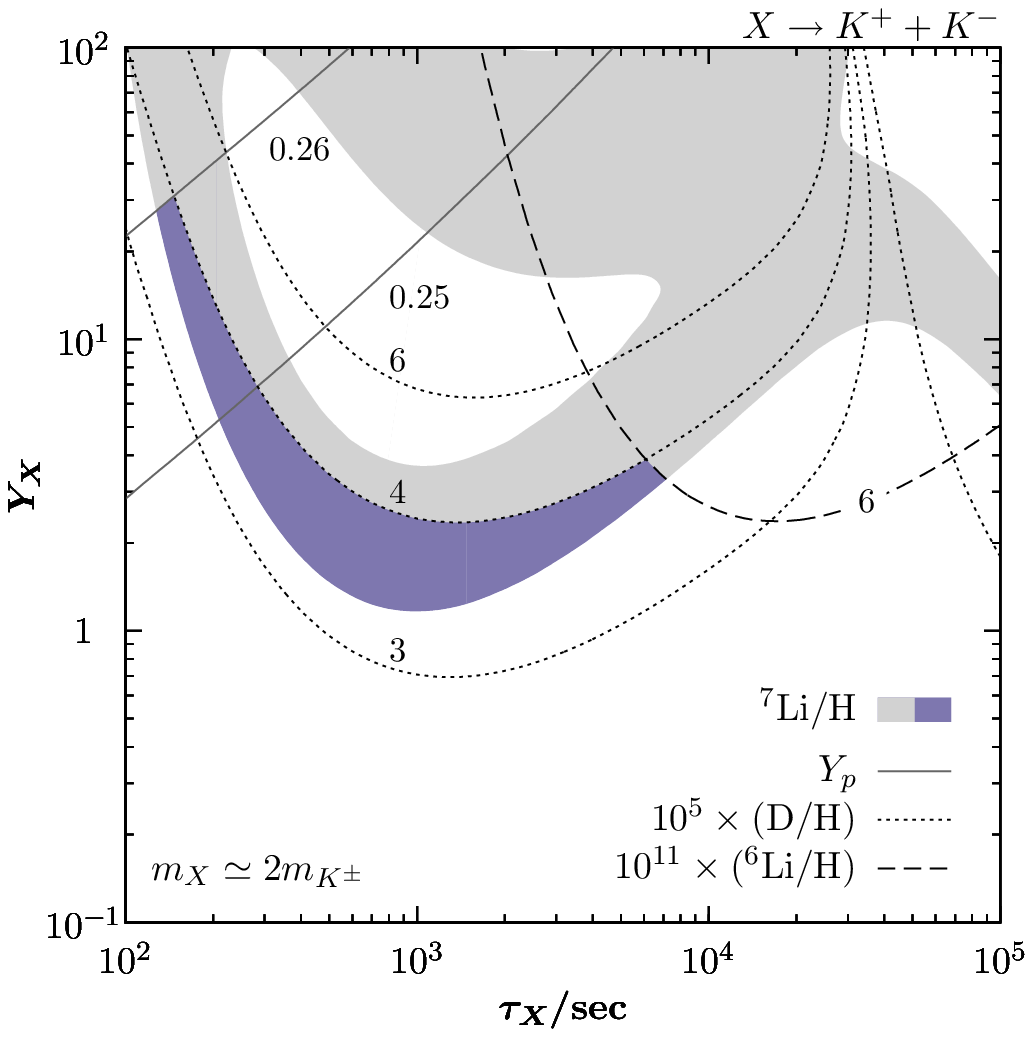}
  \caption{\small Same as in Fig.~\ref{figure1}, but for the case of
    $\Kpm$ injection and assuming one charged kaon pair per $X$-decay,
    $\xi^{(X)}_{\Kpm} = 1$, being injected close to the kinematic
    threshold.  Because of the larger cross sections of $K^-$ on
    nucleons and helium, one requires a fewer number of initial $X$
    decays. The trend of depleting $\bes+\lisv$ for larger values of
    $Y_X$ is eventually reverted because of the enhanced production of
    lithium in form of \lisv\ due to numerous extra neutrons.}
\label{figure2}
}

The kaon-induced solution to the cosmological lithium problem is
plotted in Figure~\ref{figure2}. In contrast to the case of primary
pions at rest (see Fig.~\ref{figure1}), a reduction of $\lisv/\Hyd$ to
observed values~(\ref{eq:LIobs}) is already possible for one injected
kaon per baryon. This is evidently due to the fact that $\Km$ cross
sections on nucleons and helium are significantly larger than for
thermal $\pim$. Moreover, pions from kaon decays only amplify any effect on
the light elements. Notice, however, that for too high of an
$n$-abundance (large $Y_X$) $\lisv + \bes$ starts to increase
again. Though $\bes$ is still being depleted, net production of
lithium in form of $\lisv$ eventually takes over. Similar to the case
of primary $\pipm$ we find that solutions corresponding to
$\tau_{X}\gtrsim 10^4\ \seconds$ are not viable because of \lisx\
overproduction. This again excludes the potentially interesting region
with $\tau_{X}=\mathrm{few}\times 10^5\ \seconds$ in which previously
formed deuterium is destroyed in photo-dissociation and brought back
to its SBBN prediction.

We now turn to the case of injected neutral kaons $K^0,\bar
K^0$---propagating as $K_L$ and $K_S$. Whereas $K_S$ definitely decay before
interacting with the nuclear background, the relatively large $K_L$
lifetime $\tau_{K_L} = 5.1\times 10^{-8}\ \seconds$ may render its
impact on the BBN output even more drastic then in the $\Kpm$ case. In
contrast to $\Kpm$, however, $K_L$ are not stopped by electromagnetic
interactions so that one should take into account the energy dependence
of their cross sections with nucleons. Again, we avoid this complication by assuming the 
injection of kaons close to the threshold. 

For $K_L$-nucleon scattering we exploit charge indepence and use
isospin relations of the strong interaction processes. Indeed,
conservation of strangeness implies that the inelastic scattering of
$K_L$ on nucleons essentially resembles the one of $\bar K^0$, as 
 $K_L = 2^{-1/2}(K^0-\bar K^{0})$ up to small
$CP$-violating corrections. Thus, up to corrections due to $\Kpm$, $\bar K^0$,
and $n$, $p$ mass differences and other isospin-violating effects, one finds
\begin{align}
\label{eq:KLrct1}
  \sigma(K_L p\to Y \pi) &\simeq \frac{1}{2} \sigma(\bar K^0 p\to Y \pi
  ) \simeq \frac{1}{2} \sigma(K^{-} n \to \tilde Y \tilde\pi)
\\
\label{eq:KLrct2}
  \sigma(K_L n\to Y' \pi') &\simeq \frac{1}{2} \sigma(\bar K^0 n\to Y' \pi'
  ) \simeq \frac{1}{2} \sigma(K^{-} p \to \tilde Y' \tilde\pi')
\end{align}
where $Y,Y'(\pi,\pi')$ stands for a hyperon (pion) and $\tilde
Y,\tilde Y'(\tilde\pi,\tilde\pi')$ are the associated states with
flipped third isospin component---the final states
of~(\ref{eq:kprct2}) and (\ref{eq:kprct}), respectively. The energy
dependence of the cross sections not far from the threshold is
inferred from~(\ref{eq:isospin-abs-cs}).  Indeed, since
(\ref{eq:KLrct1}) is a pure $I=1$ process, $ \sigma(K_L p\to Y \pi) =
\sigma_1/2$.  With the branching fractions
(\ref{eq:Hyperon-branchings}) the inclusive cross sections for $p\to n
$ as well as for $n\to p$ conversion are then readily found from
(\ref{eq:KLrct1}) and (\ref{eq:KLrct2}), respectively.

In addition to the reactions listed above, the following processes are
possible: $K_{L} n \to \Km p$, $K_{L} p \to \Kp n$ and $K_{L} p \to
K_{S} p$. The cross sections for the former two processes are
approximately equal, with $K_L$ scattering being mediated by its
$\bar{K}^0$ or $K^0$ component, so that they can be inferred from an
expression for $\Km p \to \bar{K}^0 n$ in~\cite{Martin:1970je} by
detailed balancing.  Though the above processes potentially contribute
to the loss of $K_{L}$ and/or generation of $\Kpm$, the probability of
such processes within a kaon lifetime is very small so that they can
be safely neglected.
In addition to the processes discussed above, one should again account
for the induced charged pion population from to $K_L$ and $K_S$
decays. The respective branching fractions read
\begin{align}
  K_L \to \pip X, ~  \pim  X  :\quad 46.3\% ,
\qquad 
  K_S \to \pip \pim :\quad 69.2\% ,
\end{align}
where $X$ is the inclusive particle final state without charged pions
(not to be confused with $X$-relics). As in the case of charged kaons,
the kinetic energy of outgoing pions can be as large as $m_K/2 - m_\pi
\simeq 110$ MeV, and already in the regime where $p+\pim$ cross
sections is enhanced.  Finally, we would like to make a comment on
\hef-$K_L$ reactions. As we have just seen, $K_L$ scattering on
nucleons can be related to $\Kpm$-reactions using charge independence
and isospin symmetry. On these grounds there is no obvious reason to
suspect that $K_L$-reactions on helium will be drastically different
in magnitude when compared to $\Km$.

Instead of attempting to infer the \hef-$K_L$ cross sections we shall
rather outline some general features connected to the $K_L$ case. By
turning off the $\Km$ reactions on helium in the Boltzmann code we can
directly assess the difference between the charged and neutral kaon
case. Whenever the $K_L$ energy in the final state is very small,
\emph{i.e.} on the order of a few MeV, the results are essentially
identical to Fig.~\ref{figure2}.
To conclude the kaon section, we would like to comment that several
avenues for improvement still exist after our analysis.  For example,
the treatment of pions from kaon decays is done via assigning them a
simplistic average energy, while of course a continuous energy
spectrum would be required. Furthermore, the effects of incomplete
stopping of $\Km$ and energetic $K_L$ would also need to be included
if one intends to explore the BBN sensitivity to a wider range of
$m_X$.
Nevertheless, we also remind the reader that the effect of in-flight
pions is not as drastic as naively expected so that the corrections
due to continuously distributed $\pipm$ from kaon decays will be only
very moderate. Moreover, we close this section by remarking that, in
principle, no further conceptual complications arise when considering
more energetic injections of $\Kpm$ and $K_L$ and that such an
analysis can readily be performed by employing a treatment along the
lines of the previous section.

\subsection{Decays to muons and neutrinos}

Stopped muons in the final state of the $X$-decay are sourcing
energetic muon- and electron neutrinos. 
In order to include these neutrinos in the set of Boltzmann equations
we shall use the fact that $\Gamma^{\nu}_{\rm stop}$ is slower than
the Hubble rate. We account for the continuous neutrino injection and
energy redshifting by calculating the neutrino phase space
distribution function $f(T,E_\nu)$. Previous analysis of BBN modified
in the presence of the energetic neutrinos \cite{Kanzaki:2006hm}
concentrated on the effects of energy deposition and neglected the
direct nuclear-chemical impact of
neutrinos~\cite{1984MNRAS.210..359S}, which turns out to be a more
important effect at early times.

The calculation of $f(T,E_\nu)$ depends on the injection rate
$\Gamma_{\mathrm{inj}}$, and on the primary spectrum of neutrinos at
the time of injection,
\begin{eqnarray}
\label{eq:primary-nu-spectrum}
F^0_{e,\, \mu}(E,E_0) = \left\{ 
\begin{array}{rll} 
\nu_e,\,\overline\nu_e: &12E^2(E_0-E)E_0^{-4} &{\rm for}\quad 0<E<E_0,\\
\nu_{\mu},\,\overline\nu_{\mu}: &2E^2(3E_0-2E)E_0^{-4} &{\rm for}\quad 0<E<E_0,\\
&0 &{\rm for}\quad E>E_0.
\end{array}
\right.
\end{eqnarray}
where $E_0 \simeq m_{\mu}/2$ is the neutrino end-point energy in the
muon decay, and $F^0_{e,\, \mu}$ is normalized to unity, $\int F^0_{e,\,
  \mu} dE = 1$. Once injected, neutrinos are subject to flavor
oscillations so that the primary energy spectrum will be
``distorted''.
Apart from vacuum oscillations, the neutrino propagation is affected by
the coherent neutral- and charged-current interactions with particles in primordial 
plasma. To assess the importance of neutrino-refraction we can
compare the vacuum contributions in the neutrino Hamiltonian,
\begin{align}
\label{eq:vac-osc}
  \frac{\Delta m_{\mathrm{sol}}^2}{4E} \gtrsim 10^{-13}\ \eV , \qquad \frac{|\Delta
     m_{\mathrm{atm}}^2|}{4E} \gtrsim 10^{-11}\ \eV,
\end{align}
with the matter-induced potential $V_{\mathrm{M}}$; $E < E_0$. Here,
$\Delta m_{\mathrm{sol}}^2 \simeq 7.7\times 10^{-5}\ \eV^2$ and
$|\Delta m_{\mathrm{atm}}^2| \simeq 2.4 \times 10^{-3}\ \eV^2$ are the
respective mass-squared differences responsible for solar and
atmospheric neutrino mixing~\cite{Amsler:2008zzb}. For $V_{\mathrm{M}}$
one can write (see \textit{e.g.} \cite{Raffelt:1996wa})
\begin{align}
\label{eq:mat-osc}
  V_{\mathrm{M}} 
 & = \pm \left( 8\times 10^{-19} \ \eV \right) T_9^3  \times
  \begin{cases}
    ( - 1 + 4 Y_{\nu_e}  + 3 Y_e) & \mathrm{for}\ \nu_e (\overline\nu_e) \\
    (  - 1 + 2 Y_{\nu_e} + Y_e)   & \mathrm{for}\ \nu_{\mu,\tau} (\overline\nu_{\mu,\tau})
  \end{cases} ,
\end{align}
where the $Y_i$ denote the particle-antiparticle asymmetries
normalized to baryons. The numerical factor in front of (\ref{eq:mat-osc}) is $G_F n_b/\sqrt{2}$ with
the overall $+(-)$ sign for $\nu (\overline{\nu})$. Unlike the case of charged leptons 
where  $Y_e = \Orderof{1}$, the asymmetry in the neutrino sector could be large.
In this paper we do not consider such  scenarios and limit the asymmetry
in the neutrino sector by requiring 
$|Y_{\nu}| < 10^4$.  The last condition corresponds to $|\xi_{\nu}|
\equiv |\mu_{\nu}|/T < 5\times 10^{-5}$ with $\mu_{\nu}$ being the
neutrino chemical potential. Comparing (\ref{eq:vac-osc}) with
(\ref{eq:mat-osc}) then implies that the flavor-evolution of injected
neutrinos is given by their vacuum oscillations. Note that the
restriction on the neutrino chemical potentials also renders
neutrino-neutrino self-interactions unimportant which usually have the
effect of locking the neutrino modes to each other, leading to a
coherent oscillatory behavior~\cite{Pastor:2001iu}.  

A major simplification in calculation 
of $f(T,E_\nu)$ occurs due to the large rates for vacuum
oscillations,
\begin{align}
  \Gamma_{i,\,\mathrm{osc}} = 1.2\times 10^8 \left(\frac{1\
      \MeV}{E}\right)\left(\frac{|\Delta m_i^2|}{1\ \eV^{2}}\right) \gtrsim
  \begin{cases} 
    175\ \seconds^{-1} & i = \mathrm{sol} \\5.5\times 10^3\ \seconds^{-1}
    &i = \mathrm{atm}
  \end{cases}
\end{align}
compared to Hubble rate and neutrino injection rate,
 $  \Gamma_{i,\,\mathrm{osc}} \gg H, \Gamma_X.$
This allows us to replace the primary injection spectrum $F^0_{e}$ in
(\ref{eq:primary-nu-spectrum}) by an effective one,
\begin{align}
\label{eq:primary-effective}
  F_e &= \VEV{P_{ee}} F_e^0 + \VEV{P_{\mu e}}F_{\mu}^{0} 
\end{align}
where the $\VEV{P_{ee}}$ and $\VEV{P_{\mu e}}$ are the
${\nu_e}(\overline\nu_e)$-survival and $\nu_{\mu}(\overline\nu_{\mu})
\to \nu_e (\overline\nu_e)$-appearance probabilities, averaged over
oscillations,
\begin{align}
  \VEV{P_{ee}} & = 1 - \frac{1}{2} \sin^2{2\theta_{12}} \simeq 0.57 , \\
  \VEV{P_{\mu e}} &= \VEV{P_{ e\mu}} =
  \frac{1}{2}\sin^2{2\theta_{12}}\cos^2(\theta_{23}) \simeq 0.23 . 
\end{align}
The vacuum mixing angles are given by $\sin^2\theta_{12} = 0.312$ and
$\sin^2\theta_{23} = 0.466$~\cite{Amsler:2008zzb} and we have assumed
$\theta_{13} = 0 $.  Note that (\ref{eq:primary-effective}) already
accounts for the appropriate reduction of the electron neutrino flux
due to $\nu_e(\overline\nu_e)\to\nu_{\mu}(\overline\nu_{\mu})$ and
$\nu_e(\overline{\nu}_e)\to\nu_{\tau}(\overline{\nu}_{\tau})$
disappearances.%
\footnote{In a similar fashion, $F_{\mu} = \VEV{P_{\mu\mu}} F_\mu^0 +
  \VEV{P_{e \mu}}F_{e}^{0}$ and $F_{\tau} = \VEV{P_{e\tau }} F_e^0 +
  \VEV{P_{\mu \tau}}F_{\mu}^{0}$ with $\int dE (F_e + F_{\mu}
  +F_{\tau}) = 2$.}

Defining $f(T,E_\nu)$ in such a way that $\int f(T,E_\nu)dE_\nu $ is
equal to the total number of energetic neutrinos per baryon, we arrive
at the neutrino distribution function,
\begin{align}
\label{eq:neut-distro}
  f_{e}(T,E_\nu) = \int_{T}^{\infty} \fr{dT_1\Gamma_{\rm inj
  }Y_X(T_1)}{H(T_1)T_1}  F_{e}\left(E_{\nu},\fr{E_0 T}{T_1}\right) ,
\end{align}
where $\Gamma_{\mathrm{inj}} = \Gamma_X$, since we assume that the
neutrino-sourcing muons result from the decays of the $X$-relics.

Electron antineutrinos will have the largest effect on the BBN
network. Integrated with the weak cross-section over energy, the
distribution function gives the energetic neutrino-induced rate of
$p\to n$ conversion:
\begin{align}
\label{eq:avg-cs-nubar}
\overline{\nu}_{e} + p &\to n + e^{+} : \quad \Gamma^\nu_{pn} = n_b(T)
\int_{0}^{E_0} \sigma_{pn}^{\bar\nu}(E_{\nu}) f_e(T,E_\nu) dE_\nu
\end{align}
where $n_b(T)$ is the number density of baryons at temperature $T$.
The cross section for quasi-elastic neutrino nucleon scattering reads
(see e.g.~\cite{Bemporad:2001qy})
\begin{align}
\label{eq:cs-nubar}
  \sigma_{pn}^{\bar\nu}(E_{\nu}) =
  0.0952\times 10^{-42} \left(\frac{p_e E_e}{1\ \MeV^2}\right)
  S(E_{\nu})\ \cm^{2} ,\quad Q \simeq -1.8\ \MeV ,
\end{align}
where $E_{\nu}$ and $E_e = E_{\nu} - \Delta m_{np}$ are the energies
of $\overline{\nu}_e$ and $e^{+}$ in the rest frame of the proton,
respectively; $\Delta m_{np} \simeq 1.293\ \MeV$ is the neutron-proton
mass difference and $p_e$ is the positron momentum. We introduce a
correction factor $S(E_{\nu}) = (1-0.0063 {E_{\nu}}/{\MeV})$ which
improves the agreement between the simple formula (\ref{eq:cs-nubar})
and a precise evaluation of $\sigma_{pn}^{\bar\nu}$
in~\cite{Strumia:2003zx} to better than $1\%$ in the $E_{\nu}$-regime
of interest.

In our code we also account for exoergic $ n\to p $ conversions via
$\nu_e + n \to p + e^{-}$. This process, however, is only important
for $\tau_X\lesssim 180\ \seconds$, \textit{i.e.} prior to $n$
consumption by \hef. The associated cross section $
\sigma_{np}^{\nu}(E_{\nu})$ is obtained from (\ref{eq:cs-nubar}) with
$E_e = E_{\nu} + \Delta m_{np}$ and $S(E_{\nu})= 1$. As in
(\ref{eq:avg-cs-nubar}) we average $ \sigma_{np}^{\nu}(E_{\nu})$ over
$f_e$ in order to obtain the rate $\Gamma^\nu_{np}$ for
$ n\to p $ conversion.

\FIGURE[t]{\includegraphics[width=0.6\textwidth]{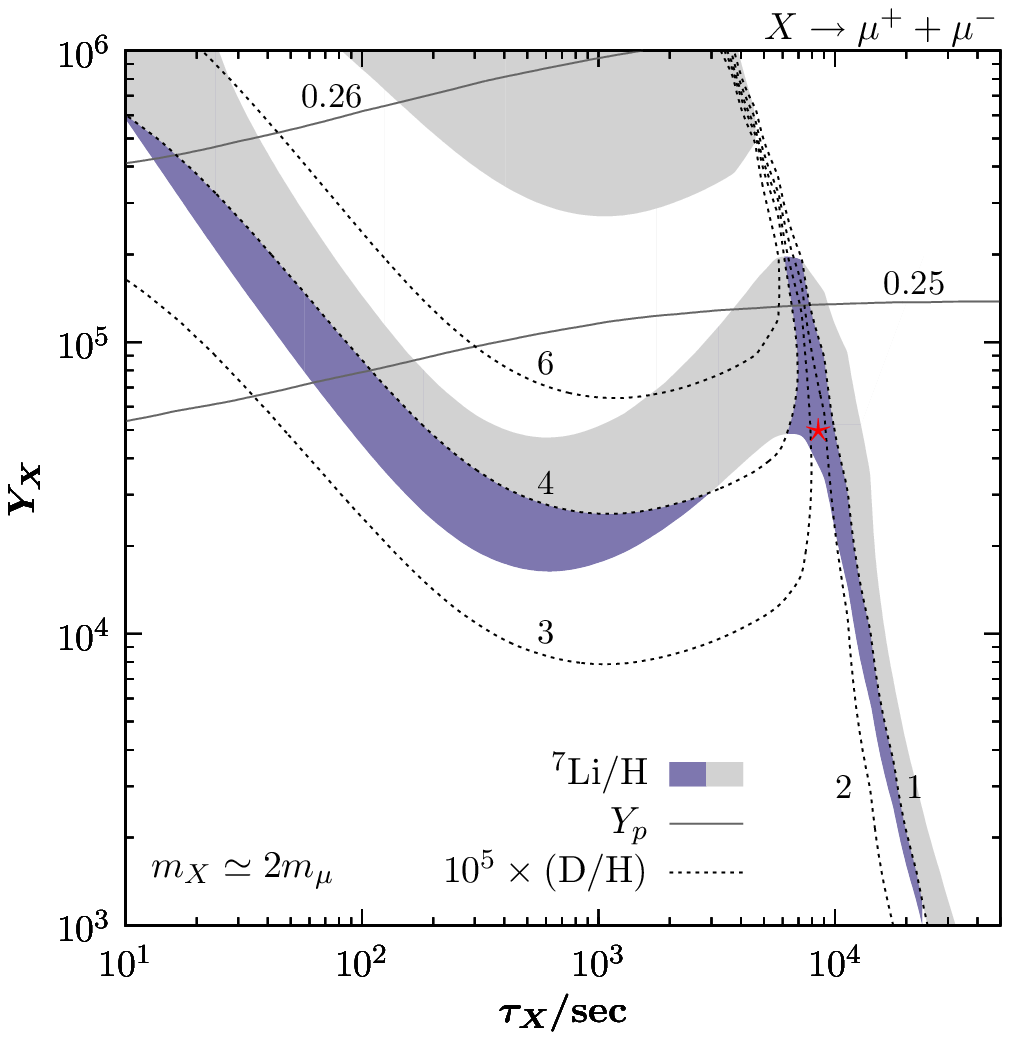}
\caption{\small Same as in Fig.~\ref{figure1}, but for $X$ decaying
  into $\mu^{+}\mu^{-}$ pairs which eventually decay into neutrinos
  (and $e^{\pm}$).  Note the smaller values $\tau_{X}$ in comparison
  to the previous figures. Energetic neutrinos are accumulating over
  time so that $X$-decays as early in time as $t=10\ \seconds$ can
  affect the BBN network.  Noteworthy is also the increased \hef\
  abundance for large values of $Y_X$ over the whole $\tau_{X}$-range
  since the $X$-matter density is contributing to $H$. In this figure,
  $m_{X} \simeq 2m_{\mu}$. A novel feature is also the appearance of a
  second observationally favored region for $\tau_X\simeq 10^4\
  \seconds$. No additional \lisx\ in excess of SBBN values is produced
  and electromagnetic energy injection depletes previously formed
  D/H. For the parameters marked by the star the temporal evolution of
  the light elements is shown in Figure~\ref{evolution2}.}
\label{neutrinos}
}

Muon neutrions $\nu_{\mu}$ and $\overline{\nu}_{\mu}$ which are
sourced from muon decays (at rest) are not capable of interconverting
protons and neutrons as their maximum injection energy $m_{\mu}/2$
lies below the reaction threshold $\Orderof{100\ \MeV}$. However,
${\nu}_{\mu}({\overline\nu}_{\mu})$ along with
${\nu}_{e,\tau}({\overline\nu}_{e,\tau})$ are in principle capable of
dissocia\-ting \hef\ via their neutral current interactions. Among the
possible final states, only the ones with mass-3 elements have an
appreciable cross section in the low energy regime
$E_{\nu}<m_{\mu}/2$~\cite{2008ApJ...686..448Y}. We include the
following neutral current (NC) and charged current (CC) reactions into
our Boltzmann network
\begin{align}
\label{eq:nu-he4-rct}
  \mathrm{NC:\ } 
  \begin{cases}
    \hef + \stackrel{(-)}{\nu}_{\!\!\!e,\mu,\tau} \to \stackrel{(-)}{\nu}_{\!\!\!e,\mu,\tau}  + p +\trit \\
    \hef +  \stackrel{(-)}{\nu}_{\!\!\!e,\mu,\tau} \to  \stackrel{(-)}{\nu}_{\!\!\!e,\mu,\tau}+ n +\het
  \end{cases},
\quad
  \mathrm{CC:\ } 
  \begin{cases}
\vphantom{ \stackrel{(-)}{\nu}_{e,\mu,\tau}}
    \hef + \nu_e \to e^{-} + p +\het \\
\vphantom{ \stackrel{(-)}{\nu}_{e,\mu,\tau}}
    \hef + \overline\nu_{e} \to e^{+}+ n +\trit
  \end{cases}.
\end{align}
For the NC reactions of muon and tau neutrinos we infer the reaction
rates averaged over $f_{\mu,\tau}$ by following the same logic from
(\ref{eq:primary-effective}) through (\ref{eq:neut-distro}).  For the
cross sections we use fitting formulas interpolating the tables
provided in~\cite{2008ApJ...686..448Y}.

\FIGURE[t]{\includegraphics[width=0.6\textwidth]{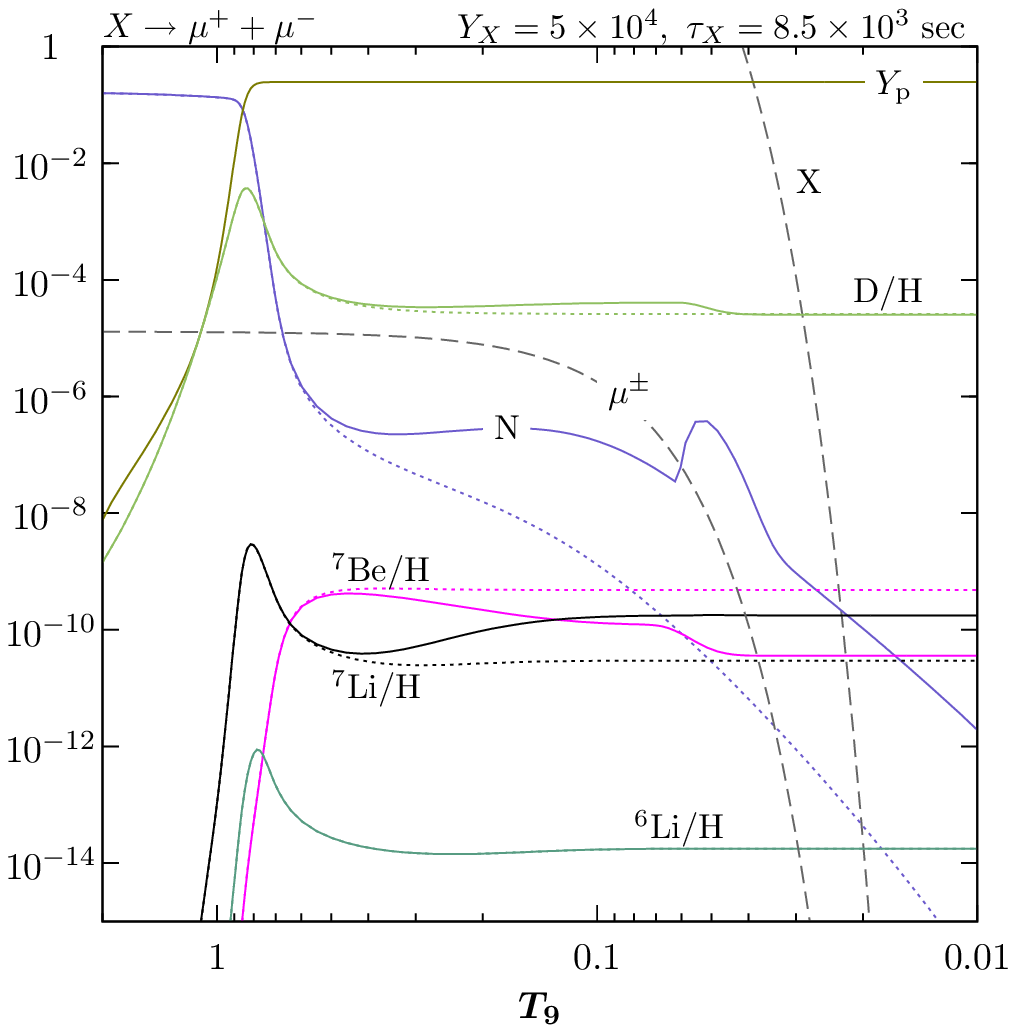}
\caption{\small Temperature evolution of light nuclei, meta-stable
  parent $X$ particles and daughter $\mu^{\pm}$ for $Y_X = 5\times
  10^4$ prior to decay and $\tau_{X} = 8.5\times 10^3\ \seconds$.
  Starting off with the previously observed effects which are induced
  by an elevated neutron abundance (suppressed \bes/H, elevated
  \deut/H), electromagnetic energy injection eventually dissolves
  deuterium and beryllium for $T_{9}\lesssim 0.06$. This not only
  reinstates the SBBN D/H prediction (corresponding dotted line) but
  also implies that lithium is mainly produced in form of \lisv.}
\label{evolution2}
}

Given that the efficiency for neutron-to-proton conversion per each
injected neutrino is so low [cf. (\ref{estimate2})], a high initial
$X$-abundance (in comparison to baryons) is needed in order to achieve
an appreciable reduction of $\lisv+\bes$. In our code we take the
extra contribution of $\rho_X$ to the total energy density into
account. By comparing $\rho_X$ (prior to decay) with the radiation
energy density~$\rho_{\rad}$,
\begin{align}
  \frac{\rho_X}{\rho_{\rad}} \simeq \frac{10^{-6} Y_{X}}{T_9} \left(
    \frac{m_{X}}{1 \GeV} \right) ,
\end{align}
we see that a GeV-scale relic $X$ starts contributing to the Hubble
rate appreciably during BBN for $Y_X\gtrsim 10^5$.

In Figure~\ref{neutrinos} we show the results of our computation for
$X\to \mu^{+}\mu^{-}$ in the usual $(\tau_X,\xi^{(X)}_{\mu} Y_X)$ plane
for $\xi^{(X)}_{\mu} =1$. For simplicity, again, we assume the muon injection close to 
the kinematic threshold, $m_{X} = 2m_{\mu}$, although this assumption 
is less crucial compared to the hadronic case. Note that this time we have extended the
$\tau_X$ range to lifetime values as small as $10\ \seconds$. This is
because the usual temporal correlation between $X$-decay and its
effect on the BBN yields breaks down.  For example, neutrinos injected at 
$t=10\ \seconds$ may still induce $p\to n $ conversion at $t=100\
\seconds$. As expected, the results show a similar pattern as in the
pion and kaon cases with $Y_{X}$ scaled to larger values.  In the neutrino case,
however, one finds an increased \hef\ abundance for $Y_X\gtrsim 10^5$
over the whole $\tau_{X}$-range. This is because the $X$ energy
density is contributing to the Hubble rate, leading to an earlier
$n/p$ freezeout and thereby to a higher \hef\ abundance. Though this
makes results sensitive to $m_{X}$, it is nevertheless only a mild
dependence for $m_{X}\lesssim 1\ \GeV$ in the interesting region
$Y_{X} \sim 10^4$, in which lithium is reconciled with observations.
 As it further turns out, the neutrino
reactions dissociating $\hef$ are giving at most a marginal correction
to the BBN yields even for $X$-abundances as high as $Y_X = 10^6$. The
reasons are that the processes of (\ref{eq:nu-he4-rct}) possess
threshold energies $E_{\mathrm{th}}\sim 20\ \MeV$ and because the
associated cross sections are significantly smaller in comparison
to~(\ref{eq:cs-nubar}). This has another interesting consequence: in
contrast to the pion- and kaon-scenarios essentially no \lisx\ is
produced for $\tau_X\gtrsim 10^4\ \seconds$, allowing for a second
region of cosmologically favored \lisv\ abundances in which also high
D/H is photo-dissociated.  As discussed earlier, this region is
necessarily fine-tuned in $\tau_X$ because of the exponential
sensitivity in injected electromagnetic energy. Indeed, for even
larger lifetimes $\tau_X\gtrsim 10^6\ \seconds$ one would observe what
could be called the ``photo-dissociation catastrophe'' with \hef\
along with all other elements being dissociated (for large enough
$Y_X$). Nevertheless, we find this scenario rather distinct if not remarkable. 
The injection of muons with relic $X$ decays with lifetimes of $10^4$ seconds
is capable of reducing \lisv\ abundance while keeping other abundances close to 
the their SBBN-predicted values. This occurs because of the 
non-monotonic evolution of D/H: first the increase due to neutrino-induced neutron enrichment, followed by
the decrease due to energy injection. 

The temporal evolution of elemental abundances for a single point in the parameter space marked by the star in
Figure~\ref{neutrinos} is illustrated in Figure~\ref{evolution2}.
 The most prominent feature is the ``bump''
in the neutron abundance at $T_9\sim 0.06$ marking the point in time
in which deuterium is subjected to photo-destruction. The associated
decrease in $\deut/\Hyd$ from its elevated values down to its original
SBBN prediction (dotted lines) are clearly seen in this Figure. Also \bes\ is
dissociated around the same time so that the overall lithium abundance
comes essentially in the form of \lisv.

\section{Metastable GeV-scale states and the lithium abundance}
\label{sec:metastable-gev-scale}

In this section we consider some specific models of GeV-scale relics
and their impact on the BBN predictions. All models can be subdivided
into broad categories of WIMPs and super-WIMPs, where "super-" refers
to the superweak interaction strength with SM-states.  WIMPs are
thermally excited above their mass scale, then depleting their number
density via annihilation at $T<m_{\mathrm{WIMP}}$, and---in our
case---decaying after the start of BBN. Superweakly interacting
particles typically have very small production rates throughout the
whole history of the Universe. This justifies the assumption of
super-WIMPs being initially absent as the Universe enters its thermal
radiation-dominated stage {\em e.g.}  after inflation and reheating,
so that only some small super-WIMP abundance develops due to the
thermal leakage from the SM states. Let us stress again that the
notion of a (super-)WIMP here is not to be confused with a dark matter
state (which has to be stable on cosmological timescales.) Of course,
some of metastable particles presented in this section are motivated
by their potential as mediators of the interactions between dark
matter particles and the SM sector~\cite{Pospelov:2007mp,Finkbeiner:2007kk}.

The simplest WIMP model without any need of UV completion
\cite{McDonald:1993ex,Burgess:2000yq} uses the so-called Higgs
portal. In this model, a singlet scalar field $S$ interacts with the
rest of the SM via its coupling to the Higgs doublet $H$,
\begin{equation} {\cal L}_{\rm H-portal} = \fr12(\partial_\mu S)^2 -
  V(S) - (\lambda SS + A S )(H^\dagger H).
\label{S-model}  
\end{equation}
Here we assume an approximate $Z_2$ symmetry, $S\to -S$, broken only
by the last trilinear term $ A S (H^\dagger H)$, and we further take
$A$ to be very small; $V(S)$ is the scalar potential in the secluded
sector.  If the $\lambda$-coupling is above $O(10^{-10})$, the
$S$-particles are guaranteed to be in thermal equilibrium with the
SM-states as soon as the plasma temperature is around the electroweak
scale. Smaller values of $\lambda$ make $S$ to a
super-WIMP~\cite{McDonald:2001vt}. The physical mass of $S$ particles
comes from the potential term $m_0^2S^2/2 $ and the electroweak vacuum
expectation value (VEV) $v\simeq 246\ \GeV$, $m_S^2 = m_0^2 + \lambda
v^2$. Having
\begin{align}
\label{Smodel-param}
    A,\ \lambda,\ \mathrm{and}\ m_S^2\qquad  \text{(S-portal)}
\end{align}
at our disposal, we can always choose the region of parameter space
where the lithium abundance is reduced along the lines described in
the previous section.  Of course, in this model the GeV scale is not
special, and metastable $S$-particles at the electroweak scale can
also be used for the same purposes, exploiting nucleons in the decay
products of $S$.

Going away from the simplest possibility, we introduce a \us\ vector
portal model~\cite{Holdom:1985ag},
\begin{align}
  \label{Vportal} {\cal L}_{\rm V-portal} = -\frac{1}{4} V_{\mu\nu}^2
  -\frac{\kappa}{2}\, F^Y_{\mu\nu}V^{\mu\nu} + |D_\mu \phi |^2 -V(\phi),
\end{align}
where the connection between the \us\ field strength $V_{\mu\nu}$ and
the hypercharge field strength $F^Y_{\mu\nu}$ is mediated via the
kinetic mixing parameter $\kappa$. The \us\ covariant
derivative is given by $D_{\mu}= \partial_{\mu}+i e' V_{\mu}$, with
$V_{\mu}$ being the new vector-state associated with $V_{\mu\nu}$ and
$e'$ the gauge coupling strength in the secluded sector.

Since we are going to consider GeV-scale phenomenology, one can substitute
the hypercharge with the photon field strength, $F^Y_{\mu\nu}\to
F_{\mu\nu}$ and absorb the cosine of the Weinberg angle into the new definition of $\kappa$.  
After the spontaneous breaking of the \us\ gauge group
by a Higgs$'$ field $\phi$, the low-energy Lagrangian can be written
as
\begin{equation} {\cal L}=-\frac{1}{4} V_{\mu\nu}^2+\frac{1}{2}m_V^2
  V_\mu^2 +\frac{1}{2}(\partial_\mu h')^2-\frac{1}{2}m_{h'}^2 h'^2
  +{\cal L}_{\rm int},
\end{equation}
where $m_V = e'v'$ becomes the mass of $V_{\mu}$ and $m_{h'}$ is the
mass of the physical Higgs$'$ field $h'$, $\VEV{\phi} =
v'/\sqrt{2}$. Assuming a standard Higgs potential in the \us\ sector,
the interaction terms are given by
\begin{equation} {\cal
    L}_{\rm int}=-\frac{\kappa}{2}\,V_{\mu\nu}F^{\mu\nu}+\frac{m_V^2}{v'}
  h' V_\mu^2 +\frac{m_V^2}{v'^2}\,h'^2 V_\mu^2 - \frac{m_{h'}^2 }{2
    v'} h'^3 - \frac{m_{h'}^2}{8 v'^2} h'^4.
\end{equation}

Thus, like in the previous example, the model is characterized by only
a handful of free parameters:
\begin{align}
\label{Vportal-param}
 \alpha',\ \kappa,\ m_{h'},\ \mathrm{and}\ m_V\qquad  \text{(V-portal)},
\end{align}
where, as usual, $\alpha'=e'^2/4\pi$. 
The strength of the mixing angle $\kappa$ is undoubtedly a very
important parameter, as it allows to make the link between the SM- and
\us-sector arbitrarily weak. Indeed, for $\kappa< 10^{-12}$ any
production rate of $h'$ and $V$ particles is smaller than the Hubble
rate, and the model becomes a good candidate for the super-weak
regime. It is also important to notice that long lifetimes of
particles from the \us\ sector can be achieved {\em without} requiring
exceedingly small~$\kappa$. This happens rather naturally in the
regime $m_{h'} < m_V$, where the decay amplitude of the Higgs from the secluded
sector is suppressed by $\kappa^2$
\cite{Batell:2009di,Pospelov:2008zw}. As mentioned in the
introduction, the (sub)-GeV mass scales for \us\ states are motivated
by the possible enhancement of annihilation of weak-scale WIMPs into
leptons \cite{ArkaniHamed:2008qn,Pospelov:2008jd}.

\subsection{WIMP regime}

Here we would like to determine the values of parameters for both
models, (\ref{Smodel-param}) and (\ref{Vportal-param}), that lead to
the desirable depletion of the \lisv\ abundance.  Since in the WIMP
regime the coupling constants are not necessarily small, the results
of this subsection might be relevant for the direct searches of new
physics at GeV energy scales.

For the Higgs portal model in the WIMP regime there are several
generic choices of parameters that lead to the desirable range of the
lifetime--mass--abundance "islands" that suppress the overall lithium
abundance. These regions can be readily found using the results for
the lifetime \cite{Batell:2009jf} and freeze-out abundance of
$S$-particles \cite{Burgess:2000yq,Bird:2004ts,Bird:2006jd}.  For
example, the following choice of parameters,
\be Y_S \simeq 3\times 10^4, ~~ \tau_X \sim 500\ {\rm sec},~~m_S = 250
~{\rm MeV}~~\Longrightarrow~~ \lambda^2 \simeq 2\times 10^{-2}, ~~ A =
4\times 10^{-8}~ {\rm GeV},
\label{Sest}
\ee
leads to the depletion of lithium via the decays of $S$-particles to
muon pairs, generating electron anti\-neutrinos. Higher masses of $S$
particles, $m_S > 2 m_\pi$ that generate hadrons in the decay product
would typically require $\lambda \sim O(1)$ in order to have $Y_S \la
10^2$. The estimate (\ref{Sest}) is performed for a Higgs mass value
of 100 GeV but can be easily rescaled for heavier values.  Notice that
the $A$ parameter does not enter in the abundance calculation, and can
always be adjusted to obtain a desirable lifetime.

Interestingly enough, the mass range of $m_S \la 2$ GeV is being
probed through the search of missing energy decays of $B$-mesons, $B
\to K^{(*)}SS$, \cite{Bird:2004ts,Bird:2006jd,Badin:2010uh} as the
eventual decays of $S$ particles occur far away from the $B$-decay
vertex.  When the phase space suppression can be neglected, the
prediction for the missing energy branching ratio at $m_h =100 $ GeV
is ${\rm Br}_{B\to K E\!\!\!/} \sim 4\times 10^{-6} + 3\times 10^{-4}
\lambda^2 $ \cite{Bird:2004ts}, where the first term stands for the SM
contribution via $B\to K\nu\bar\nu$. The current upper bound on the
branching ratio stands at $1.5 \times 10^{-5}$, excluding models with
$\lambda\sim O(1)$.  The choice of parameters (\ref{Sest}) is still
(barely) allowed, both by the $B$ and $K$ decays with missing energy.
We conclude that only this low $m_S$ option (or alternatively $m_S \ge
2$ GeV) is capable of solving the lithium problem without enhancing
the missing energy branching ratios of $B$-mesons beyond what is
observed. Perhaps the most interesting consequence of large values for
$\lambda$ required by the solution to the lithium problem are the
implications for Higgs physics. For Higgs masses of 150 GeV and
lighter the missing energy decays $h\to SS$ will dominate over the
Standard Model width and lead to a significant enhancement of the
missing energy channel and suppression of "visible" decay modes
\cite{Burgess:2000yq}. In short, the viability of the proposed
scenario will be tested at the LHC, (super-)$B$-factories and new kaon
facilities.

Metastable GeV-scale particles from a \us\ sector is another
interesting possibility. In the WIMP-regime we are drawn to consider
$h'$ since the vectors $V$ decay well before BBN through their now
appreciable kinetic mixing $\kappa \gg 10^{-10}$ [see
Eq.~(\ref{GammaVlep}) below],
\begin{align}
\label{eq:lifetime-vector}
  \tau_V \leq 0.05\ \seconds\times \left( \frac{ 10^{-10}}{\kappa}
  \right)^{2} \left(\frac{500\ \MeV}{m_V}\right) \quad
  \mathrm{for}\quad m_V \gtrsim m_{e}.
\end{align}
Indeed, the longevity of $h'$ particles can be achieved rather
naturally in the regime $m_{h'} < m_V$.  The lifetime of the $h'$
particles due to the four-body decays and one-loop induced amplitudes
was calculated in Ref.~\cite{Batell:2009yf}. For example, the decay
width to muons in the regime $2 m_\mu < m_{h'} \ll m_V$ is dominated
by loop effects and is given by
\begin{align}
  \tau_{h'} \sim (10^{3}\div 10^4)\, \seconds \times \left(
    \frac{\alpha}{\alpha'} \right) \left( \frac{3.4\times
      10^{-5}}{\kappa} \right)^{4} \left( \frac{{250~\rm MeV}}{m_{h'}}
  \right) \left( \frac{m_V}{{500~\rm MeV}} \right)^2.
\end{align}
This formula remains approximately valid in the regime $m_{h'} \sim
m_V$, and breaks down at $m_V - m_{h'} \ll m_{h'}$. With $m_V\sim 500\
\MeV$ and $\alpha'\sim \alpha$ it is easy to see that for $m_{h'} =
250$ MeV and $\kappa \simeq (3\div 10)\times 10^{-5}$ the decay rate
of $h'$ is in the right ballpark, while for the same value of
couplings and masses the decay of vectors happens very fast, $\tau_V <
10^{-16}\ \seconds$.

Since we take the $h'$ particle to be the lightest in the \us\ sector
and long-lived, what physical mechanisms could possibly deplete its
abundance to an acceptable level? It turns out that possible {\em
  endothermic} excitations into the $V$ state, $h'\to V$ followed by
$V$ decays, although exponentially sensitive to the mass parameters in
the \us\ sector, are in principle sufficient for the $h'$
depletion\footnote{MP would like to thank Neal Weiner for a very
  useful discussion of that point.}. In particular, the following
processes are capable of suppressing the $h'$ abundance:
\begin{eqnarray}
  h' + h' \to V+V, &~~& \Gamma_1 \propto (\alpha')^2 \kappa^0 \exp(-m_{h'}/T-2\Delta m/T) \\
  h' + V \to l^+l^-, &~~& \Gamma_2 \propto \alpha'\alpha \kappa^2 \exp(-m_{h'}/T-\Delta m/T)\\
  \label{depletion}
  h'+ l^\pm \to V + l^\pm, &~~& \Gamma_3 \propto \alpha'\alpha \kappa^2 \exp(-\Delta m/T), 
\end{eqnarray}
where $\Delta m = m_V - m_{h'}$, and $l^\pm$ are the charged particles
of the SM.  The last process in (\ref{depletion}) is especially
important because it comes with the least amount of exponential
suppression. We now estimate the freeze-out abundance of $h'$ due to
this process.

In the norelativistic regime, the equilbrium number density
$n_{h'}^{\mathrm{eq}}$ of $h'$ follows the well-known curve, and the
freeze-out abundance can be estimated by equating the depletion rate
$\Gamma_3$ with the Hubble rate, $H(T_f) = \Gamma_3(T_f)$,
\begin{align}
Y_{h'} &\simeq  n^{eq}(T_f)/n_b(T_f),\\
  n_{h'}^{\mathrm{eq}}(T) &= \left(\fr{ m_{h'}T}{2\pi}
  \right)^{3/2}\exp(-m_{h'}/T).
\end{align}

We would like to determine the dependence of the freeze-out
temperature $T_f$ on $\Delta m$, and specialize it to two cases
considered in the previous section: decays to muons and decays to
pions/kaons.  Considering the exponential dependence on the mass
splitting $\Delta m$ we are making a series of simplifying
approximations: $m_e\ll T_f \ll \Delta m \ll m_V,m_{h'}$. In that case
the cross section for the $h' + e^\pm \to V + e^\pm$ process is given
by
\be \sigma = \frac{16\pi\kappa^2\alpha\alpha'}{m_V^4} \frac{\Delta m
  (E_e-\Delta m)^2}{E_e} ~~\Longrightarrow ~~ \Gamma_3(T) \simeq
\frac{64\kappa^2\alpha\alpha'T^3(\Delta m)^2}{\pi m_V^4} \exp(-\Delta
m/T), \ee
where $E_{e}$ is the electron energy.  For $T_f$ larger than 100~MeV,
other charged SM states would have to be included in the
coannihilation process~(\ref{depletion}). However, since we are in
need of a suppression of the $h'$ abundance by a large factor, the
freeze out of a GeV-scale $m_h'$ has to happen below 100~MeV, so that
only $e^{\pm}$ will contribute to depletion process.

Introducing two dimensionless variables $x_f= m_{h'}/T_f$ and $\delta
= \Delta m /m_{h'} = m_V/m_{h'}-1$, we arrive at the following
analytic estimate of the freeze-out temperature and the resulting
abundance as a function of mass splitting in the \us\ sector:
\begin{align}
  x_f &\simeq \fr{1}{\delta}\left\{13+
    \ln\left[\frac{\alpha'}{\alpha} \fr{\delta^2}{(1+\delta)^4}
      ~\fr{15}{x_f}\left(\frac{\kappa}{3\times 10^{-5}}\right)^2
      \fr{250~{\rm MeV}}{m_{h'}} \right] \right\},
  \\
\label{eq:hWimpabd}
  Y_{h'} &\simeq 4.2\times 10^8 \times x_f^{3/2}\exp(-x_f).
\end{align}
It is easy to see that for mass splittings $\delta \simeq 0.3\div 0.5$ the
abundance varies in the interval $10^{-1}\div 10^2$, which will be
suitable for the solution of the lithium problem using the decays of
$h'$ to pions and kaons. Mass splittings of order $\delta \simeq
0.8\div 1.4$ correspond to abundances on the order $10^4\div 10^6$, which are
suitable for suppressing lithium via the decays of $h'$ to muons. We
remark in passing that~(\ref{eq:hWimpabd}) represents the fraction of
$h'$ relative baryons after $e^{\pm}$-annihilation but prior to their
decay.

Since the realization that a new attractive \us\ gauge force can be
used for the explanation of the PAMELA anomaly
\cite{ArkaniHamed:2008qn,Pospelov:2008jd}, a lot of dedicated work has
been done in order to understand the prospects of searching/detecting
particles from a putative \us\ sector.  The most promising avenues for
the discovery of new GeV-scale particles are high-luminosity
medium-to-low energy experiments
\cite{Batell:2009yf,Reece:2009un,Essig:2009nc}, although some
prospects of discovering such particles in the fragmentation of heavy
exotic particles produced in the high-energy collisions have also been
investigated~\cite{Baumgart:2009tn}.  The mixing angles deduced in
this section, $\kappa \sim 3\times 10^{-5}$, and vector masses in the
range of $400$~MeV and larger, represents one of the most challenging
corners in the $m_V$-$\kappa$ parameter space. These angles are small
enough so that the search of $V$ in this range at even the highest
luminosity $e^+e^-$ machines is not possible. Therefore, the search
for $V$ in this regime with fixed target experiments is perhaps the
only realistic option. Vector particles are relatively short lived,
$c\tau \le 1$~cm, so that in the set-up with a detector at some
macroscopic distance behind the target all the decays will happen
before reaching the detector.  With proton beams, where the detector
is typically tens or hundreds of meters behind the target, such a
search might prove to be very challenging.  Therefore, the best
discovery potential for $V$ in this parameter range would probably be
a high-intensity electron beam on a thin
target~\cite{Bjorken:2009mm}. Lastly, the lifetimes of mediators in
excess of milliseconds lead to interesting effects in the annihilation
of dark matter, when the decays of mediators occur away from the point
where the annihilation occur
\cite{Batell:2009zp,Schuster:2009au,Rothstein:2009pm}. This leads to
novel signatures in the indirect detection of dark matter in models
with light mediators. Finally, for overdensity constraints on some
variants of these models see, \emph{e.g}~\cite{Chen:2009ab}.

\subsection{GeV-scale super-WIMPs} 

In the previous examples we considered metastable particles that
initially had thermal abundances.  This is not the only possibility,
and in the following we will look at the abundances of vectors $V$ and
scalars $S$ in the super-WIMP regime.  Our assumption is that the link
between the SM and the secluded sectors is extremely weak, and given
by a single parameter, $\kappa$ for (\ref{Vportal}), and $A$ for
(\ref{S-model}). The states $X=S$ or $V$ are then produced with
sub-Hubble rates from the thermal scattering of SM particles. In this
case, both, the production and the decay rates are proportional to the
square of the small parameters $\kappa$ or $A$.  Consequently, the
product of abundance and lifetime, $Y_X\tau_X$ is independent on $A$
or $\kappa$ and is only a function of the mass $m_X$ and the SM
couplings/masses.  This scaling holds for both cases despite the fact
that the physical production mechanisms of $S$ and $V$ are vastly
different. An extra vector particle that appears just as another
massive photon is mostly produced at temperatures comparable to its
mass $T \ge m_V$ whereas a scalar particle with mixing to the Higgs
boson is most efficiently produced at $T\sim m_W$, where the
$T/m_h$-dependence of the production rate relative to the Hubble rate
is maximized.

We begin by analyzing the \us\ model~(\ref{Vportal}) in the super-WIMP
regime. The decay widths of the vector particles to leptons and
hadrons are given by~\cite{Batell:2009yf}
\begin{align}
\label{GammaVlep}
\Gamma_{ V \rightarrow \overline{l}l } &=\frac{1}{3} \alpha \kappa^2 m_V \sqrt{1-\frac{4 m_l^2}{m_V^2}}
\left(1+\frac{2 m_l^2}{m_V^2}\right),\\
\label{GammaVhad}
\Gamma_{ V \rightarrow {\rm hadrons} } &=\frac{1}{3} \alpha \kappa^2
m_V \sqrt{1-\frac{4 m_\mu^2}{m_V^2}} \left(1+\frac{2
    m_\mu^2}{m_V^2}\right) R(s= m_V^2),
\end{align}
where $R$ is the experimentally measured ratio of the inclusive
hadroproduction in $e^+e^-$ collisions to the direct muon production,
$ R= \sigma_{e^+ e^- \rightarrow {\rm hadrons}}/ \sigma_{e^+ e^-
  \rightarrow \mu^+ \mu^-}$.  It follows immediately [see also
Eq.~(\ref{eq:lifetime-vector})] that a desirable lifetime range
$\tau_{V}\gtrsim \Orderof{100\ \seconds}$ requires $\kappa$ in the
ballpark of $10^{-12}$.

The production of GeV-scale $V$-bosons in the early Universe cannot be
calculated "exactly" because of the unsurmountable difficulty in
treating hadrons in the intermediate regime $T\sim \Lambda_{\rm QCD}$.
In particular, all production modes involving free quark and gluons,
$q\bar q \to V$, $q + g \to q +V$ etc are not calculable because of
the strong QCD coupling.  Cases that can be reliably calculated are
$m_V \ll 100$~MeV, and $m_V \gg 1$~GeV, which are, unfortunately,
exactly opposite to the regime we are interested in.  Nevertheless, we
can obtain an order-of-magnitude estimate that captures the main
behavior of the production mechanism as a function of $m_V$.  The main
feature of the production mechanism is that it receives an exponential
cutoff $\exp(-m_V/T)$ in regime $T<m_V$, and this property is not
modified by strong dynamics. Therefore, the production effectively
stops at $T \sim m_V$ with the residual abundance of $V$ particles
parametrically dependent on the ratio of the production rate over the
Hubble rate,
\begin{align}
  \label{YV}
  Y_V = \fr{s}{n_b}\left.\fr{n_V}{s}\right|_{f} \simeq (1.2 \div 4.9)
  \times\fr{s}{n_b} \fr{1}{h_{\mathrm{eff}}(m_V)} \fr{\Gamma_{V\to
      e^{+}e^{-}}}{H(m_V)} \sim 0.3\times
  \left(\fr{10^3\ \seconds}{\tau_V}\right)\left(
    \fr{\GeV}{m_V}\right)^{2} \left( \fr{40}{g_{\mathrm{eff}}}
  \right)^{3/2} .
\end{align}
The estimated range follows from a computation of the collision
integral for the inverse decays of Fermi-Dirac distributed pairs
$e^{\pm}$ (lower value) and massless
$e^{\pm},\,\mu^{\pm},\,u\bar{u},\,d\bar{d},\,s\bar{s}$ (upper value);
for a related calculation see also~\cite{Redondo:2008ec}.  The upper
value is saturated for ``freeze-out'' temperatures above the QCD
hadronization scale, $\Lambda_{\mathrm{QCD}}<T_f\lesssim\Orderof{1\
  \GeV}$, where the perturbative treatment of the light quarks becomes
accessible. On the other hand, lower mass vectors whose production
ceases later are guaranteed to receive a contribution to their
abundance from electron-positron pairs. This limits $Y_V$ from below.
In the last relation we have normalized the effective number of
entropy degrees of freedom $h_{\mathrm{eff}}\simeq g_{\mathrm{eff}}$
on a typical value during the QCD hadronization epoch and further made
the assumption that $\tau_V$ is dominated by the channels which are
also responsible for the inverse decay.

The results suggest that a sub-GeV state with a lifetime of the order
of a thousand seconds is abundant enough to produce $\Orderof{1}$
pions/kaons per baryon, and thus capable of reducing the lithium
abundance by a factor of a few.  In the most optimal range with $m_V
\sim m_\rho$, the pion branching dominates over lepton branching,
while the energy of injected pions is close to the
delta-resonance. This enhances the the efficiency of $p\to n$
conversion, Fig. \ref{figure1b}, so that $Y_V\sim 2$ is required.
Since $Y_V$ is not calculable exactly, it is unfortunately not
possible to give a final judgement whether the required abundance can
be achieved or falls marginally short.
We also notice that $m_V$ below the di-pion threshold, $2m_\mu < m_V <
2m_\pi$ can only be created in abundances less than $\Orderof{10^2}$
which renders this parameter range incapable of reducing lithium via
$V$-decays into muons. On the other hand, a heavier mass range for a
super-WIMP $V$ is also interesting. For example, $V$-bosons in the
mass range of 10 GeV or more are less abundant than protons, but will
contain nucleons among their decay products. In this case one can
easily suppress the lithium abundance via direct injection of extra
nucleons along the lines of Refs.~\cite{Reno:1987qw,Jedamzik:2004er}.

The model with the singlet $S$ mixed to the Higgs boson via the
$ASH^\dagger H$ coupling can be treated very similarly.
Interestingly, it turns out that in this model the abundance of $S$
particles can be calculated without significant QCD uncertainties
because it dominatly occurs at electroweak scale temperatures, whereas
the lifetime of the $S$ boson with a mass in the GeV range is
notoriously difficult to handle.

In principle many SM scattering processes may contribute to the
emission of the $S$ boson. These include $f_1\bar f_2 \to VS$, $f_1V
\to f_2 S$, $hh\to SV$, $VV\to SV$, $tg\to tS$, $t\bar t \to gS$ etc,
where $f_1,f_2$ stand for the SM fermions, $V=W,Z$ for massive gauge
bosons, and $g$, $t$ are gluons and top quark.  An exact treatment of
the production mechanisms in this model falls outside the scope of
this paper. However, despite this rather large number of production
channels, the asymptotic behavior of the production mechanisms are
quite obvious.  In the regime $T \gg v$ the production rate has to
scale as $\Gamma \propto A^2/T$, and in the low-energy regime it is
$\propto A^2 T^3 m_h^{-4}$. This implies that $\Gamma$ is naturally
peaked at temperatures around the electroweak scale.

In order to avoid straightforward but tedious calculations of the $S$
abundance in the most minimal model (\ref{S-model}), we go to the
two-Higgs doublet model analogue of (\ref{S-model}), and couple the
$S$ scalar to a "mixed" Higgs portal: 
\be ASH^\dagger H \to \fr{1}{2}AS(H_1H_2 + h.c.).  \ee 
Assuming a mild hierarchy of scales in the Higgs sector together with
a ratio of the Higgs VEVs $ \tan \beta = v_2/v_1 \gg 1$, $H_2$
represents the SM Higgs doublet while $H_1$ contains heavi(er)
physical scalars $H$, $A$, $H^\pm$ with common mass scale $m_H$.  The
production of scalars $S$ is peaked around $m_H$. In what follows we
calculate the abundance of $S$ resulting from $H_1 \to H_2 S$ decays
using $m_H \gg m_h, m_{W(Z)}$.  In exact analogy to
(\ref{eq:meson-decay-rate}) the decay rate can be written as
$\Gamma_H\langle m_H/E \rangle$, where \be \Gamma_H = \fr{A^2}{16\pi
  m_H} \ee is the decay rate of $H_1$ particles at rest.  Accounting
for that $H_1$ carries four degrees of freedom we arrive at the
following integral formula for the freeze-out (or rather "freeze-in")
abundance of $S$ bosons realtive to baryons.
\begin{align}
\label{YS}
Y_S = \fr{s}{n_b} \int \fr{m_H dT}{H(T) T s(T)} \int
\fr{4d^3p}{E(2\pi)^3} \fr{\Gamma_{H}}{\exp(E/T) -1}
\end{align}
To good approximation we take $h_{\mathrm{eff}}(T)$ in $s = 2\pi^2
h_{\mathrm{eff}} T^3/45$ to be that of the SM, $h_{\mathrm{eff}} =
427/4 $. The integrals in (\ref{YS}) can be taken exactly, resulting
in \be Y_S = \fr{s}{n_b} \fr{135\zeta(5)}{2\pi^3}
\frac{\Gamma_H}{NH(m_H)}\simeq 3.8\times 10^5 \times
\fr{A^2M_{P}}{m_H^3} \ee where $H(m_H)$ is the value of the Hubble
constant at $T=m_H$ and $\zeta(x)$ is the Riemann Zeta-function.

The decay of the $S$-bosons at late times proceed to muons and hadrons
depending on what channels are kinematically allowed.  At $m_S \ll 1$
GeV the decay rates can be obtained within the framework of chiral
perturbation theory and by low-energy
theorems~\cite{Voloshin:1985tc,Truong:1989my}. For $m_{S}\gtrsim
1.5$~GeV or so, perturbative QCD starts getting applicable. Around
these energies the decay rate is dominated by $S$ decaying into
$s$-quarks with a $\sim 25\%$ contribution to muons:
\be \Gamma_S \simeq \fr{3m_S}{ 8 \pi} \left( \fr{A m_s
    \tan\beta}{m_H^2} \right)^2 \,\left(1 + \fr{m_\mu^2}{3m_s^2}
\right)
\label{GS}
\ee
Notice that heavy-quark-mediated $S\to gg$ decays are suppressed
relative to the SM by a factor of 9 because in the large $\tan\beta$
regime only $b$-quarks contribute, and $t,c$ contributions are
subdominant. Since $m_s \ge m_\mu$, the decays to strange-antistrange
pairs happen in at least $75\%$ of the cases. When $m_S$ is below the
di-hyperon threshold, each of these decays lead to either $K^-$ or
$K_L$ in the final state. The last step in checking the suitability of
this model involves trading $A^2$ in the formula for the abundance
(\ref{GS}) for the lifetime of $S$. Taking $m_s \sim 100$ MeV, we get
\be Y_S = 3.4 \times \left(\fr{10^3\ \seconds}{\tau_X}\right)
\left(\fr{2~{\rm GeV}}{m_S}\right) \left( \fr{m_H
    \tan^{-2}\beta}{5~{\rm GeV}}\right), \ee
where all parameters are normalized on their "natural" values.  It is
in a way remarkable that in these models many numbers of very
different orders of magnitude conspire to yield abundances in the
right ballpark necessary for a suppresion of the \lisv\ abundance via
the emission of kaons (and pions).

\section{Conclusions}
\label{Sec:Conclusions}

The current discrepancy between the prediction of the lithium
abundance in SBBN and its observation at the Spite-plateau value
prompted a number of particle-physics related explanations.  Previous
works have mainly concentrated on the effects of heavy, electroweak
scale particles on the \lisv\ abundance. Such models typically involve
decays of electroweak scale WIMPs with nucleons among the decay
products, or the catalytic suppression of \lisv\ by metastable charged
particles, that have their masses again in the electroweak range.  In
this paper we have considered a related but nevertheless distinct
possibility: decays of (sub-)GeV-scale neutral relic
particles. Because of their kinematic constraints, neither of these
particles can be charged nor can their decay products contain nucleons
in the final state.  Nevertheless, the reduction of the lithium
abundance may occur simply because pions and kaons in the final state
have a significant probability of interacting with protons and
creating a neutron-excess. We have shown that
$\Orderof{1}-\Orderof{100}$ abundances of metastable particles with
respect to baryons and with lifetimes around a few hundred to ten
thousand seconds can easily reduce the lithium abundance by a factor
of a few and still be consistent with measurements of primordial
deuterium.  We have also shown that if the decays to the charged
hadrons are kinematically not possible, the muons in the final state
can also reduce the overall lithium abundance. In this case the main
effect comes from electron antineutrinos that induce $p \to n$
interconversions. It turns out that $\Orderof{10^5}$ muon decays per
proton around a few hundred seconds are needed in order to reduce the
lithium abundance to observationally favored levels. Moreover,
injection of pions and muons not far from thresholds also leave less
"damage" in form of \lisx, compared to scenarios with electroweak-mass
decaying relics.  This is because the \lisx\ production requires
significant amounts of energy to split \hef\ and create energetic
$A=3$ nuclei, which in case of the muons/pion injection is far less
efficient. This broadens the acceptable lifetime range to up to
$O(10^4)$ seconds.

We have done a careful numerical investigation of this problem by
including charge exchange and absorption reactions of pions, kaons and
(anti)neutrinos on nucleons and \hef. In doing so we have updated the
values for the most relevant reactions, and integrated them into a
Kawano-based nuclesynthesis code. We have also accounted for the
change in the neutrino spectrum due to the Hubble expansion and due to
neutrino oscillations. The results are presented in the form of
abundance-lifetime plots, Figures~\ref{figure1}-\ref{neutrinos}, where
we have assumed a maximal branching into each species. We can see
that the elevated deuterium abundance is an inevitable consequence of
all such scenarios, resulting from the "removal" of extra neutrons by
the $np\to d\gamma$ reaction. The only exception is the case of relic particles
decaying to muons at $\sim 10^4$ seconds. In this case, the late energy injection,
combined with early neutrino-induced $p\to n$ interconversion can lead to the 
SBBN value of deuterium, and achieve the suppression of \lisv\ by a factor of a few. 

In order to relate this proposal to some concrete model realizations,
we have considered two minimalistic extensions of the SM. We chose to
extend the SM by a singlet scalar $S$ coupled via the so-called Higgs
portal, or a new \us\ group that has kinetic mixing with the photon
field strength.  Both models, in principle, possess new long-lived
states, and in the case of the \us\ model this is either the vector
$V$ or the Higgs$'$ particle $h'$ of the \us-breaking sector.  The
cosmological history in these models then depends on the strength of
the couplings that connect the secluded sectors to the SM
particles.

If the couplings to the Standard Model are large enough, the new
GeV-scale states $X$ are thermally excited (WIMP regime). An
acceptable abundance of $X=S,\ h'$ prior to decay is then either
achieved via annihilation, {\em e.g.}  $SS\to f\bar f$, or via the
excitation to the unstable state, $ h' e \to V e \to e\bar e e$. We
have shown that in the WIMP regime both models provide viable
solutions to the lithium problem, and that both, the $\pi,K$-mediated
or the antineutrino-mediated suppression of \lisv\ indeed work.  We
have identified the plausible choice for the parameters where such
reduction will occur. The same parameters may be chosen to fit the
PAMELA and FGST anomalies, when the mediator models are complemented
with electroweak-scale WIMPs.  In addition to cosmological probes,
there is a certain chance of detecting these new GeV-scale states in
laboratory experiments, such as in rare decays of mesons, in Higgs
decays, and in fixed target experiments. Last but not least, we remark
that a $m_{h'} < m_V$ mass pattern naturally arises in straightforward
supersymmetric generalizations of the \us\ model.

We have further shown that the super-WIMP regime of these models also
have a potential to provide a viable resolution to the lithium
problem. In this case, the metastable states are never in thermal
equilibrium and are only produced via an extremely small thermal
rate. Such models are relatively predictive because the mass and the
coupling of $X$ can be translated directly into lifetime- and
abundance-predictions.  Unfortunately, the extreme smallness of the
coupling constants prevents direct experimental searches of such
particles in the laboratory. On the other hand, the sensitivity that
the early Universe exhibits to new physics is once again emphasized by
the fact the lithium and deuterium abundances are capable of probing
effective coupling constants as small as $\alpha \kappa^2 \sim
10^{-26}$.  It remains to be seen whether the \lisv\ reduction can be
achieved in another well-motivated model of super-WIMPs in the form of
heavy sterile neutrinos.

To close up, the problem of the \lisv\ abundance definitely deserves
special attention, as it hints at one of the very few inconsistencies
in the standard cosmological picture.  Of course, at this point it is
difficult to insist that particle physics must necessarily be key for
solving the current lithium problem. Yet, as shown in this paper,
several particle physics scenarios that reduce the \lisv\ abundance are
quite natural and perhaps deserve further detailed considerations. As
to the lithium problem, only future progress in cosmology,
astrophysics, particle, and nuclear physics may help to resolve this
intriguing problem.

{\bf Acknowledgments} Research at the Perimeter Institute is supported
in part by the Government of Canada through NSERC and by the Province
of Ontario through MEDT. JP wants to thank the Galileo Galilei
Institute for Theoretical Physics for the hospitality and the INFN for
partial support during the final stages of this work.

\appendix
\section{Appendix}
\label{sec:appendix}
\setcounter{equation}{0}

\subsection{Some details on the Boltzmann code}
\label{sec:some-details-boltzm}

The Boltzmann code that we use is based on the one of
Ref.~\cite{Kawano:1992ua}, in which we incorporate some significant
improvements and updates: physical constants, isotope masses, and
conversion factors are determined from the
evaluations~\cite{Audi:2002rp,Mohr:2005kk}. For all important SBBN
reactions (\textit{i.e.} up to A=7) we employ the results
of~\cite{Descouvemont:2004cw} with the exceptions of the
$n(p,\gamma)\deut$ and $\het(\alpha,\gamma)\bes$ reactions for which
we follow~\cite{Ando:2005cz} and \cite{Cyburt:2008up},
respectively. All other $\Orderof{80}$ remaining reactions we update
by following the recommendations of~\cite{2010ApJS..189..240C}.

In order to arrive at an accurate prediction of \hef\ we numerically
integrate all weak rates at each time step for which zero temperature
radiative corrections are taken into account
following~\cite{Lopez:1998vk}.  When applicable, we further take into
account relativistic Coulomb corrections to the weak rates by
multiplying the integrands with the Fermi function
\begin{align}
\label{eq:fermi-function}
  F(Z,v) = 2(1+S) (2 p R)^{-2(1-S)} \frac{\left| \Gamma(S + i \eta )
    \right|^{2}}{\Gamma(2S+2)^{2}} e^{\pi \eta} 
\end{align}
where $Z$ is the nuclear charge, $S = ({1-\alpha^2Z^2}$)$^{1/2}$, $p$
($v = p/E$) is the momentum (velocity) of the relative motion, and $R
\simeq 1\ \fm$ is the proton radius; $\eta$ is given in
(\ref{eq:coulomb-correction-cs}). In our code we approximate $F(Z,v)$
by~\cite{Wilkinson:1982hu}
\begin{align}
\label{eq:coulomb-relativistic-approx}
F(Z,v) \simeq \sum_{n=0}^{3} a_n (\alpha Z)^n , \quad\text{with}\hspace*{4cm} \nonumber\\
a_0 = 1 ,\quad a_1 = \frac{\pi}{v},\quad a_2 = \frac{11}{4} -\gamma_E
- \ln{(2 p R)} + \frac{\pi^2}{3v^{2}} , \quad a_3 = \frac{\pi}{v}
\left[ \frac{11}{4} - \gamma_E - \ln{(2pR)} \right]
\end{align}
where $\gamma_E \simeq 0.5772$ is the Euler-Mascheroni
constant. Note, however, that in the non-relativistic case, $F(Z,v)$
is better approximated by~(\ref{eq:non-rel-coulomb-factor}). As the
dividing line we choose $v=0.045$ which guarantees that
(\ref{eq:coulomb-relativistic-approx}) and
(\ref{eq:non-rel-coulomb-factor}) represent (\ref{eq:fermi-function})
to better than $0.1\%$. In the non-relativistic limit the familiar
factor reads
\begin{align}
\label{eq:non-rel-coulomb-factor}
  F(Z,v) \simeq \frac{ 2\pi \eta }{1-e^{-2\pi\eta}},\quad \eta = \frac{
    Z \alpha}{ v}
\end{align}
where $Z$ is the nuclear charge; $\eta$ is called the Sommerfeld
parameter. Evidently, note that for Coulomb corrections of equally
charged particles one has to replace $Z$ by $-Z$, \emph{e.g.} in
$\pip$+\hef\ reactions. When obtaining Coulomb corrected cross
sections, we make the approximation
\begin{align}
\label{eq:coulomb-correction-cs}
  \sigmavof{}_i \simeq F(Z,\VEV{v}) (\sigma v)_{i},\quad \VEV{v}
  = \sqrt{{2 T}/{\mu}}
\end{align}
where $\mu$ is the reduced mass of the incoming particles.

Finally, as assessed in \cite{Lopez:1998vk}, we apply a slight
upward-shift of $0.72\%$ of the resulting \hef\ abundance in order to
account for more subtle, subleading corrections to the BBN reaction
network. We then find very good agreement of our light element
abundance predictions with the ones presented in the recent
work~\cite{Cyburt:2008kw} at the WMAP value of $\eta_{b}= 6.23\times
10^{-10}$ and with a neutron lifetime of $\tau_n=885.7\,$s.

\subsection{Pion and kaon capture on helium}
\label{sec:pion-kaon-capture}

Experimental data on absorption of stopped charged mesons on various
target materials are usually cast in the form of absorption rates and
branching fractions of the reaction fragments. After being stopped,
\pim\ are captured into atomic orbits from which they emit X-rays (and
Auger electrons) until they come close enough to interact strongly
with the nucleus. The cascading time is significantly smaller than
$\tau_{\pipm}$ and pions reach a 1S orbit prior to absorption.  A
measurement of the ground state level width $\Gamma_{1S} = (45\pm 3)\
\eV $~\cite{Backenstoss1974519} of pionic helium then allows us to
obtain the low-energy in-flight cross section $(\sigma v)$ of a free
\hef-\pim pair via
\begin{align}
  \label{eq:1S-absorption}
 \Gamma_{1S}^{\mathrm{abs}} =  |\psi_{1S}(0)|^{2} (\sigma v) ,
\end{align}
where $|\psi_{1S}(0)|^{2} v $ is the flux density at the origin.
Assuming a $100\%$ absorption from the 1S state, we find for the total
\pim-\hef\ reaction cross section
\begin{align}
  (\sigma_{\hef}^{\pi} v) \simeq 7.3\ \mbarn .
\end{align}

To extract the cross sections for the $\Km + \hef$ reaction we can
proceed in a similar fashion as for $\pim + \hef$. Interestingly,
however, measurements of kaonic X-rays from \hef\ show no radiative
transitions from the $2P\to 1S$ state~\cite{PhysRevLett.27.1410} which
tells us that $K^{-}$ on the $2P$-level prefers to react strongly with
the alpha nucleus rather than undergoing another radiative transition
to the ground state. Since $\sim 16\%$ of captured $\Km$ indeed make
it to the $2P$-state, this implies a lower bound on the $2P$
absorption rate~\cite{PhysRevLett.27.1410}
\begin{align}
\label{eq:lower-limit-2P}
  \Gamma_{2P}^{\mathrm{abs}} \geq 3\times 10^{14}\ \seconds^{-1} .
\end{align}
Using that value will, however, only yield the (momentum-dependent)
p-wave contribution to the total in-flight absorption cross section.
Therefore, and since we could not locate further experimental data on
s-wave absorption, we have to rely on theoretical calculations.

Reference \cite{Onaga:1989dn} finds for the absorption rate from the
$2S$-state $\Gamma_{2S}^{\mathrm{abs}} = 48.1\times 10^{18}\
\seconds^{-1}$ which points towards a very large cross section $
(\sigma v) \simeq 1.2\ \barn $ [using $\psi_{2S}$ instead of
$\psi_{1S}$ in (\ref{eq:1S-absorption})]. This value, however, is in
conflict with the s-wave unitarity bound $\sigmavof{\mathrm{max}}
\simeq 0.4\barn/\sqrt{T_9}$ so that we refrain from using it.
We further observe that the quoted value for the $2P$-absorption rate
in~\cite{Onaga:1989dn} is a factor of $\sim 20$ larger than the
experimental lower bound~(\ref{eq:lower-limit-2P}).
Employing the (somewhat ad-hoc) prescription to scale down
$\Gamma_{2S}$ of~\cite{Onaga:1989dn} by the same factor $20$ (which
saturates the lower bound for the $2P$-case), we arrive at a total
in-flight threshold cross section of
\begin{align}
\label{eq:sigma-Km-hef}
  (\sigma_{\hef}^{\Km} v) \simeq 60\ \mbarn .  
\end{align}
Indeed, this cross section compares well with the values inferred from
another calculation~\cite{Common1964465} for \Km\ absorption from the
1S-state which gives us some confidence that~(\ref{eq:sigma-Km-hef})
lies in the right ballpark.

\subsection{Pion reactions across the Delta-resonance}
\label{sec:inflight}

\FIGURE[t]{
\includegraphics[width=0.5\textwidth]{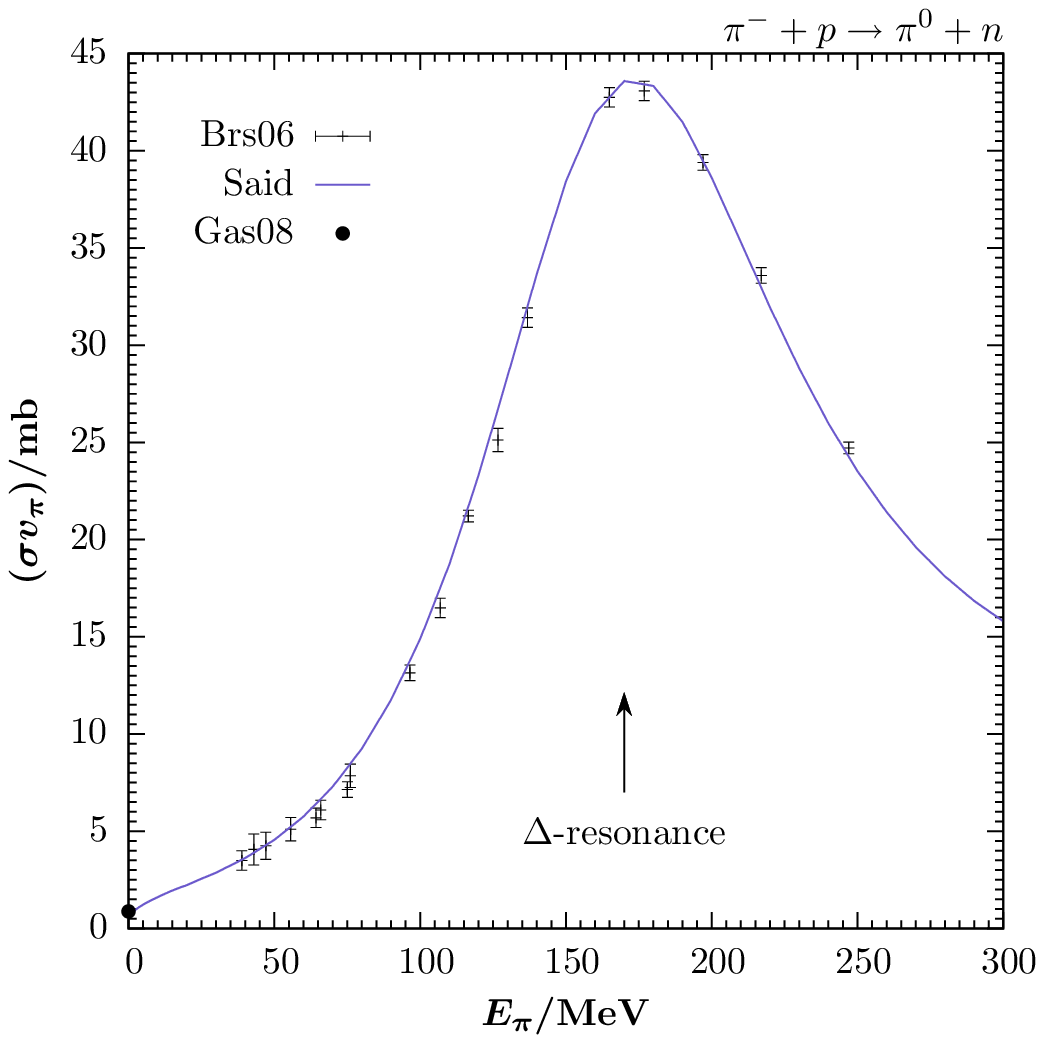}%
\includegraphics[width=0.5\textwidth]{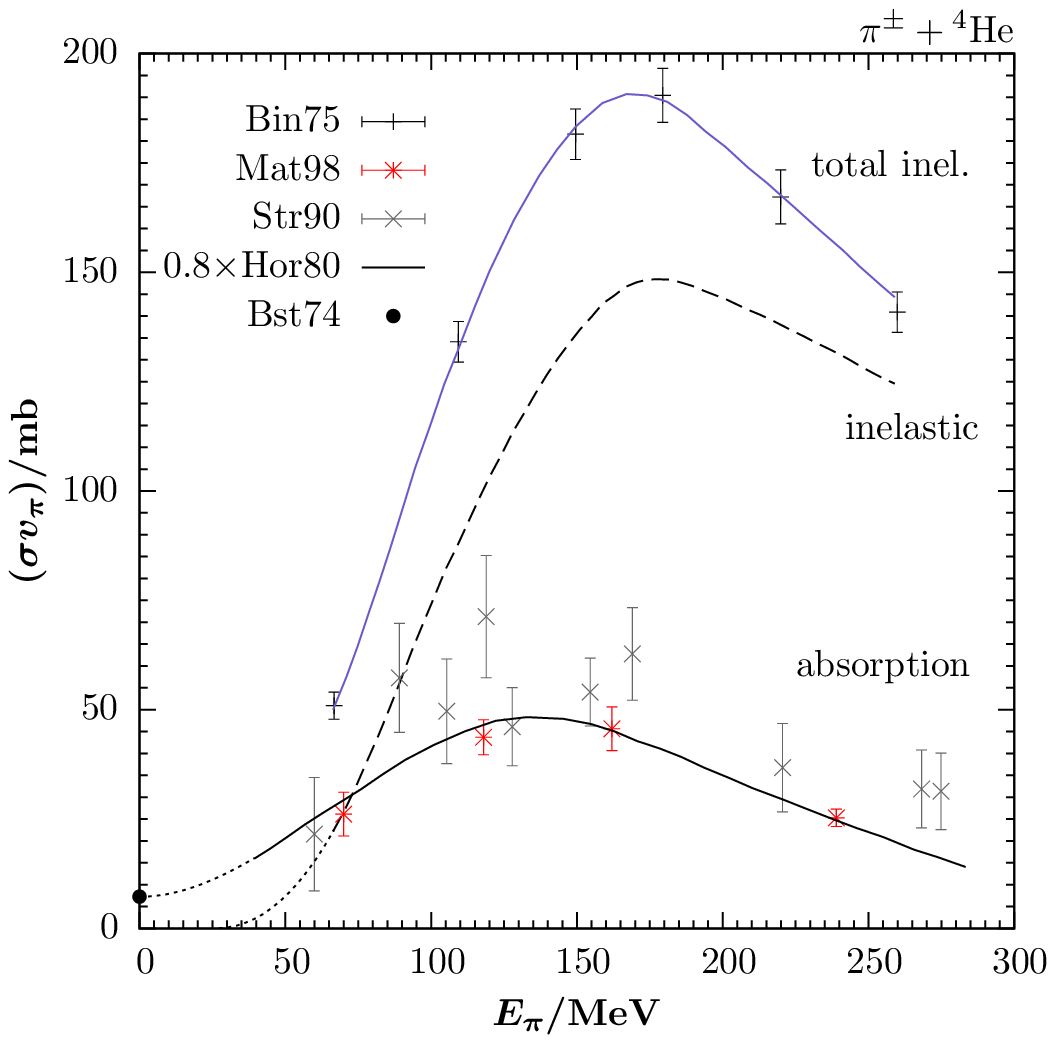}
\caption{\small \textit{Left}: Cross section for pion-induced proton
  conversion, $\pim+p\to\pizero+ n$, across the $\Delta^0$-resonance
  as a function of pion kinetic energy in the LAB-frame. The solid
  line shows a pion-nucleon partial-wave model
  calculation~\cite{Arndt:2003fj,Bellaachia:2002qe}. In-flight
  experimental data labeled ``Brs06'' are from
  \cite{Breitschopf:2006gn}, the point at threshold labeled ``Gas08''
  is inferred from the 1S level-width as quoted in
  \cite{Gasser:2007zt}. \textit{Right}: The top curve shows the total
  \pipm-\hef\ inelastic scattering cross section as given
  in~\cite{Binon:1975zc} and labeled as ``Bin75''. The data labeled
  ``Mat98''~\cite{Mateos:1998bc} is the absorptive part; older data
  ``Str90'' on pion absorption is collected in
  \cite{Steinacher1990413}.  We employ the model calculation
  of~\cite{Horikawa:1980cv}, scaled by a factor 0.8 to fit ``Bin75''.
  The dashed line shows the difference between the latter two lines
  and thus provides the remaining part of the inelastic cross
  section. The dotted parts of the lines indicate the regions in which
  we extrapolate without data. The dot at threshold is again inferred
  from the pionic helium 1S level-width given
  in~\cite{Backenstoss1974519}.}
\label{deltares}
}

The predominant feature in low-energy pion collisons with nucleons and
nuclei is the formation of $\Delta$-resonances. For reactions with nucleons,
we only need to consider $\pim+p\to\pizero+ n$ across the resonance.
Though at at cosmic times $t\lesssim 180\ \seconds$ neutrons are still
an abundant target, \pip\ are stopped before interacting with
them.  For the charge exchange reaction of $\pim+p$ we employ the
partial wave analysis of~\cite{Arndt:2003fj,Bellaachia:2002qe} as
shown in the left panel of Fig.~\ref{deltares} as a function of pion
kinetic energy $E_{\pi}$ in the rest frame of the proton. It agrees
well with the experimental results of~\cite{Breitschopf:2006gn}. The
dot at threshold corresponds to~(\ref{pi-on-p-chx}). The alternative
exit channel $n+\gamma$ is only relevant at threshold and becomes
quickly negligible with increasing pion energy. We have verified this
from a model calculation~\cite{Arndt:2003fj,Bellaachia:2002qe} of the
inverse process of photo-pion production by employing the principle of
detailed balance.

The right panel of Fig.~\ref{deltares} shows the various contributions
to \pipm-\hef\ scattering. The top curve and data ``Bin75`` represent
the total inelastic cross section as taken
from~\cite{Binon:1975zc}. The cross section breaks down into an
absorptive part with no pion in the final state and the remaining
inelastic part. We fit the in-flight absorption data
``Mat98''~\cite{Mateos:1998bc} by scaling down the calculation
of~\cite{Horikawa:1980cv} by a factor of 0.8; older data ``Str90'' on
absorption is collected in~\cite{Steinacher1990413}. The dotted parts
of the lines show the regions in which we had to extrapolate without
data. In particular, note that inelastic pion scattering (without
absorption) has thresholds in the range $\sim(15\div 30)\ \MeV$
depending on the final state. For simplicity, and because absorption
dominates for low energies, we have extrapolated the inelastic cross
section to a common threshold value $E^{\pipm}_{4,\mathrm{th}}\sim
25\ \MeV$.

\TABLE[t]{ 
  \caption{Adopted final state multiplicities for the various
    \pipm\hef\ reactions. Values are rounded such that baryon number
    remains conserved,
    $3\xi_{\het}+3\xi_{\trit}+2\xi_{\deut}+\xi_n+\xi_p=4$.}
\begin{tabular}{lccccc}
\toprule
 process  & $\xi_{\het}$ & $\xi_{\trit}$ & $\xi_{\deut}$ & $\xi_{n}$ & $\xi_{p}$ \\
\midrule
\pip\ absorption, threshold & 0.17  & --   & 0.63 & 0.2  & 2.03\\
\pim\ absorption, threshold & --    & 0.17 & 0.63 & 2.03 & 0.2 \\
\pip\ absorption, in-flight & 0.01  & --   & 0.1  & 0.87 & 2.9 \\
\pim\ absorption, in-flight & --    & 0.01 & 0.1  & 2.9  & 0.87\\
\pip\ inelastic, \SCX &       0.4   & --   & 0.2  & 0.4  & 2.0 \\
\pim\ inelastic, \SCX &       --    & 0.4  & 0.2  & 2.0  & 0.4 \\
\pip\ inelastic, \NCX &       0.4   & 0.2  & 0.4  & 0.8  & 0.6 \\
\pim\ inelastic, \NCX &       0.4   & 0.2  & 0.4  & 0.8  & 0.6 \\
\pip\ inelastic, \DCX &       0     & 0    & 0    & 0    &  4  \\
\pim\ inelastic, \DCX &       0     & 0    & 0    & 4    &  0  \\
\bottomrule
\label{table1}
\end{tabular}
}
In particular, the output of \het\ and \trit\ from absorption and
inelastic \pipm-\hef\ scatterings is important for the non-thermal
\lisx\ abundance. The adopted nuclear/nucleon final-state
multiplicities are summarized in Table~\ref{table1}.  For absorption,
one finds that the branchings into the various final states differ
significantly depending on whether the process occurs at threshold or
not.  At threshold the fraction of $A=3$ elements is $\sim 20\%$ but
for $E_{\pi}=120\ \MeV$ the respective multiplicities, as inferred
from the measurements of~\cite{Steinacher1990413}, are already very
small. Since, by nature, absorption yields the largest recoil energies
such a behavior is not unexpected and between $E_{\pi}=0$ and 30~\MeV\
we interpolate between the threshold and in-flight $\xi$-values.  For
larger pion energies we switch to the in-flight data.

Above the threshold of inelastic \pipm-\hef\ scattering the
energy-dependent multiplicities $\xi_i$ of elements
$i=\het,\,\trit,\,\deut,\,n$ and $p$ are given by
\begin{align}
  \xi_{i,\mathrm{inel}}^{\pipm} = f_{\SCX} \xi_{i,\SCX}^{\pipm} + (1-f_{\SCX})
  \big[ (1-g_{\DCX}) \xi_{i,\NCX}^{\pipm} + g_{\DCX}\xi_{i,\DCX}  \big] .
\end{align}
The contribution of single charge exchange (\SCX) $f_{\SCX} \equiv
\sigma_{\SCX} / \sigma_{\mathrm{inel}} $ is infered from the
measurements of~\cite{PhysRevC.60.024603} by simple linear regression,
$f_{\SCX}= 0.11 + 6\times 10^{-4} (\Tpi/\MeV )$. We further assume a
constant ratio of double charge exchange (\DCX) to scatterings
preserving the incident the pion charge (\NCX), $g_{\DCX} \equiv
\sigma_{\DCX}/\sigma_{\NCX} = 0.1$. For the multiplicities $
\xi_{i,\alpha}^{\pipm}$ of the individual processes
$\alpha=\SCX,\,\NCX$, and \DCX\ we resort to data of inelastic
\hef-$p$ scattering. To this end, one observes that for incident beam
kinetic energies in excess of $ \sim(50\div 80)\ \MeV$ (above which
also pion inelastic scattering starts dominating over absorpion) the
branchings into $\het n p$, $\trit 2p$, $\deut n2p$ and $\deut\deut p$
become approximately energy independent with respective values $\sim
0.4,\, 0.2,\, 0.2$ and 0.1 (for moderate energies $\lesssim
300\MeV$). Adopting those values yield the multiplicities for
\hef-\pipm\ reactions as listed in Table~\ref{table1}.  Note that the
$\xi$-values imply a quantitative difference between \pim-\hef\ and
\pip-\hef\ scattering.

In a similar manner as in Section~\ref{sec:treatment-pions} where we
introduced the correction factor $\kappa$ one can account for the
incomplete stopping of pions in reactions on \hef\ by defining an
effective cross section for $\pi^{j}\hef\to i\, $,
$i=\het,\,\trit,\,\deut,\,n$ and $p$, averaged over the
pion-trajectory,
\begin{align}
  \VEV{\sigma_{\pi^{j}\hef\to i\, } v}_{\mathrm{traj.}} =
  \tau_{\pi}^{-1} \int_0^{\infty} dt\, P_{\mathrm{surv}}(t) \,
  F_{\pi^j\hef}\, \sum_{\alpha=\mathrm{inel,abs}}
  (\sigma_{\pipm\hef,\alpha} v) \times \xi_{i,\alpha}^{\pi^j}
\end{align}
where 
\begin{align}
  P_{\mathrm{surv}}(t) = \exp{\left(-\int_0^t
      \frac{m_{\pipm}}{\tau_{\pipm}(m_{\pipm}+\Tpi(t'))} dt'\right)}
\end{align}
is the time-dilatated survival probability of the pion. Note that it
was not possible to define a correction factor $\kappa$ as in
(\ref{kappaET}) because the effective rates for $\pim\hef\to \trit $
and $\pip\hef\to \het $ vanish below the threshold of inelastic
\pipm-\hef\ scattering.

The output of secondary \lisx, triggered by \pipm-\hef\ reactions and
prior to potential destruction on protons, can be tracked via
\begin{align}
\label{eq:nonthermli6}
\left.  \frac{d n_{\lisx}}{dt} \right|_{\mathrm{sec}} = n_{\hef}
 \sum_{{i=\het,\trit} \atop j=+,- }
\left\{
\VEV{ \Gamma^{\mathrm{inel}}_{\pi^j\hef\to i} }*
\VEV{N_{\lisx}}_{f_i} 
+
\VEV{ \Gamma^{\mathrm{abs}}_{\pi^j\hef\to i} }*
N_{\lisx}
\right\}
\end{align}
where $\VEV{\Gamma^{\mathrm{inel/abs}}_{\pi^j\hef\to i}} = n_{\pi^j}
\VEV{\sigma^{\mathrm{inel/abs}}_{\pi^{j}\hef\to i\, }
  v}_{\mathrm{traj.}}$ are the effective rates for energetic
$i=\het,\,\trit$ production.  The dependence of $\Tpi$ on time is
suggested by the * in (\ref{eq:nonthermli6}) indicating that the
effective rates are to be ``convolved'' with the number of produced
secondary \lisx\ per injected $A=3$ nucleus, $N_{\lisx}$. For
inelastic scattering this involves an additional average over the
spectrum $f_i$ of primary mass-3 injection energies
$E_{i,\mathrm{in}}$,
\begin{align}
\label{eq:nonthermli6-2}
  \VEV{N_{\lisx}}_{f_i} = n_{\hef}
  \int_{E^i_{6,\mathrm{th}}}^{E_{i,\mathrm{max}}(E_{\pi}(t))} dE_{i,\mathrm{in}}\,
  f_i(E_{i,\mathrm{in}},E_{\pi}(t))
  \int_{E_{i,\mathrm{in}}}^{E^i_{6,\mathrm{th}}} dE_i\,
  \frac{(\sigma_{i+\hef\to \lisx +X} v_i)}{dE_i/dt} .
\end{align}
where we employ~\cite{Cyburt:2009pg}
\begin{align}
\label{eq:a3spectrum}
  f_i(E_{i,\mathrm{in}},E_{\pi}) = \mathcal N^{-1} {\sum_{k=1,2} w_{i,k} E_{i,\mathrm{in}}^{1/2} e^{-E_{i,\mathrm{in}}/T_k}
  }\times \left( \frac{E_{i,\mathrm{max}}-E_{i,\mathrm{in}}}{E_{i,\mathrm{max}}} \right) .
\end{align}
Note that we have introduced a kinematical cutoff in the last factor;
$E_{i,\mathrm{max}}(E_{\pi})$ is the maximal kinetic recoil energy of
$A=3$ in the rest frame of the mother \hef\ nucleus (which
approximately coincides with the frame of the thermal bath) and can be
found by simple kinematical considerations (see also below). The
spectra~(\ref{eq:a3spectrum}) are obtained from fits to experimental
$p$-\hef\ data with coefficients $(w_{\het})=(1,0.19)$,
$(T_{\het}/\MeV)=(1.9,4.9)$ and $(w_{\trit})=(1,0.15)$,
$(T_{\trit}/\MeV)=(1.3,3.9)$~\cite{Blinov:2006rb,Blinov:2}.  The
normalization factor $\mathcal N$ is chosen such that
${\int_0^{E_{i,\mathrm{max}}} dE_{i,\mathrm{in}}\, f_i =1}$.

Finally, we remark that for pion absorption, the resulting non-thermal
\lisx-yield is easier to track because \trit\ and \het\ are injected
mono-energetically in the CM frame so that $ \VEV{N_{\lisx}}_{f_i} \to
N_{\lisx}$ as in (\ref{eq:Lisx-efficiency}). The primary $A=3$
injection energy in the rest frame of \hef\ (the frame of the thermal
bath) then depends on the scattering angle and lies within the interval
\begin{align}
    E_{3,in}(\Tpi) \in \frac{1}{2 m_3} \left[\pm \sqrt{2\mu'
      \left( \frac{\mu_{\pi 4}}{m_{\pipm}} \Tpi +Q \right)}
    +\frac{ \sqrt{2 m_{\pipm} \Tpi}m_3}{m_{\pipm}+m_{\alpha}} \right]^2 ,
\end{align}
where $\mu_{\pi 4}$ and $\mu'$ are the respective reduced masses
before and after scattering. In our code, we adopt the central value.
Figure~\ref{NLi} shows the efficiency for production of a secondary
\lisx\ nucleus, per injected $\het$+\trit\ as a function of
temperature. Pion absorption (solid and dotted lines) \textit{a
  priori} is the most efficient way of generating \lisx\ as the $A=3$
recoil is maximized. The probabilities have to be weighted by the
mass-3 multiplicities and relative strengths of absorption and
inelastic scattering. Moreover, it has to be stressed that
Fig.~\ref{NLi} does not include \lisx-destruction on thermal protons
which ultimately determines the surviving lithium fraction.

\FIGURE[t]{
\includegraphics[width=0.6\textwidth]{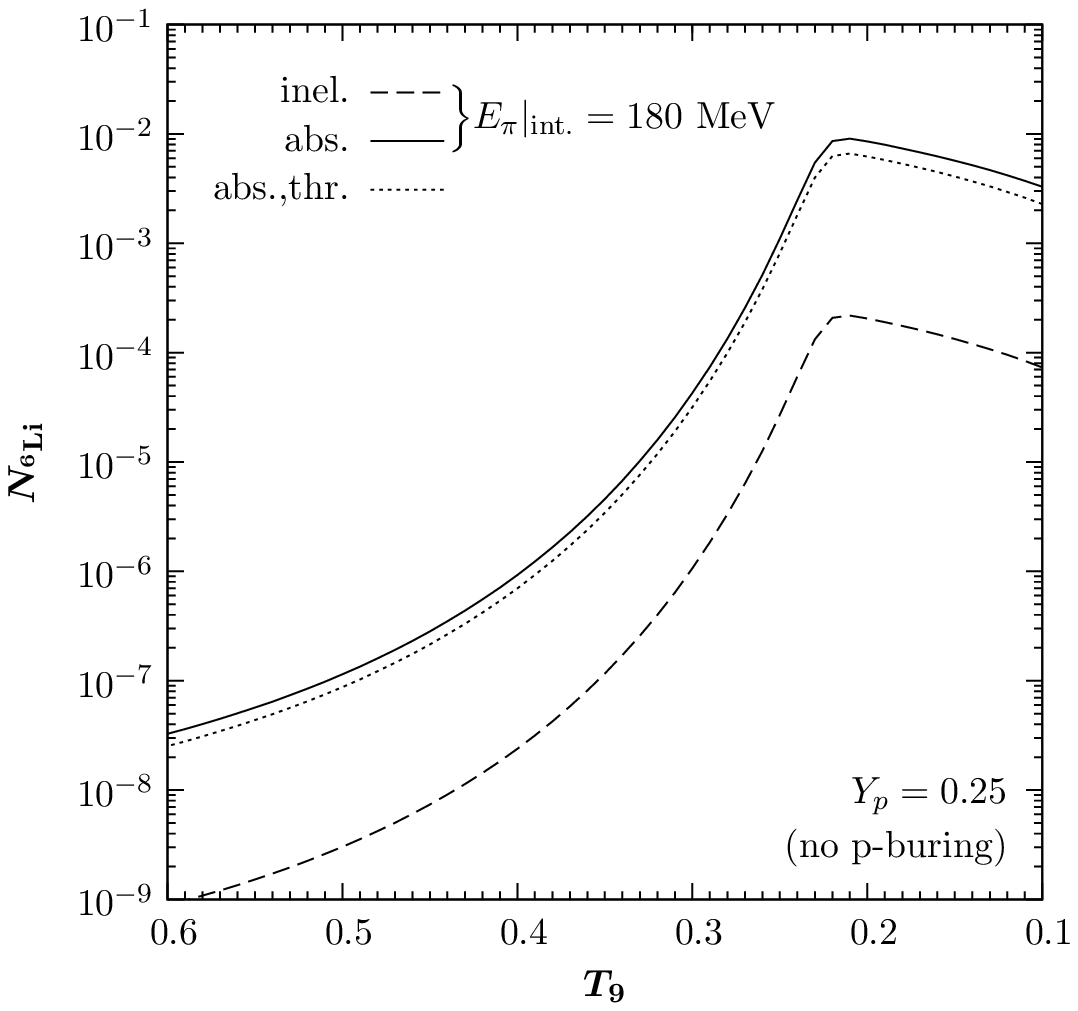}
\caption{\small Number of secondary \lisx\ produced per injected
  $\het+\trit$ \textit{prior} to potential destruction on protons. The
  dotted line shows $N_{\lisx}$ for absorption on threshold [compare
  (\ref{eq:Lisx-efficiency})]. The solid and dashed lines show the
  probability of producing \lisx\ from in-flight reactions of a pion
  with fixed kinetic energy $E_{\pi}|_{\mathrm{int.}}=180\ \MeV$ on \hef\ via
  absorption and inelastic scattering, respectively. For the latter
  compare with (\ref{eq:nonthermli6-2}). These curves have to be
  convoluted with the actual energy-dependent branching into mass-3
  elements as in (\ref{eq:nonthermli6}). Note that $p$-burning will
  have a substantial effect on the ultimately surviving fraction of
  secondary \lisx\ down to the lowest temperatures shown.}
\label{NLi}
}

\subsection{Electromagnetic energy injection}
\label{sec:electr-energy-inject}

Our treatment of the non-thermal electromagnetic component introduced
in the decay of pions, kaons, and muons is mainly along the
lines of Ref.~\cite{Cyburt:2002uv}. The injection of energetic
primaries $e^{\pm}$ and/or $\gamma$-particles leads to a quick
formation of an electromagnetic cascade via the processes of inverse
Compton scattering and $e^{\pm}$-pair production on background
photons. The resulting ``zeroth-generation'' differential photon
spectrum is given by a broken power law~\cite{Protheroe:1994dt},
\begin{align}
  p_{\gamma}(E_{\gamma}) = 
\begin{cases} 
K_0 (E_{\gamma}/E_{\mathrm{low}})^{-1.5}  &\mathrm{for}\
  E_{\gamma}<E_{\mathrm{low}} \\
 K_0 (E_{\gamma}/E_{\mathrm{low}})^{-2.0} & \mathrm{for}\
  E_{\mathrm{low}}<E_{\gamma}<E_C \\
\qquad\quad 0 & \mathrm{for}\ E_{\gamma}>E_C
\end{cases}
\label{eq:pgamma}
\end{align}
with a cut-off at the pair production threshold $E_C \simeq m_e^2/22T$
and a power break at $E_{\rm low} \simeq
m_e^2/(80T)$~\cite{Kawasaki:1994sc}.  The overall normalization of the
spectrum is determined from the primary injected energy $E_0$,
\begin{equation}
  E_0 = \int d E_\gamma E_\gamma p_{\gamma}(E_\gamma) = K_0 E_{\rm low}^2(2+\ln(E_C/E_{\rm low}).
\end{equation}

The photons in the cascade~(\ref{eq:pgamma}) undergo further
degradation via the (slower) processes of Compton scattering,
pair-production on nuclei, and elastic $\gamma$-$\gamma$ scattering so
that the total number of energetic photons is given by a competition
of injection rate $\Gamma_{\rm inj}$ and total energy loss rate
$\Gamma_{\gamma}(E_\gamma)$.  The energy spectrum can then be obtained
in form of a quasi-static equilibrium solution
\begin{equation}
  f_{\gamma}^{\rm qse} = n_X\fr{\Gamma_{\rm inj}p_\gamma(E_\gamma)}{\Gamma_\gamma(E_\gamma)},
\end{equation}
where $n_X$ is the time-dependent number density of the decaying
particles, and $\Gamma_{\rm inj} = \tau_X^{-1}$. For the
thermalization rate we put
\begin{align}
\label{eq:Gamma-degrad-rate}
  \Gamma_{\gamma} = \Gamma_{\mathrm{BH,p}} +\Gamma_{\mathrm{BH,\hef}}
  + k(E_{\gamma})\Gamma_C + 0.5\Gamma_{\gamma\gamma} ,
\end{align}
where $\Gamma_{\mathrm{BH,N}}$, $\Gamma_C$, and
$\Gamma_{\gamma\gamma}$ are the respective rates for Bethe-Heitler
scattering, Compton scattering, and $\gamma$-$\gamma$ scattering. For
$ \Gamma_{\mathrm{BH},N} = n_N \sigma_{\mathrm{BH},N} $ we employ the
cross section formul\ae\ as given in~\cite{Kawasaki:1994sc}. The rate
for Compton scattering, $\Gamma_C = n_e \sigma_{\mathrm{KN}}$, depends
on the Klein-Nishina cross section
$\sigma_{\mathrm{KN}}$~\cite{1976tper.book.....J} and on the total
electron (positron) number density $n_{e}$, with an average fractional
energy-loss per scattering~\cite{Protheroe:1994dt},
\begin{align}
  k(E_{\gamma}) \simeq 1 - \frac{4/3}{\ln{(2E_{\gamma}/m_e)} + 1/2} .
\end{align}
The rate of photon scattering on blackbody photons, obtained in
\cite{1990ApJ...349..415S}, reads
  \begin{align}
    \Gamma_{\gamma \gamma}(\varepsilon) = \frac{2^5 139 \pi^4
      \alpha^{4}}{3^{6} 5^4 } \frac{m_{e}}{2\pi} \left( \frac{T}{m_e}
    \right)^{6} \left(\frac{E_{\gamma}}{m_e} \right)^3 ,
  \end{align}
  and is valid below the $e^{\pm}$ pair production threshold. The
  factor of $1/2$ in (\ref{eq:Gamma-degrad-rate}) can be understood by
  noting that photon-photon scattering tends to split primary gamma
  rays into two gamma rays carrying on average each $50\%$ of the
  primary energy~\cite{1990ApJ...349..415S}. Taken all together,
  $\Gamma_{\gamma}^{-1}$ then defines an approximate survival time for
  the ``zeroth-generation'' photons.

  The production and destruction of elements due to electromagentic
  energy injection can then be described by the Boltzmann equations
  for nuclei $T,A,P$, ($A_T > A_A > A_P$):
\begin{eqnarray}
  -HT\frac{dY_A}{dT} =  \sum_{T} Y_T \int_0^\infty dE_\gamma f_{\gamma}^{\rm qse}(E_\gamma)
  \sigma_{\gamma +T \to A}(E_\gamma) \nonumber \\ 
  - Y_A \sum_{P} \int_0^\infty dE_\gamma f_{\gamma}^{\rm qse}(E_\gamma)
  \sigma_{\gamma +A \to P}(E_\gamma).
\end{eqnarray} 
In our Boltzmann code we include all processes determined
in~\cite{Cyburt:2002uv} from which we also take the photodissociation
cross sections~$\sigma_{\gamma + i \to j}$. 
Finally, we remark that we also take into account the secondary
production of $\lisx$ by fusion-reactions of energetic mass-3 elements
which are produced by photo-spallations of \hef. The efficiency of
producing $\lisx$ can be found in analogy
to~(\ref{eq:Lisx-efficiency}).

\bibliography{biblio}

\providecommand{\href}[2]{#2}\begingroup\raggedright\begin{thebibliography}{10%
0}

\bibitem{Dunkley:2008ie}
J.~Dunkley {\em et.~al.}, {\it {Five-Year Wilkinson Microwave Anisotropy Probe
  (WMAP) Observations: Likelihoods and Parameters from the WMAP data}},  {\em
  Astrophys. J. Suppl.} {\bf 180} (2009) 306--329
  [\href{http://arXiv.org/abs/0803.0586}{{\tt 0803.0586}}].

\bibitem{Iocco:2008va}
F.~Iocco, G.~Mangano, G.~Miele, O.~Pisanti and P.~D. Serpico, {\it {Primordial
  Nucleosynthesis: from precision cosmology to fundamental physics}},  {\em
  Phys. Rept.} {\bf 472} (2009) 1--76
  [\href{http://arXiv.org/abs/0809.0631}{{\tt 0809.0631}}].

\bibitem{Cyburt:2008kw}
R.~H. Cyburt, B.~D. Fields and K.~A. Olive, {\it {A Bitter Pill: The Primordial
  Lithium Problem Worsens}},  {\em JCAP} {\bf 0811} (2008) 012
  [\href{http://arXiv.org/abs/0808.2818}{{\tt 0808.2818}}].

\bibitem{Steigman:2007xt}
G.~Steigman, {\it {Primordial Nucleosynthesis in the Precision Cosmology Era}},
   {\em Ann. Rev. Nucl. Part. Sci.} {\bf 57} (2007) 463--491
  [\href{http://arXiv.org/abs/0712.1100}{{\tt 0712.1100}}].

\bibitem{Jedamzik:2009uy}
K.~Jedamzik and M.~Pospelov, {\it {Big Bang Nucleosynthesis and Particle Dark
  Matter}},  {\em New J. Phys.} {\bf 11} (2009) 105028
  [\href{http://arXiv.org/abs/0906.2087}{{\tt 0906.2087}}].

\bibitem{Spite:1982dd}
F.~Spite and M.~Spite, {\it {Abundance of lithium in unevolved halo stars and
  old disk stars: Interpretation and consequences}},  {\em Astron. Astrophys.}
  {\bf 115} (1982) 357--366.

\bibitem{Reno:1987qw}
M.~H. Reno and D.~Seckel, {\it {Primordial Nucleosynthesis: The Effects of
  Injecting Hadrons}},  {\em Phys. Rev.} {\bf D37} (1988) 3441.

\bibitem{Jedamzik:2004er}
K.~Jedamzik, {\it {Did something decay, evaporate, or annihilate during big
  bang nucleosynthesis?}},  {\em Phys. Rev.} {\bf D70} (2004) 063524
  [\href{http://arXiv.org/abs/astro-ph/0402344}{{\tt astro-ph/0402344}}].

\bibitem{Kawasaki:2004yh}
M.~Kawasaki, K.~Kohri and T.~Moroi, {\it {Hadronic decay of late-decaying
  particles and big-bang nucleosynthesis}},  {\em Phys. Lett.} {\bf B625}
  (2005) 7--12 [\href{http://arXiv.org/abs/astro-ph/0402490}{{\tt
  astro-ph/0402490}}].

\bibitem{Kawasaki:2004qu}
M.~Kawasaki, K.~Kohri and T.~Moroi, {\it {Big-bang nucleosynthesis and hadronic
  decay of long-lived massive particles}},  {\em Phys. Rev.} {\bf D71} (2005)
  083502 [\href{http://arXiv.org/abs/astro-ph/0408426}{{\tt
  astro-ph/0408426}}].

\bibitem{Jedamzik:2006xz}
K.~Jedamzik, {\it {Big bang nucleosynthesis constraints on hadronically and
  electromagnetically decaying relic neutral particles}},  {\em Phys. Rev.}
  {\bf D74} (2006) 103509 [\href{http://arXiv.org/abs/hep-ph/0604251}{{\tt
  hep-ph/0604251}}].

\bibitem{Steffen:2006hw}
F.~D. Steffen, {\it {Gravitino dark matter and cosmological constraints}},
  {\em JCAP} {\bf 0609} (2006) 001
  [\href{http://arXiv.org/abs/hep-ph/0605306}{{\tt hep-ph/0605306}}].

\bibitem{Cyburt:2006uv}
R.~H. Cyburt, J.~R. Ellis, B.~D. Fields, K.~A. Olive and V.~C. Spanos, {\it
  Bound-state effects on light-element abundances in gravitino dark matter
  scenarios},  {\em JCAP} {\bf 0611} (2006) 014
  [\href{http://arXiv.org/abs/astro-ph/0608562}{{\tt astro-ph/0608562}}].

\bibitem{Cyburt:2009pg}
R.~H. Cyburt {\em et.~al.}, {\it {Nucleosynthesis Constraints on a Massive
  Gravitino in Neutralino Dark Matter Scenarios}},  {\em JCAP} {\bf 0910}
  (2009) 021 [\href{http://arXiv.org/abs/0907.5003}{{\tt 0907.5003}}].

\bibitem{Freitas:2009jb}
A.~Freitas, F.~D. Steffen, N.~Tajuddin and D.~Wyler, {\it {Late Energy
  Injection and Cosmological Constraints in Axino Dark Matter Scenarios}},
  {\em Phys. Lett.} {\bf B682} (2009) 193--199
  [\href{http://arXiv.org/abs/0909.3293}{{\tt 0909.3293}}].

\bibitem{Jedamzik:2004ip}
K.~Jedamzik, {\it {Neutralinos and Big Bang nucleosynthesis}},  {\em Phys.
  Rev.} {\bf D70} (2004) 083510
  [\href{http://arXiv.org/abs/astro-ph/0405583}{{\tt astro-ph/0405583}}].

\bibitem{Jedamzik:2005dh}
K.~Jedamzik, K.-Y. Choi, L.~Roszkowski and R.~Ruiz~de Austri, {\it {Solving the
  cosmic lithium problems with gravitino dark matter in the CMSSM}},  {\em
  JCAP} {\bf 0607} (2006) 007 [\href{http://arXiv.org/abs/hep-ph/0512044}{{\tt
  hep-ph/0512044}}].

\bibitem{Cumberbatch:2007me}
D.~Cumberbatch {\em et.~al.}, {\it Solving the cosmic lithium problems with
  primordial late- decaying particles},  {\em Phys. Rev.} {\bf D76} (2007)
  123005 [\href{http://arXiv.org/abs/0708.0095}{{\tt 0708.0095}}].

\bibitem{Pospelov:2006sc}
M.~Pospelov, {\it Particle physics catalysis of thermal big bang
  nucleosynthesis},  {\em Phys. Rev. Lett.} {\bf 98} (2007) 231301
  [\href{http://arXiv.org/abs/hep-ph/0605215}{{\tt hep-ph/0605215}}].

\bibitem{Bird:2007ge}
C.~Bird, K.~Koopmans and M.~Pospelov, {\it {Primordial Lithium Abundance in
  Catalyzed Big Bang Nucleosynthesis}},  {\em Phys. Rev.} {\bf D78} (2008)
  083010 [\href{http://arXiv.org/abs/hep-ph/0703096}{{\tt hep-ph/0703096}}].

\bibitem{Jittoh:2007fr}
T.~Jittoh {\em et.~al.}, {\it {Possible solution to the $^7$Li problem by the
  long lived stau}},  {\em Phys. Rev.} {\bf D76} (2007) 125023
  [\href{http://arXiv.org/abs/0704.2914}{{\tt 0704.2914}}].

\bibitem{Kusakabe:2007fv}
M.~Kusakabe, T.~Kajino, R.~N. Boyd, T.~Yoshida and G.~J. Mathews, {\it {The
  $X^-$ Solution to the $^6$Li and $^7$Li Big Bang Nucleosynthesis Problems}},
  \href{http://arXiv.org/abs/0711.3858}{{\tt 0711.3858}}.

\bibitem{ArkaniHamed:2008qn}
N.~Arkani-Hamed, D.~P. Finkbeiner, T.~R. Slatyer and N.~Weiner, {\it {A Theory
  of Dark Matter}},  {\em Phys. Rev.} {\bf D79} (2009) 015014
  [\href{http://arXiv.org/abs/0810.0713}{{\tt 0810.0713}}].

\bibitem{Pospelov:2008jd}
M.~Pospelov and A.~Ritz, {\it {Astrophysical Signatures of Secluded Dark
  Matter}},  {\em Phys. Lett.} {\bf B671} (2009) 391--397
  [\href{http://arXiv.org/abs/0810.1502}{{\tt 0810.1502}}].

\bibitem{Adriani:2008zr}
{\bf PAMELA} Collaboration, O.~Adriani {\em et.~al.}, {\it {An anomalous
  positron abundance in cosmic rays with energies 1.5.100 GeV}},  {\em Nature}
  {\bf 458} (2009) 607--609 [\href{http://arXiv.org/abs/0810.4995}{{\tt
  0810.4995}}].

\bibitem{Abdo:2009zk}
{\bf The Fermi LAT} Collaboration, A.~A. Abdo {\em et.~al.}, {\it {Measurement
  of the Cosmic Ray e+ plus e- spectrum from 20 GeV to 1 TeV with the Fermi
  Large Area Telescope}},  {\em Phys. Rev. Lett.} {\bf 102} (2009) 181101
  [\href{http://arXiv.org/abs/0905.0025}{{\tt 0905.0025}}].

\bibitem{Pospelov:2007mp}
M.~Pospelov, A.~Ritz and M.~B. Voloshin, {\it {Secluded WIMP Dark Matter}},
  {\em Phys. Lett.} {\bf B662} (2008) 53--61
  [\href{http://arXiv.org/abs/0711.4866}{{\tt 0711.4866}}].

\bibitem{Feng:2010zp}
J.~L. Feng, M.~Kaplinghat and H.-B. Yu, {\it {Sommerfeld Enhancements for
  Thermal Relic Dark Matter}},  \href{http://arXiv.org/abs/1005.4678}{{\tt
  1005.4678}}.

\bibitem{Cirelli:2010nh}
M.~Cirelli and J.~M. Cline, {\it {Can multistate dark matter annihilation
  explain the high- energy cosmic ray lepton anomalies?}},
  \href{http://arXiv.org/abs/1005.1779}{{\tt 1005.1779}}.

\bibitem{Fox:2008kb}
P.~J. Fox and E.~Poppitz, {\it {Leptophilic Dark Matter}},  {\em Phys. Rev.}
  {\bf D79} (2009) 083528 [\href{http://arXiv.org/abs/0811.0399}{{\tt
  0811.0399}}].

\bibitem{Nomura:2008ru}
Y.~Nomura and J.~Thaler, {\it {Dark Matter through the Axion Portal}},  {\em
  Phys. Rev.} {\bf D79} (2009) 075008
  [\href{http://arXiv.org/abs/0810.5397}{{\tt 0810.5397}}].

\bibitem{Batell:2009di}
B.~Batell, M.~Pospelov and A.~Ritz, {\it {Exploring Portals to a Hidden Sector
  Through Fixed Targets}},  {\em Phys. Rev.} {\bf D80} (2009) 095024
  [\href{http://arXiv.org/abs/0906.5614}{{\tt 0906.5614}}].

\bibitem{Amsler:2008zzb}
{\bf Particle Data Group} Collaboration, C.~Amsler {\em et.~al.}, {\it {Review
  of particle physics}},  {\em Phys. Lett.} {\bf B667} (2008) 1.

\bibitem{Gasser:2007zt}
J.~Gasser, V.~E. Lyubovitskij and A.~Rusetsky, {\it {Hadronic atoms in QCD +
  QED}},  {\em Phys. Rept.} {\bf 456} (2008) 167--251
  [\href{http://arXiv.org/abs/0711.3522}{{\tt 0711.3522}}].

\bibitem{Flugel:1999gr}
T.~Flugel, {\it {The pion beta decay experiment and a remeasurement of the
  Panofsky ratio}}, . DISS-ETH-13105.

\bibitem{Backenstoss1974519}
G.~Backenstoss, J.~Egger, T.~von Egidy, R.~Hagelberg, C.~J. Herrlander,
  H.~Koch, H.~P. Povel, A.~Schwitter and L.~Tauscher, {\it Pionic and muonic
  x-ray transitions in liquid helium},  {\em Nuclear Physics A} {\bf 232}
  (1974), no.~2 519 -- 532.

\bibitem{Daum1995553}
E.~Daum, S.~Vinzelberg, D.~Gotta, H.~Ullrich, G.~Backenstoss, P.~Weber, H.~J.
  Weyer, M.~Furic and T.~Petkovic, {\it Pion absorption at rest in 4he},  {\em
  Nuclear Physics A} {\bf 589} (1995), no.~4 553 -- 584.

\bibitem{Binon:1975zc}
{\bf Brussels-Orsay} Collaboration, F.~G. Binon and et~al., {\it {Pion He-4
  Scattering Around the 3/2, 3/2 Resonance}},  {\em Phys. Rev. Lett.} {\bf 35}
  (1975) 145.

\bibitem{Steinacher1990413}
M.~Steinacher, G.~Backenstoss, M.~Izycki, P.~Salvisberg, P.~Weber, H.~J. Weyer,
  A.~Hoffart, B.~Rzehorz, H.~Ullrich, M.~Dzemidzic, M.~Furic and T.~Petkovic,
  {\it Pion absorption in flight on 4he},  {\em Nuclear Physics A} {\bf 517}
  (1990), no.~3-4 413 -- 454.

\bibitem{Mateos:1998bc}
{\bf LADS} Collaboration, A.~O. Mateos {\em et.~al.}, {\it {Total and partial
  pion absorption cross sections on He-4 in the Delta resonance region}},  {\em
  Phys. Rev.} {\bf C58} (1998) 942--952.

\bibitem{Dimopoulos:1988ue}
S.~Dimopoulos, R.~Esmailzadeh, L.~J. Hall and G.~D. Starkman, {\it {Limits on
  late decaying particles from nucleosynthesis}},  {\em Nucl. Phys.} {\bf B311}
  (1989) 699.

\bibitem{Kawasaki:1994sc}
M.~Kawasaki and T.~Moroi, {\it {Electromagnetic cascade in the early {U}niverse
  and its application to the big bang nucleosynthesis}},  {\em Astrophys. J.}
  {\bf 452} (1995) 506 [\href{http://arXiv.org/abs/astro-ph/9412055}{{\tt
  astro-ph/9412055}}].

\bibitem{2002A&A...390...91B}
P.~{Bonifacio}, L.~{Pasquini}, F.~{Spite}, A.~{Bragaglia}, E.~{Carretta},
  V.~{Castellani}, M.~{Centuri{\`o}n}, A.~{Chieffi}, R.~{Claudi},
  G.~{Clementini}, F.~{D'Antona}, S.~{Desidera}, P.~{Fran{\c c}ois}, R.~G.
  {Gratton}, F.~{Grundahl}, G.~{James}, S.~{Lucatello}, C.~{Sneden} and
  O.~{Straniero}, {\it {The lithium content of the globular cluster NGC 6397}},
   {\em \aap} {\bf 390} (July, 2002) 91--101
  [\href{http://arXiv.org/abs/arXiv:astro-ph/0204332}{{\tt
  arXiv:astro-ph/0204332}}].

\bibitem{Melendez:2004ni}
J.~Melendez and I.~Ramirez, {\it {Reappraising the Spite Lithium Plateau:
  Extremely Thin and Marginally Consistent with WMAP}},  {\em Astrophys. J.}
  {\bf 615} (2004) L33 [\href{http://arXiv.org/abs/astro-ph/0409383}{{\tt
  astro-ph/0409383}}].

\bibitem{Aoki:2009ce}
W.~Aoki {\em et.~al.}, {\it {Lithium Abundances of Extremely Metal-Poor
  Turn-off Stars}},  {\em Astrophys. J.} {\bf 698} (2009) 1803--1812
  [\href{http://arXiv.org/abs/0904.1448}{{\tt 0904.1448}}].

\bibitem{2010arXiv1003.4510S}
L.~{Sbordone}, P.~{Bonifacio}, E.~{Caffau}, H.~{Ludwig}, N.~T. {Behara}, J.~I.
  {Gonzalez Hernandez}, M.~{Steffen}, R.~{Cayrel}, B.~{Freytag}, C.~{Van't
  Veer}, P.~{Molaro}, B.~{Plez}, T.~{Sivarani}, M.~{Spite}, F.~{Spite}, T.~C.
  {Beers}, N.~{Christlieb}, P.~{Francois} and V.~{Hill}, {\it {The metal-poor
  end of the Spite plateau. 1: Stellar parameters, metallicities and lithium
  abundances}},  {\em ArXiv e-prints} (Mar., 2010)
  [\href{http://arXiv.org/abs/1003.4510}{{\tt 1003.4510}}].

\bibitem{Melendez:2010kw}
J.~Melendez, L.~Casagrande, I.~Ramirez, M.~Asplund and W.~Schuster, {\it
  {Observational evidence for a broken Li Spite plateau and mass-dependent Li
  depletion}},  \href{http://arXiv.org/abs/1005.2944}{{\tt 1005.2944}}.

\bibitem{Burles:1997fa}
S.~Burles and D.~Tytler, {\it {The Deuterium Abundance Towards QSO 1009+2956}},
   {\em Astrophys. J.} {\bf 507} (1998) 732--744
  [\href{http://arXiv.org/abs/astro-ph/9712109}{{\tt astro-ph/9712109}}].

\bibitem{Kirkman:2003uv}
D.~Kirkman, D.~Tytler, N.~Suzuki, J.~M. O'Meara and D.~Lubin, {\it {The
  cosmological baryon density from the deuterium to hydrogen ratio towards QSO
  absorption systems: D/H towards Q1243+3047}},  {\em Astrophys. J. Suppl.}
  {\bf 149} (2003) 1 [\href{http://arXiv.org/abs/astro-ph/0302006}{{\tt
  astro-ph/0302006}}].

\bibitem{Izotov:2010ca}
Y.~I. Izotov and T.~X. Thuan, {\it {The primordial abundance of 4He: evidence
  for non-standard big bang nucleosynthesis}},  {\em Astrophys. J.} {\bf 710}
  (2010) L67--L71 [\href{http://arXiv.org/abs/1001.4440}{{\tt 1001.4440}}].

\bibitem{Aver:2010wq}
E.~Aver, K.~A. Olive and E.~D. Skillman, {\it {A New Approach to Systematic
  Uncertainties and Self- Consistency in Helium Abundance Determinations}},
  \href{http://arXiv.org/abs/1001.5218}{{\tt 1001.5218}}.

\bibitem{Asplund:2005yt}
M.~Asplund, D.~L. Lambert, P.~E. Nissen, F.~Primas and V.~V. Smith, {\it
  {Lithium isotopic abundances in metal-poor halo stars}},  {\em Astrophys. J.}
  {\bf 644} (2006) 229--259 [\href{http://arXiv.org/abs/astro-ph/0510636}{{\tt
  astro-ph/0510636}}].

\bibitem{Steffen:2010pe}
M.~Steffen, R.~Cayrel, P.~Bonifacio, H.~G. Ludwig and E.~Caffau, {\it
  {Convection and 6Li in the atmospheres of metal-poor halo stars}},
  \href{http://arXiv.org/abs/1001.3274}{{\tt 1001.3274}}.

\bibitem{1993oee..conf.....P}
N.~{Prantzos}, E.~{Vangioni-Flam} and M.~{Casse}, eds., {\em {Origin and
  evolution of the elements}}, Jan., 1993.

\bibitem{Dalitz:1960du}
R.~H. Dalitz and S.~F. Tuan, {\it {The phenomenological description of -K
  -nucleon reaction processes}},  {\em Annals Phys.} {\bf 10} (1960) 307.

\bibitem{Martin:1970je}
A.~D. Martin and G.~G. Ross, {\it {K matrix analysis of the low-energy data for
  k- p and k02 p reactions}},  {\em Nucl. Phys.} {\bf B16} (1970) 479--502.

\bibitem{PhysRevD.1.1267}
P.~A. Katz, K.~Bunnell, M.~Derrick, T.~Fields, L.~G. Hyman and G.~Keyes, {\it
  Reactions of stopping $k-$ in helium},  {\em Phys. Rev. D} {\bf 1} (Mar,
  1970) 1267--1276.

\bibitem{Kanzaki:2006hm}
T.~Kanzaki, M.~Kawasaki, K.~Kohri and T.~Moroi, {\it Cosmological constraints
  on gravitino lsp scenario with sneutrino nlsp},  {\em Phys. Rev.} {\bf D75}
  (2007) 025011 [\href{http://arXiv.org/abs/hep-ph/0609246}{{\tt
  hep-ph/0609246}}].

\bibitem{1984MNRAS.210..359S}
R.~J. {Scherrer}, {\it {Deuterium and helium-3 production from massive neutrino
  decay}},  {\em \mnras} {\bf 210} (Sept., 1984) 359--371.

\bibitem{Raffelt:1996wa}
G.~G. Raffelt, {\em {Stars as laboratories for fundamental physics: The
  astrophysics of neutrinos, axions, and other weakly interacting particles}}.
\newblock Chicago, USA: Univ. Pr. 664 p, 1996.

\bibitem{Pastor:2001iu}
S.~Pastor, G.~G. Raffelt and D.~V. Semikoz, {\it {Physics of synchronized
  neutrino oscillations caused by self-interactions}},  {\em Phys. Rev.} {\bf
  D65} (2002) 053011 [\href{http://arXiv.org/abs/hep-ph/0109035}{{\tt
  hep-ph/0109035}}].

\bibitem{Bemporad:2001qy}
C.~Bemporad, G.~Gratta and P.~Vogel, {\it {Reactor-based neutrino oscillation
  experiments}},  {\em Rev. Mod. Phys.} {\bf 74} (2002) 297
  [\href{http://arXiv.org/abs/hep-ph/0107277}{{\tt hep-ph/0107277}}].

\bibitem{Strumia:2003zx}
A.~Strumia and F.~Vissani, {\it {Precise quasielastic neutrino nucleon cross
  section}},  {\em Phys. Lett.} {\bf B564} (2003) 42--54
  [\href{http://arXiv.org/abs/astro-ph/0302055}{{\tt astro-ph/0302055}}].

\bibitem{2008ApJ...686..448Y}
T.~{Yoshida}, T.~{Suzuki}, S.~{Chiba}, T.~{Kajino}, H.~{Yokomakura},
  K.~{Kimura}, A.~{Takamura} and D.~H. {Hartmann}, {\it {Neutrino-Nucleus
  Reaction Cross Sections for Light Element Synthesis in Supernova
  Explosions}},  {\em \apj} {\bf 686} (Oct., 2008) 448--466
  [\href{http://arXiv.org/abs/0807.2723}{{\tt 0807.2723}}].

\bibitem{Finkbeiner:2007kk}
D.~P. Finkbeiner and N.~Weiner, {\it {Exciting Dark Matter and the INTEGRAL/SPI
  511 keV signal}},  {\em Phys. Rev.} {\bf D76} (2007) 083519
  [\href{http://arXiv.org/abs/astro-ph/0702587}{{\tt astro-ph/0702587}}].

\bibitem{McDonald:1993ex}
J.~McDonald, {\it {Gauge Singlet Scalars as Cold Dark Matter}},  {\em Phys.
  Rev.} {\bf D50} (1994) 3637--3649
  [\href{http://arXiv.org/abs/hep-ph/0702143}{{\tt hep-ph/0702143}}].

\bibitem{Burgess:2000yq}
C.~P. Burgess, M.~Pospelov and T.~ter Veldhuis, {\it {The minimal model of
  nonbaryonic dark matter: A singlet scalar}},  {\em Nucl. Phys.} {\bf B619}
  (2001) 709--728 [\href{http://arXiv.org/abs/hep-ph/0011335}{{\tt
  hep-ph/0011335}}].

\bibitem{McDonald:2001vt}
J.~McDonald, {\it {Thermally generated gauge singlet scalars as self-
  interacting dark matter}},  {\em Phys. Rev. Lett.} {\bf 88} (2002) 091304
  [\href{http://arXiv.org/abs/hep-ph/0106249}{{\tt hep-ph/0106249}}].

\bibitem{Holdom:1985ag}
B.~Holdom, {\it {Two U(1)'s and Epsilon Charge Shifts}},  {\em Phys. Lett.}
  {\bf B166} (1986) 196.

\bibitem{Pospelov:2008zw}
M.~Pospelov, {\it {Secluded U(1) below the weak scale}},  {\em Phys. Rev.} {\bf
  D80} (2009) 095002 [\href{http://arXiv.org/abs/0811.1030}{{\tt 0811.1030}}].

\bibitem{Batell:2009jf}
B.~Batell, M.~Pospelov and A.~Ritz, {\it {Multi-lepton Signatures of a Hidden
  Sector in Rare B Decays}},  \href{http://arXiv.org/abs/0911.4938}{{\tt
  0911.4938}}.

\bibitem{Bird:2004ts}
C.~Bird, P.~Jackson, R.~V. Kowalewski and M.~Pospelov, {\it {Search for dark
  matter in b {$\to$} s transitions with missing energy}},  {\em Phys. Rev.
  Lett.} {\bf 93} (2004) 201803
  [\href{http://arXiv.org/abs/hep-ph/0401195}{{\tt hep-ph/0401195}}].

\bibitem{Bird:2006jd}
C.~Bird, R.~V. Kowalewski and M.~Pospelov, {\it {Dark matter pair-production in
  b {$\to$} s transitions}},  {\em Mod. Phys. Lett.} {\bf A21} (2006) 457--478
  [\href{http://arXiv.org/abs/hep-ph/0601090}{{\tt hep-ph/0601090}}].

\bibitem{Badin:2010uh}
A.~Badin and A.~A. Petrov, {\it {Searching for light Dark Matter in heavy meson
  decays}},  \href{http://arXiv.org/abs/1005.1277}{{\tt 1005.1277}}.

\bibitem{Batell:2009yf}
B.~Batell, M.~Pospelov and A.~Ritz, {\it {Probing a Secluded U(1) at
  B-factories}},  {\em Phys. Rev.} {\bf D79} (2009) 115008
  [\href{http://arXiv.org/abs/0903.0363}{{\tt 0903.0363}}].

\bibitem{Reece:2009un}
M.~Reece and L.-T. Wang, {\it {Searching for the light dark gauge boson in
  GeV-scale experiments}},  {\em JHEP} {\bf 07} (2009) 051
  [\href{http://arXiv.org/abs/0904.1743}{{\tt 0904.1743}}].

\bibitem{Essig:2009nc}
R.~Essig, P.~Schuster and N.~Toro, {\it {Probing Dark Forces and Light Hidden
  Sectors at Low-Energy e+e- Colliders}},  {\em Phys. Rev.} {\bf D80} (2009)
  015003 [\href{http://arXiv.org/abs/0903.3941}{{\tt 0903.3941}}].

\bibitem{Baumgart:2009tn}
M.~Baumgart, C.~Cheung, J.~T. Ruderman, L.-T. Wang and I.~Yavin, {\it
  {Non-Abelian Dark Sectors and Their Collider Signatures}},  {\em JHEP} {\bf
  04} (2009) 014 [\href{http://arXiv.org/abs/0901.0283}{{\tt 0901.0283}}].

\bibitem{Bjorken:2009mm}
J.~D. Bjorken, R.~Essig, P.~Schuster and N.~Toro, {\it {New Fixed-Target
  Experiments to Search for Dark Gauge Forces}},  {\em Phys. Rev.} {\bf D80}
  (2009) 075018 [\href{http://arXiv.org/abs/0906.0580}{{\tt 0906.0580}}].

\bibitem{Batell:2009zp}
B.~Batell, M.~Pospelov, A.~Ritz and Y.~Shang, {\it {Solar Gamma Rays Powered by
  Secluded Dark Matter}},  {\em Phys. Rev.} {\bf D81} (2010) 075004
  [\href{http://arXiv.org/abs/0910.1567}{{\tt 0910.1567}}].

\bibitem{Schuster:2009au}
P.~Schuster, N.~Toro and I.~Yavin, {\it {Terrestrial and Solar Limits on
  Long-Lived Particles in a Dark Sector}},  {\em Phys. Rev.} {\bf D81} (2010)
  016002 [\href{http://arXiv.org/abs/0910.1602}{{\tt 0910.1602}}].

\bibitem{Rothstein:2009pm}
I.~Z. Rothstein, T.~Schwetz and J.~Zupan, {\it {Phenomenology of Dark Matter
  annihilation into a long- lived intermediate state}},  {\em JCAP} {\bf 0907}
  (2009) 018 [\href{http://arXiv.org/abs/0903.3116}{{\tt 0903.3116}}].

\bibitem{Chen:2009ab}
F.~Chen, J.~M. Cline and A.~R. Frey, {\it {Nonabelian dark matter: models and
  constraints}},  {\em Phys. Rev.} {\bf D80} (2009) 083516
  [\href{http://arXiv.org/abs/0907.4746}{{\tt 0907.4746}}].

\bibitem{Redondo:2008ec}
J.~Redondo and M.~Postma, {\it {Massive hidden photons as lukewarm dark
  matter}},  {\em JCAP} {\bf 0902} (2009) 005
  [\href{http://arXiv.org/abs/0811.0326}{{\tt 0811.0326}}].

\bibitem{Voloshin:1985tc}
M.~B. Voloshin, {\it {Once Again About the Role of Gluonic Mechanism in
  Interaction of Light Higgs Boson with Hadrons}},  {\em Sov. J. Nucl. Phys.}
  {\bf 44} (1986) 478.

\bibitem{Truong:1989my}
T.~N. Truong and R.~S. Willey, {\it {Branching ratios for decays of light higgs
  bosons}},  {\em Phys. Rev.} {\bf D40} (1989) 3635.

\bibitem{Kawano:1992ua}
L.~Kawano, {\it {Let's go: Early {U}niverse. 2. Primordial nucleosynthesis: The
  Computer way}}, . FERMILAB-PUB-92-004-A.

\bibitem{Audi:2002rp}
G.~Audi, A.~H. Wapstra and C.~Thibault, {\it {The Ame2003 atomic mass
  evaluation (II). Tables, graphs and references}},  {\em Nucl. Phys.} {\bf
  A729} (2002) 337--676.

\bibitem{Mohr:2005kk}
P.~J. Mohr and B.~N. Taylor, {\it {CODATA-2006 recommended values of the
  fundamental physical constants: 2002}},  {\em Rev. Mod. Phys.} {\bf 77}
  (2005) 1--107.

\bibitem{Descouvemont:2004cw}
P.~Descouvemont, A.~Adahchour, C.~Angulo, A.~Coc and E.~Vangioni-Flam, {\it
  {Compilation and R-matrix analysis of big bang nuclear reaction rates}},
  \href{http://arXiv.org/abs/astro-ph/0407101}{{\tt astro-ph/0407101}}.

\bibitem{Ando:2005cz}
S.~Ando, R.~H. Cyburt, S.~W. Hong and C.~H. Hyun, {\it {Radiative neutron
  capture on a proton at BBN energies}},  {\em Phys. Rev.} {\bf C74} (2006)
  025809 [\href{http://arXiv.org/abs/nucl-th/0511074}{{\tt nucl-th/0511074}}].

\bibitem{Cyburt:2008up}
R.~H. Cyburt and B.~Davids, {\it {Evaluation of Modern
  {${}^3He(\alpha,\gamma){}^7Be$} Data}},  {\em Phys. Rev.} {\bf C78} (2008)
  064614 [\href{http://arXiv.org/abs/0809.3240}{{\tt 0809.3240}}].

\bibitem{2010ApJS..189..240C}
R.~H. {Cyburt}, A.~M. {Amthor}, R.~{Ferguson}, Z.~{Meisel}, K.~{Smith},
  S.~{Warren}, A.~{Heger}, R.~D. {Hoffman}, T.~{Rauscher}, A.~{Sakharuk},
  H.~{Schatz}, F.~K. {Thielemann} and M.~{Wiescher}, {\it {The JINA REACLIB
  Database: Its Recent Updates and Impact on Type-I X-ray Bursts}},  {\em
  \apjs} {\bf 189} (July, 2010) 240--252.

\bibitem{Lopez:1998vk}
R.~E. Lopez and M.~S. Turner, {\it {An accurate calculation of the big-bang
  prediction for the abundance of primordial helium}},  {\em Phys. Rev.} {\bf
  D59} (1999) 103502 [\href{http://arXiv.org/abs/astro-ph/9807279}{{\tt
  astro-ph/9807279}}].

\bibitem{Wilkinson:1982hu}
D.~H. Wilkinson, {\it Analysis of neutron beta decay},  {\em Nucl. Phys.} {\bf
  A377} (1982) 474--504.

\bibitem{PhysRevLett.27.1410}
C.~E. Wiegand and R.~H. Pehl, {\it Measurement of kaonic x rays from $^{4}he$},
   {\em Phys. Rev. Lett.} {\bf 27} (Nov, 1971) 1410--1412.

\bibitem{Onaga:1989dn}
T.~Onaga, H.~Narumi and T.~Kohmura, {\it Hyperon production from kaon
  absorption by he-4},  {\em Prog. Theor. Phys.} {\bf 82} (1989) 222--225.

\bibitem{Common1964465}
A.~K. Common and K.~Higgins, {\it The non-mesic interactions of k- mesons in
  helium},  {\em Nuclear Physics} {\bf 60} (1964), no.~3 465 -- 482.

\bibitem{Arndt:2003fj}
R.~A. Arndt, I.~I. Strakovsky and R.~L. Workman, {\it {The SAID PWA program}},
  {\em Int. J. Mod. Phys.} {\bf A18} (2003) 449--455.

\bibitem{Bellaachia:2002qe}
A.~Bellaachia {\em et.~al.}, {\it {ISAID: A web-based implementation of the
  SAID system}}, . Prepared for NSTAR 2002 Workshop on the Physics of Excited
  Nucleons, Pittsburgh, Pennsylvania, 9-12 Oct 2002.

\bibitem{Breitschopf:2006gn}
J.~Breitschopf {\em et.~al.}, {\it {Pionic charge exchange on the proton from
  40-MeV to 250- MeV}},  {\em Phys. Lett.} {\bf B639} (2006) 424--428
  [\href{http://arXiv.org/abs/nucl-ex/0605017}{{\tt nucl-ex/0605017}}].

\bibitem{Horikawa:1980cv}
Y.~Horikawa, M.~Thies and F.~Lenz, {\it {The delta nucleus spin orbit
  interaction in pi nucleus scattering}},  {\em Nucl. Phys.} {\bf A345} (1980)
  386--408.

\bibitem{PhysRevC.60.024603}
A.~Lehmann, D.~Androi\ifmmode~\acute{c}\else \'{c}\fi{}, G.~Backenstoss,
  D.~Bosnar, T.~Dooling, M.~Furi\ifmmode~\acute{c}\else \'{c}\fi{}, P.~A.~M.
  Gram, N.~K. Gregory, A.~Hoffart, C.~H.~Q. Ingram, A.~Klein, K.~Koch,
  J.~K\"ohler, B.~Kotli\ifmmode~\acute{n}\else \'{n}\fi{}ski, M.~Kroedel,
  G.~Kyle, A.~O. Mateos, K.~Michaelian, T.~Petkovi\ifmmode~\acute{c}\else
  \'{c}\fi{}, M.~Planini\ifmmode~\acute{c}\else \'{c}\fi{}, R.~P. Redwine,
  D.~Rowntree, N.~\ifmmode \check{S}\else \v{S}\fi{}imi\ifmmode \check{c}\else
  \v{c}\fi{}evi\ifmmode~\acute{c}\else \'{c}\fi{}, R.~Trezeciak, H.~Ullrich,
  H.~J. Weyer and M.~Wildi, {\it Total cross sections of the charge exchange
  reaction $(\pi{}+,\pi{})$ on $2$h, $3$he, and $4$he across the
  $\delta{}(1232)$ resonance},  {\em Phys. Rev. C} {\bf 60} (Jun, 1999) 024603.

\bibitem{Blinov:2006rb}
A.~V. Blinov and M.~V. Chadeeva, {\it {Cumulative production of nucleons and
  extremely light nuclei in He-4 p interactions at an incident momentum of 5
  GeV/c}},  {\em Phys. Atom. Nucl.} {\bf 69} (2006) 1439--1447.

\bibitem{Blinov:2}
A.~V. Blinov and M.~V. Chadeeva, {\it {Interactions between 4He nuclei and
  protons at intermediate energies}},  {\em Physics of Particles and Nuclei}
  {\bf 39} (2008) 526--559.

\bibitem{Cyburt:2002uv}
R.~H. Cyburt, J.~R. Ellis, B.~D. Fields and K.~A. Olive, {\it {Updated
  nucleosynthesis constraints on unstable relic particles}},  {\em Phys. Rev.}
  {\bf D67} (2003) 103521 [\href{http://arXiv.org/abs/astro-ph/0211258}{{\tt
  astro-ph/0211258}}].

\bibitem{Protheroe:1994dt}
R.~J. Protheroe, T.~Stanev and V.~S. Berezinsky, {\it {Electromagnetic cascades
  and cascade nucleosynthesis in the early universe}},  {\em Phys. Rev.} {\bf
  D51} (1995) 4134--4144 [\href{http://arXiv.org/abs/astro-ph/9409004}{{\tt
  astro-ph/9409004}}].

\bibitem{1976tper.book.....J}
J.~M. {Jauch} and F.~{Rohrlich}, {\em {The theory of photons and electrons. The
  relativistic quantum field theory of charged particles with spin one-half}}.
\newblock Texts and Monographs in Physics, New York: Springer, 2nd ed., 1976.

\bibitem{1990ApJ...349..415S}
R.~{Svensson} and A.~{Zdziarski}, {\it {Photon-photon scattering of gamma rays
  at cosmological distances}},  {\em \apj} {\bf 349} (Feb., 1990) 415--428.

\end{thebibliography}\endgroup

\end{document}